\documentclass{aastex62}

\usepackage{amsmath}
\usepackage{xcolor}
\usepackage{footmisc}
\usepackage{rotating,longtable}

\received{----, 2019}
\revised{----, 2019}
\accepted{----, 2019}

\shorttitle{Sh 2-305}
\shortauthors{Pandey et. al.}

\begin{document}

\title{Stellar cores in the Sh 2-305 H\,{\sc ii} region}

\correspondingauthor{Rakesh Pandey}
\email{rakesh.pandey@aries.res.in}

\author{Rakesh Pandey}
\affil{Aryabhatta Research Institute of Observational Sciences (ARIES),
Manora Peak, Nainital 263 002, India}
\author[0000-0001-5731-3057]{Saurabh Sharma}
\affil{Aryabhatta Research Institute of Observational Sciences (ARIES),
Manora Peak, Nainital 263 002, India}
\author{Neelam Panwar}
\affil{Aryabhatta Research Institute of Observational Sciences (ARIES),
Manora Peak, Nainital 263 002, India}
\author{Lokesh K. Dewangan}
\affil{Physical Research Laboratory, Navrangpura, Ahmedabad - 380 009, India}
\author{Devendra K. Ojha}
\affil{Tata Institute of Fundamental Research (TIFR),
Homi Bhabha Road, Colaba, Mumbai - 400 005, India}
\author{D. P. Bisen}
\affil{School of Studies in Physics and Astrophysics, 
Pt. Ravishankar Shukla University, Raipur, (C.G.), 492010, India}
\author{Tirthendu Sinha}
\affil{Aryabhatta Research Institute of Observational Sciences (ARIES),
Manora Peak, Nainital 263 002, India}
\author{Arpan Ghosh}
\affil{Aryabhatta Research Institute of Observational Sciences (ARIES),
Manora Peak, Nainital 263 002, India}
\author{Anil K. Pandey}
\affil{Aryabhatta Research Institute of Observational Sciences (ARIES),
Manora Peak, Nainital 263 002, India}



\begin{abstract}
{
Using our deep optical and near-infrared photometry along with multiwavelength archival data, we here present a detailed study of the  Galactic H\,{\sc ii} region Sh 2-305, to understand the
star/star-cluster formation. On the basis of excess infra-red emission, we have identified 116 young stellar objects (YSOs) within a field of view of $\sim18^\prime.5\times18^\prime.5$ around Sh 2-305. The average age, mass and extinction ($A_V$) for this sample of YSOs are 1.8 Myr, 2.9 M$_\odot$ and 7.1 mag, respectively. The density distribution of stellar sources along with minimal spanning tree calculations on the location of YSOs reveals at least three stellar sub-clusterings in Sh 2-305. One cluster is seen toward the center
(i.e, Mayer 3), while the other two are distributed toward the north and south directions. 
Two massive O-type stars (VM2 and VM4; ages $\sim$ 5 Myr) are located at the center of the Sh 2-305 H\,{\sc ii} region.
The analysis of the infrared and radio maps traces the photon dominant regions (PDRs) in the Sh 2-305. Association of younger generation of stars with the PDRs is also investigated in the Sh 2-305. This result suggests that these two massive stars might have influenced the star formation history in the Sh 2-305. This argument is also supported with the calculation of various pressures driven by massive stars, slope of mass function/$K$-band luminosity function, star formation efficiency, fraction of Class\,{\sc i} sources, and mass of the dense gas toward the sub-clusterings in the Sh 2-305.
}        
\end{abstract}




\section{Introduction}

It is believed that most of the stars form in some sorts of clusters or associations of various sizes and 
masses within the giant molecular clouds (GMCs) \citep{2003ARAA..41...57L}. Though the smallest groups are more frequent, however about 70\% - 90\% of all young stars are found in embedded young clusters and groups that are
found in the largest clusters  \citep{2003ARAA..41...57L, 2007prpl.conf..361A, 2017ApJ...842...25G, 2018MNRAS.481.1016G}.  
This hierarchical distribution of star clusters is governed by the fragmentation of the dense gas  
under the influence of gravitational collapse and/or turbulence, dynamical motions of young stars, 
and  other feedback processes. Hence, the distribution of embedded clusters imprints the fractal structure of the GMCs from which they 
born \citep{1978PAZh....4..125E,1986FCPh...11....1S,1996ApJ...471..816E, 2010ApJ...720..541S}.

As star clusters form at the densest part of the hierarchy, they can provide a direct observational signature 
of the star formation process. However, the most important observational constraints in the formation and early evolution of star clusters are structure of the clusters and the molecular gas, the initial mass function (IMF),
 and the star formation history. The structure of the clusters may be analyzed 
 with the spatial distribution of the complete and unbiased sample of member stars \citep{2008MNRAS.389.1209S}. The spacing of the 
 member stars in young clusters can be characterized by the Jeans scale which suggests that a Jeans-like fragmentation process is responsible for the formation of a stellar cluster from a massive  dense core. 
Since the density and temperature (which determine the Jeans length and mass) likely vary among regions, the variation in the characteristic stellar mass of clusters is also expected.  
However, the characteristic mass  of the stellar IMF seems invariant among clusters and even stars in the field, suggesting a mass scale for star formation that is consistent with thermal Jeans fragmentation \citep{2007RPPh...70..337L}.
Though, the low-mass regime of the IMF has been the subject of numerous observational 
and theoretical studies over the past decade \citep[see][]{2014ApJ...784...61O}, 
the universality of the IMF is a question yet to be answered \citep{2008AJ....135.1934S, 2010fsgc.confE.128B, 2012PASJ...64..107S, 2017MNRAS.467.2943S}.  

The feedback processes from the young massive stars also affect the evolution of the young embedded clusters by exhausting the remaining dust and gas, thus slowing down further star formation and the gravitational binding energy.
This feedback limits the star formation efficiency (SFE) and leaves many embedded clusters unbound, with their member stars likely to disperse \citep{2003ARAA..41...57L,2010ApJ...710L.142F, 2014prpl.conf..243K, 2018ApJ...859...68K}. 
In our Galaxy, the embedded-cluster phase lasts only 2-4 Myr and the vast majority of 
young star clusters (YSCs)  
which form in molecular clouds dissolve within 10 Myr or less of their birth. 
This early mortality of YSCs is likely a result of the low SFE 
that characterizes the massive molecular cloud cores within which the clusters form. 
Hence, observing low to modest final SFEs 
are key to understand the early dynamical evolution and infant mortality as well as mass distribution 
of member stars of such objects. \citet{2009ApJS..181..321E} found higher SFE ($\sim$ 30\%) for young stellar objects (YSOs) in the clusters with higher surface density. 
However, in the case of the W5 H\,{\sc ii} region, \citet{2008ApJ...688.1142K} found that the SFE is  $>$10\%-17\% for high surface
density clustering. 
Two of the best probes of these formation and disruption processes are comparison between 
the mass functions (MFs) of molecular clumps and YSCs.  
But the similarity of the mass distribution of embedded clusters to the mass 
distribution of massive cores in GMCs \citep{2003ARAA..41...57L} 
indicates that the SFE and probability of disruption are at most weak functions of mass.
Also, the star formation history of the GMCs remains difficult to constrain due to uncertainties in establishing the ages of young stars \citep{2008ASPC..384..200H}.

With an aim to investigate the stellar clustering and their origin, star formation,
shape of the MF in YSCs, and effects of the feedback from massive stars on these processes, 
we performed a multiwavelength study of the H\,{\sc ii} region `Sh 2-305'
($\alpha$$_{2000}$ =07$^{h}$30$^{m}$03$^{s}$, $\delta$$_{2000}$ = -18$\degr$32$\arcmin$27$\arcsec$).
The size of this H\,{\sc ii} region in optical observations is $\sim 10^\prime\times10^\prime$
and it contains two spectroscopically known O-type stars (O8.5V:VM4 and O9.5:VM2), an embedded cluster ([DBS2003]5) \citep{2003A&A...400..533D}, five infrared
sources, a water maser source, a young open star cluster `Mayer 3', 
and other signatures of active star formation \citep{1975A&A....45..405V,1984A&A...139L...5C,1995A&AS..114..557R}. This region is  part of a large molecular cloud complex ($\sim6^\circ \times 3^\circ$) 
located at a distance of $\sim$4.2 kpc \citep{1995A&AS..114..557R}. Though the distance to Mayer 3 cluster and H\,{\sc ii} region varies from 2.5 to 5.2 kpc \citep[cf.][]{1975A&A....45..405V, 1984A&A...139L...5C, 1995A&AS..114..557R, 2003A&A...397..177B, 2011AJ....141..123A, 2016A&A...585A.101K}, 
in the present work, we have adopted the distance of this H\,{\sc ii} region to 3.7 kpc, 
which has been estimated in the present study (cf. Section 3.3).
Most of the previous works on this region have used only optical data \citep{2013ASInC...9Q.120S,2018AN....339..465T}
 and were focused only on the central cluster i.e., Mayer 3, and there are no detailed studies on star formation 
in this H\,{\sc ii} region.
In the present work, we study the whole H\,{\sc ii} region using deep optical and near-infrared (NIR) photometric data  
along with multiwavelength archival data sets from various surveys (e.g., Gaia, 2MASS, WISE, $Spitzer$, $Herschel$, NVSS)
to identify and characterize a census of YSOs and look for any clues on star/star-cluster formation in this region. 

The organization of the present work is as follows: 
In Section 2, we describe the optical/NIR observations and data reduction 
along with the archival data sets used in our analysis. 
In Section 3 we describe the schemes to study the stellar and YSO number densities, 
membership probability of stellar sources, distance and reddening, $K$-band Luminosity function (KLF)/MF, etc.
The main results of the present study are summarized and discussed in Section 4 and we conclude in Section 5.

\begin{figure*}
\centering
\includegraphics[width=0.75\textwidth]{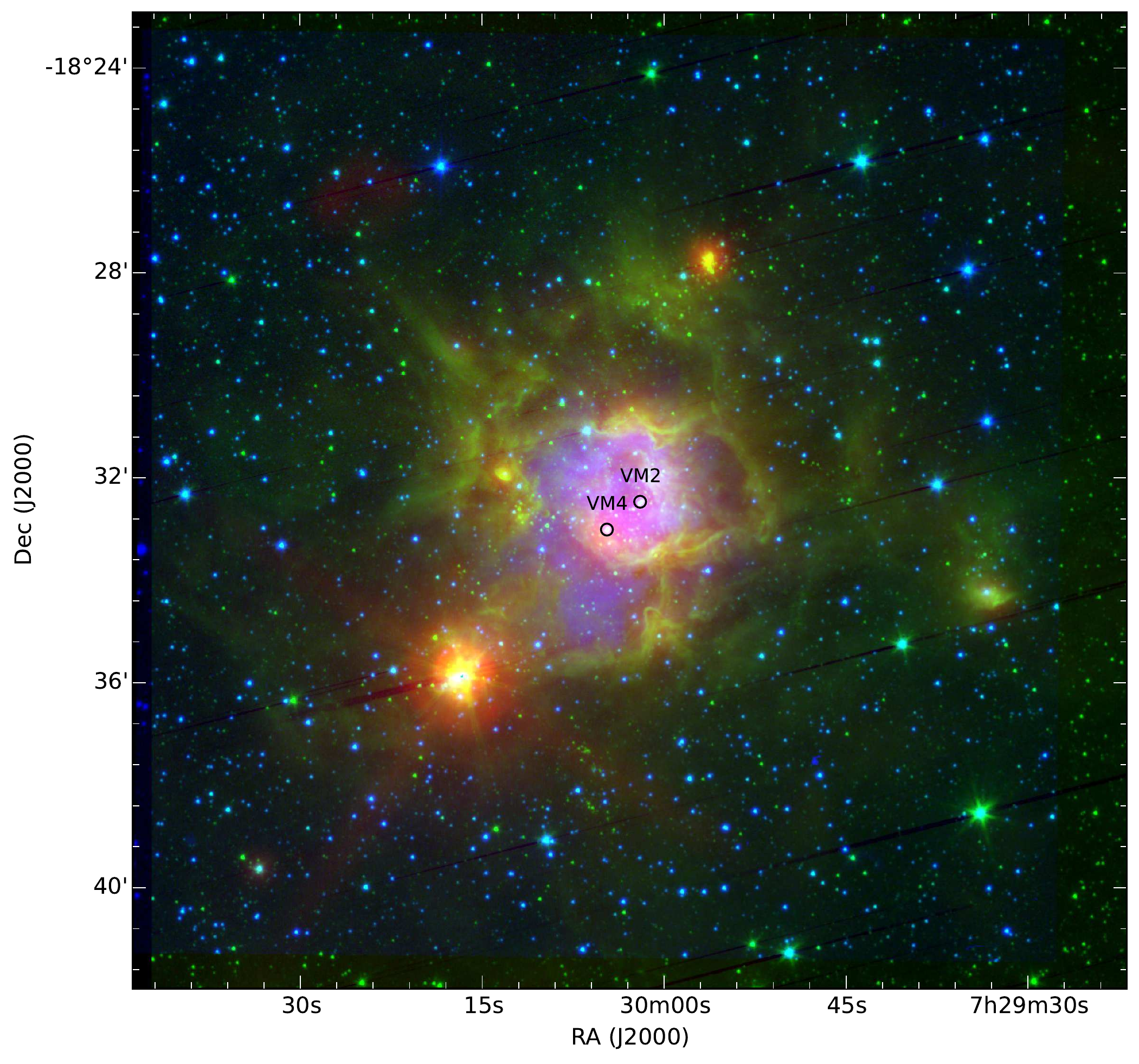}
\caption{\label{image1} Color-composite image obtained by using the  UK Schmidt telescope (UKST)
H$\alpha$ (blue), $Spitzer$ 4.5 $\mu$m (green)
and $WISE$ 22 $\mu$m (red)  images of the $\sim18^\prime.5\times 18^\prime.5$ FOV around Sh 2-305 H\,{\sc ii} region.
The locations of two massive stars VM2 (O9.5) and VM4 (O8.5) are also shown in the figure by black circles \citep{1975A&A....45..405V,1984A&A...139L...5C}.}
\end{figure*}

\section{Observation and data reduction}

\subsection{Optical photometric data}
The broad-band $UBV{(RI)}_c$ optical observations of the Sh 2-305 were  taken using 
2K$\times$2K CCD camera mounted on f/4 Cassegrain focus of the
1.3 m Devasthal fast optical telescope (DFOT) of Aryabhatta Research
Institute of Observational Sciences (ARIES), Nainital, India.
The color-composite image of the observed region of Sh 2-305 obtained by combining the  UK Schmidt telescope (UKST)\footnote{http://www-wfau.roe.ac.uk/sss/halpha/hapixel.html} 
H$\alpha$ (blue color), $Spitzer$\footnote{http://www.spitzer.caltech.edu/} 4.5 $\mu$m (green color), 
and $WISE$\footnote{https://www.nasa.gov/mission\_pages/WISE/main/index.html} 
22 $\mu$m (red color) images is shown in Figure \ref{image1}.
With a pixel size of 13.5 $\mu$m $\times$ 13.5 $\mu$m and a plate scale of  $0^{\prime\prime}$.54 pixel$^{-1}$, 
the CCD covers a field-of-view (FOV) of $\sim18^\prime.5\times18^\prime.5$ on the sky. 
The readout noise and gain of the CCD are 8.29 $e^-$ and 2.2  $e^-$/ADU, respectively. 
The average seeing during the observing nights was $\sim2^{\prime\prime}$.
The log of observations is given in Table \ref{log}.
Along with the object frames, several bias and flat frames were also taken 
during the same night. The broad-band $UBV{(RI)}_c$ observations of the Sh 2-305 were standardized by observing stars in the 
SA 98 field  \citep[$\alpha_{J2000}$: 06$^{h}$52$^{m}$14$^{s}$, $\delta_{J2000}$: -00$\degr$18$\arcmin$59$\arcsec$,][]{1992AJ....104..340L} on the same night.

Initial processing of the data frames (i.e., bias subtraction, flat fielding, etc.) was done using the IRAF\footnote{IRAF is distributed by National Optical Astronomy
Observatories, USA} and ESO-MIDAS\footnote{ ESO-MIDAS is developed and
maintained by the  European Southern Observatory.} data reduction packages. The frames in the same filter were average combined to 
increase the signal-to-noise ratio of the faint stellar sources.
Photometry of the combined frames was carried out by using DAOPHOT-II software \citep{1987PASP...99..191S}.
The point spread function (PSF) was constructed for each frame using several uncontaminated
stars. We used the DAOGROW program for construction of an aperture growth curve required for
determining the difference between the aperture and PSF magnitudes.

\begin{figure*}
\centering
\includegraphics[width=0.45\textwidth]{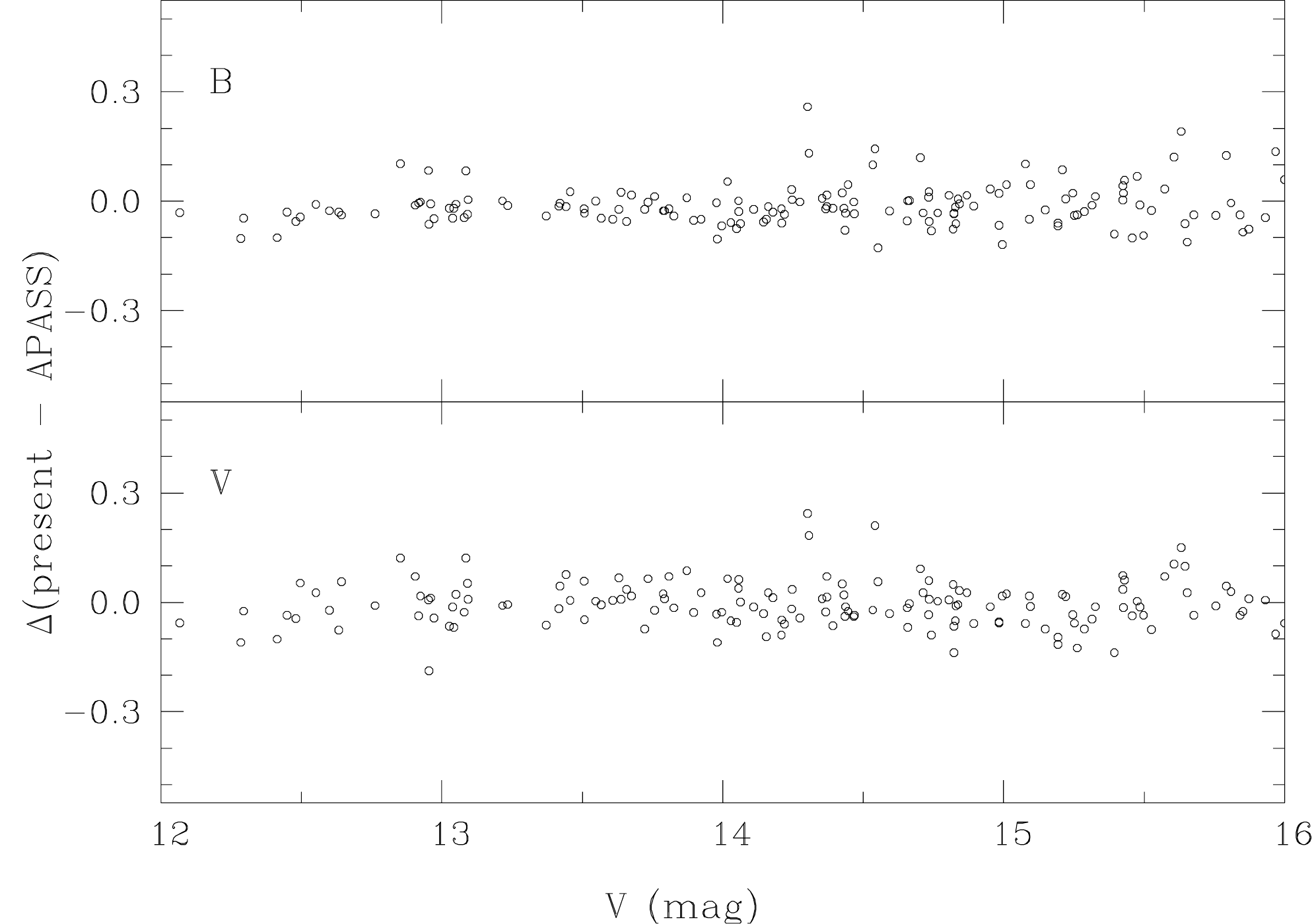}
\includegraphics[width=0.45\textwidth]{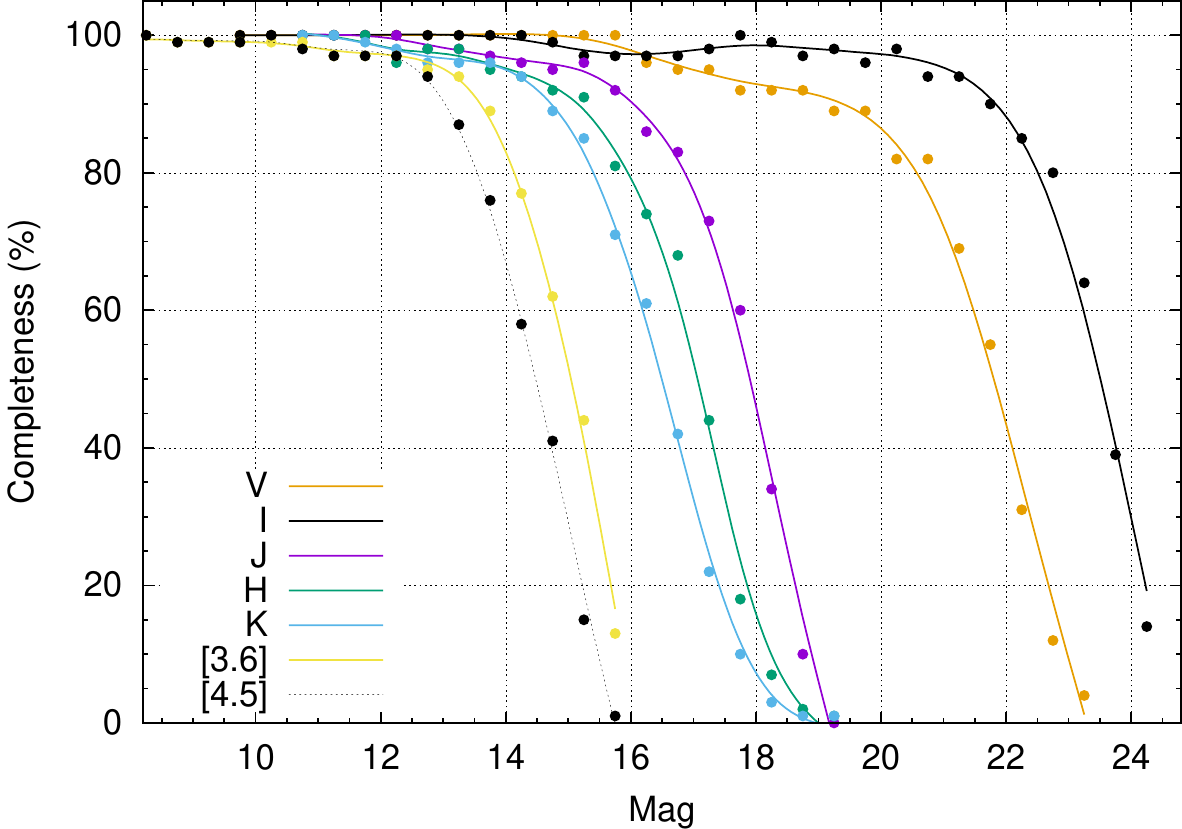}
\caption{\label{comp} Left panel: Comparison between present photometry and that from `APASS' in $V$ and $B$ bands.
Right panel: Completeness levels as a function of magnitude
derived from the artificial star experiments ({\it ADDSTAR}, see Section 2.4).}
\end{figure*}

Using the standard magnitudes of stars located in the SA 98 field, we have calibrated the stellar sources in the Sh 2-305. 
Calibration of the instrumental magnitudes to the standard system was
done by using the procedures outlined by \citet{1992ASPC...25..297S}. 
The calibration equations derived by the least-squares linear regression are as follows:

\begin{equation}
u= U + (4.776\pm0.009) -(0.072\pm0.007)(U-B) + (0.558\pm0.011)X_U,
\end{equation}
\begin{equation}
b= B + (3.01\pm0.006) -(0.147\pm0.005)(B-V) + (0.319\pm0.009)X_B,
\end{equation}
\begin{equation}
v= V + (2.467\pm0.011) +(0.099\pm0.007)(V-I_c) + (0.216\pm0.009)X_V,
\end{equation}
\begin{equation}
r_c= R_c + (2.067\pm0.008) +(0.110\pm0.012)(V-R_c) + (0.148\pm0.006)X_R,
\end{equation}
\begin{equation}
i_c= I_c + (2.592\pm0.006) -(0.032\pm0.004)(V-I_c) + (0.114\pm0.005)X_I
\end{equation}

where $U,B,V,R_c$ and $I_c$ are the standard magnitudes and $u,b,v,r_c$ and $i_c$ are the
instrumental aperture magnitudes normalized for the exposure time and $X's$
are the airmasses. 
The standard deviations of the standardization residual, $\Delta$, between standard and transformed $V$ magnitudes
and $(U-B)$, $(B-V)$, $(V-I)$ and  $(V-R)$ colors of standard stars are 0.006, 0.025, 0.015, 0.015 and 0.015 mag, respectively.
In Figure \ref{comp} (left-hand panel), we show a comparison between the final standard magnitudes from our standardization process and the magnitudes from archive `APASS'\footnote{The AAVSO Photometric All-Sky Survey, https://www.aavso.org/apass}.
It can be seen that there is almost zero difference between the magnitudes.
We have used only those stars for further analyses which are having photometric errors $<$0.1 mag. The photometry of the brightest stars that were saturated in long exposure frames, has been taken from short exposure frames. 
In total, 2646 stars were identified in the $\sim18^\prime.5\times 18^\prime.5$ FOV of Sh 2-305 with  detection limits of
21.92 mag and 19.78 mag in $V$ and $I_c$ bands, respectively.

\subsection{Near-infrared photometric data}

\begin{figure*}
\centering
\includegraphics[width=0.60\textwidth]{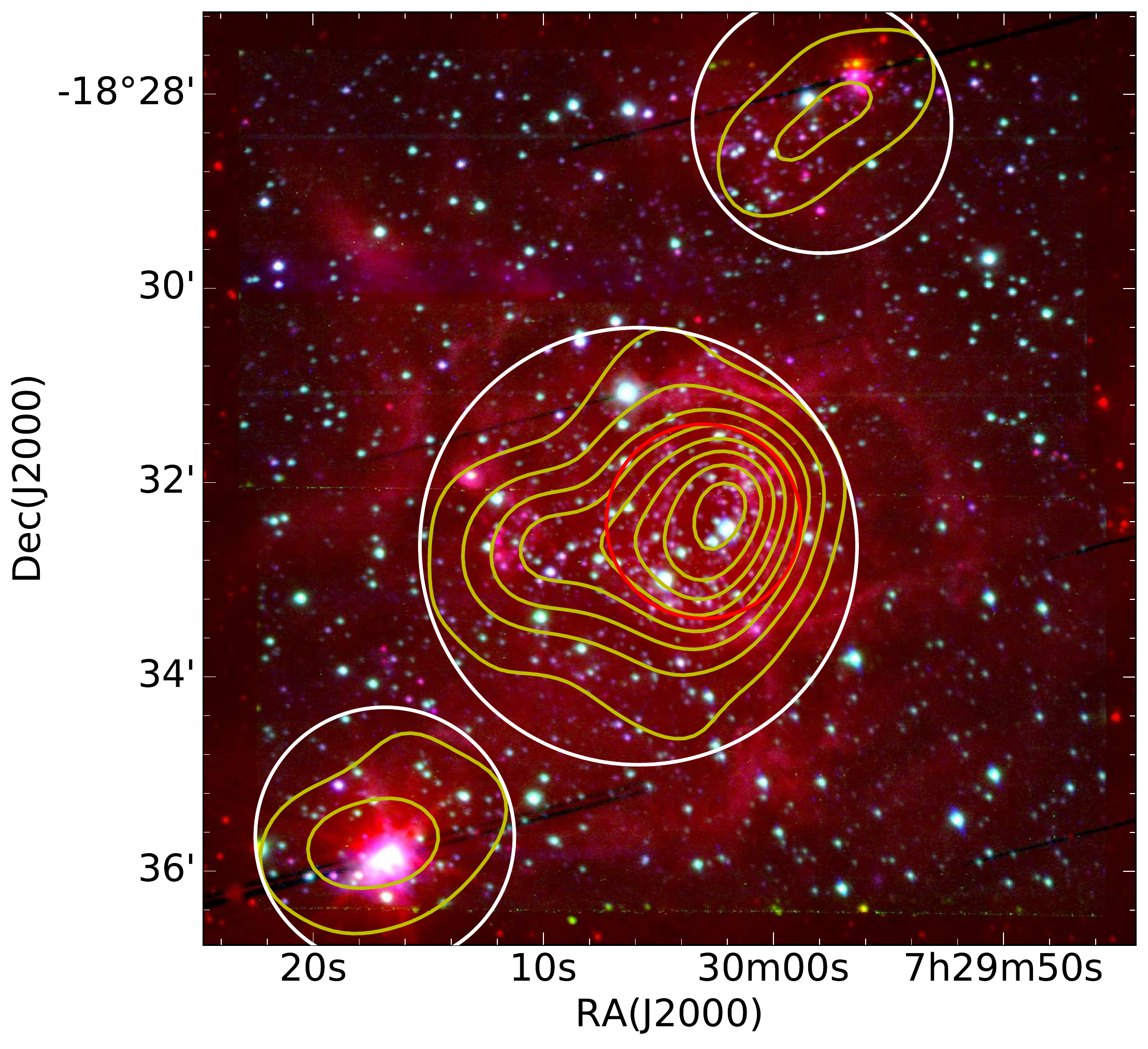}
	\caption{\label{image} Color-composite image (blue: $J$-band, green: $K$-band,
	and red: WISE $4.6$ $\mu$m) of Sh 2-305 ($\sim10^\prime\times 10^\prime$) covered in TIRSPEC NIR observations. 
The white circles represent three sub-clusterings identified in the present analysis.
The red circle is the core region of the central clustering (cf. Section 3.1).
The stellar surface density of the NIR sources  
	generated using nearest neighbor method (cf. Section 3.1) is shown with the yellow contours.}
\end{figure*}

NIR imaging data in $JHK$ bands were taken using TIFR Near Infrared Spectrometer and Imager (TIRSPEC)\footnote{http://www.tifr.res.in/~daa/tirspec/}
mounted on 2~m Himalayan $Chandra$ Telescope (HCT), Hanle, Ladakh, India. 
The detector array in the instrument is $1024 \times 1024$ Hawaii-1 array covering
1 to 2.5 $\mu$m wavelength bands. With 0$^{\prime\prime}$.3 pixel$^{-1}$ resolution,
the instrument provides a FOV of $307^{\prime\prime}\times 307^{\prime\prime}$ in the imaging mode \citep{2014JAI.....350006N}.
As this FOV is not sufficient to cover the entire H\,{\sc ii} region, we covered the entire region of our interest with
five pointings in $J,H$ and $K$ filters.
In each filter, 7 frames of 20 sec exposure were taken and each frame 
was created with 5 dithered images. 
The complete log of observation is given in Table \ref{log}. We followed the usual stpdf for NIR data: dark subtraction, flat-fielding, sky subtraction, alignment and averaging of sky-subtracted
frames for each filter separately. The sky frames were generated by median combining the dithered frames and were subtracted from the science images. 
The final instrumental magnitudes (PSF magnitudes) were determined by the same procedure as done for optical data. Calibration of instrumental magnitudes to 
the standard system was done by using 2MASS point source catalog (PSC) through following transformation equations \footnote{http://indiajoe.github.io/TIRSPEC/Pipeline/}

\begin{equation}
(J-K)= (0.90\pm0.05)\times (j-k)+ (0.69\pm0.02)
\end{equation}
\begin{equation}
(H-K)=(0.98\pm0.02)\times(h-k) + (0.68\pm0.01)
\end{equation}
\begin{equation}
(K-k)= (0.10\pm0.06)\times(H-K) +  (-4.88\pm 0.03)
\end{equation}

where, $JHK$ and $jhk$ are the standard  magnitudes of the stars taken from 2MASS catalog 
and instrumental magnitudes from HCT data, respectively. 
Because of the higher number of detected stars in $K$ and $H$ bands, we have used the $H - K$ color to calibrate the $K$ magnitude. 
Astrometry of the stars was done using the Graphical Astronomy and Image
Analysis Tool\footnote{http://star-www.dur.ac.uk/~pdraper/gaia/gaia.html} with a rms noise of the order of $\sim$0$^{\prime\prime}$.3.
We merged the sources detected in different bands with a matching radius of 1$^{\prime\prime}$. In our final NIR source catalog, we have included only those stars which are detected 
at least in $K$ and $H$ bands and have magnitude uncertainties less than $0.2$ mag. As the stars having $K$ band magnitudes less than 11 mag are saturated in our observations, 
we have taken magnitudes of those stars from the 2MASS catalog.
Our final catalog contains $1812$ stars (upto $K$$\sim$18.1 mag) located in the inner region ($\sim10^\prime\times10^\prime$ FOV) of Sh 2-305 (cf. Figure \ref{image}).

\subsection{Mid infrared photometric data}

Sh 2-305 is observed by the $Spitzer$ space telescope on  2011 August 09 (program 
ID:61071; PI: Whitney Barbara A) with Infrared Array Camera (IRAC) at 3.6 $\micron$ and 4.5 $\micron$. 
We obtained the basic calibrated data (BCD) of the region from the {\it Spitzer} data archive.
The exposure time of each BCD was 5 sec. To create final mosaicked images, 
we used 215 and 218 BCDs in 3.6 $\micron$ and 4.5 $\micron$, respectively.
Mosaicking was performed using the MOPEX software provided by the {\it Spitzer} Science Center. 
All of our mosaics were built at the native instrument resolution of 1$^{\prime\prime}$.2 pixel$^{-1}$ with the standard BCDs.
All the mosaics in different wavelengths are then aligned and trimmed to cover the region 
observed in optical wavelengths ($\sim 18^\prime.5\times18^\prime.5$ around Sh 2-305; cf. Figure \ref{image1}).
These trimmed images have been used for further analyses. 

In case of crowded and nebulous star-forming / H\,{\sc ii} regions, 
we prefer $PSF$ photometry on IRAC data as explained in 
\citet[][]{2012PASJ...64..107S,2016AJ....151..126S,2017MNRAS.467.2943S}
and \citet{2014MNRAS.443.1614P,2017MNRAS.468.2684P}. Therefore,
we used the {\it DAOPHOT} package available with the IRAF photometry routine 
to detect sources and to perform photometry in each IRAC band images. 
The full width at half maxima (FWHM) of each detection is 
measured and all detections with a FWHM $>$ 3$^{\prime\prime}$.6 are considered resolved and removed. 
The detected sources are also examined visually in each band to remove non-stellar objects and 
false detections. Aperture photometry for well isolated sources was done by using an aperture radius of 
3$^{\prime\prime}$.6 with a concentric sky annulus of the inner and outer radii 
of 3$^{\prime\prime}$.6 and 8$^{\prime\prime}$.4, respectively. 
We adopted the zero-point magnitudes for the standard aperture radius (12$^{\prime\prime}$) and 
background annulus of (12$^{\prime\prime}$-22$^{\prime\prime}$.4) of 19.67 and 18.93 in the 
3.6 $\micron$ and 4.5 $\micron$ bands, respectively. 
Aperture corrections were also made by using the values described in IRAC Data 
Handbook \citep{2006AJ....131.1479R}. In order to avoid source confusion due to crowding, the {\it PSF} photometry for all the sources 
was carried out. 
The necessary aperture corrections for
the {\it PSF} photometry were then calculated from the selected isolated sources and 
were applied to the {\it PSF} magnitudes of all the sources. 
The sources with photometric uncertainties $<0.2$ mag in each band were 
considered for further analyses. 
Present photometry is found to be comparable to those available 
in the $Spitzer$ archive\footnote{https://irsa.ipac.caltech.edu/data/SPITZER/GLIMPSE/}.
A total of 2246 sources were detected upto $16.26$ mag and $15.67$ mag in the 3.6 $\micron$ and 4.5 $\micron$ bands, respectively.
The NIR (2MASS and TIRSPEC) and optical counterparts of these IRAC sources were then searched within a matching radius of 1$^{\prime\prime}$.

\subsection{Completeness of the photometric data}

The photometric data may be incomplete due to various reasons, e.g., nebulosity, crowding of the stars, 
detection limit, etc. In particular, it is very important to know the completeness limits  in terms of mass.        
The $IRAF$ routine $ADDSTAR$  was used to determine the 
completeness factor (CF) \citep[for details, see][]{2008AJ....135.1934S}.
Briefly, in this method artificial stars of known magnitudes and positions 
are randomly added in the original frames and then these artificially generated frames are
re-reduced by the same procedure as used in the original reduction. The ratio of the
number of stars recovered to those added in each magnitude gives the CF as a function of magnitude.
The CF as a function of magnitudes in different bands are given in Figure \ref{comp} (right-hand panel).
In Table \ref{cftt}, we have listed the  number of sources detected in different wavelengths along with detection and completeness limits of the multiwavelength photometric data
collected in the present study.

\subsection{Archival data}

We have also used the 2MASS NIR (JHK$_s$) point source catalog (PSC)
\citep{2003yCat.2246....0C} for the Sh 2-305. This catalog is reported to be 99$\%$ complete down 
to the limiting magnitudes of 15.8, 15.1 and 14.3 in the $J$, $H$ and
$K_s$ band, respectively\footnote{http://tdc-www.harvard.edu/catalogs/tmpsc.html}.
We have selected only those sources which have NIR photometric accuracy $<$ 0.2 mag and detection
in at least $H$ and $K_s$ bands. 

The {\it Wide-field Infrared Survey Explorer} (WISE)
is a 40 cm telescope in low-Earth orbit that surveyed
the whole  sky in four mid-infrared bands at 3.4, 4.6, 12, and 22 $\mu$m (namely $W1$, $W2$, $W3$, and $W4$ bands)
with nominal angular resolutions of
6$^\prime$$^\prime$.1, 6$^\prime$$^\prime$.4, 6$^\prime$$^\prime$.5, and 12$^\prime$$^\prime$.0 in the respective bands \citep{2010AJ....140.1868W}.
In this paper, we make use of the AllWISE catalog of the WISE survey data \citep{2010AJ....140.1868W}.
AllWISE catalog is available via IRSA, the NASA/IPAC Infrared Science Archive.
This catalog also includes the 2MASS $JHK_s$ magnitudes of the respective WISE sources.

\section{Results and Analysis}

\subsection{Stellar clustering/groupings in the H\,{\sc ii} region}

A comparison between the stellar density distribution
and the molecular cloud structure can provide the link between star
formation, gas expulsion, and the dynamics of the clusters 
\citep{2004AJ....128.2306C,2005ApJ...632..397G,2006AJ....132.1669S}.
To study the stellar surface density distribution in the Sh 2-305, 
we have  generated  surface density maps by performing nearest neighbor (NN) method on the 2MASS 
NIR catalog. We varied the radial distance in order to encompass the 20$^{th}$ nearest star detected in 2MASS
and computed the local surface density in a grid size of 6$^{\prime\prime}$.
The density contours derived by this method are plotted in Figure \ref{image} as yellow curves
smoothened to a grid of size $3\times3$ pixels.
The lowest contour is 1$\sigma$ above the mean of stellar density (i.e. 7 stars/pc$^2$ at a distance of $3.7$ kpc, see Sec. 3.3)
and the step size is equal to  1$\sigma$ (2.5 stars/pc$^2$ at $3.7$ kpc).
The isodensity contours easily isolate the central sub-clustering (i.e., Mayer 3) of stars along with 
two highly elongated sub-structures  in north-west and south-east directions of the central cluster.

The three sub-clusterings identified are encircled in Figure \ref{image}. 
The radius of the central sub-clustering i.e., `Mayer 3 cluster' is 2$^{\prime}$.25, 
whereas for the other two sub-clusterings it is 1$^{\prime}$.3. 
The stellar density in the  central cluster `Mayer 3' region 
is significantly higher than for the other two sub-clusterings
and its peak seems to be slightly off-center to the cluster. The stellar core  region of  `Mayer 3' cluster is also
shown by a red circle of $\sim1^{\prime}$ radius in  Figure \ref{image}.
The central coordinates of the circular area for the north-west, center and south-east sub-clusterings are 
$\alpha_{2000}$: 07$^{h}$29$^{m}$57$^{s}$.9, $\delta_{J2000}$: -18$\degr$28$\arcmin$18$\arcsec$; 
$\alpha_{2000}$: 07$^{h}$30$^{m}$05$^{s}$.9, $\delta_{J2000}$: -18$\degr$32$\arcmin$39$\arcsec$; and
$\alpha_{2000}$: 07$^{h}$30$^{m}$16$^{s}$.9, $\delta_{J2000}$: -18$\degr$35$\arcmin$39$\arcsec$, respectively.
We also define a bigger circular area of radius 5$^{\prime}$.65 centered at
$\alpha_{2000}$: 07$^{h}$30$^{m}$05$^{s}$.9, $\delta_{J2000}$: -18$\degr$32$\arcmin$15$\arcsec$
as a boundary of the Sh 2-305 H\,{\sc ii} region (whole region), which encloses all of these sub-clusterings.

\begin{figure*}
\centering
\includegraphics[width=9cm]{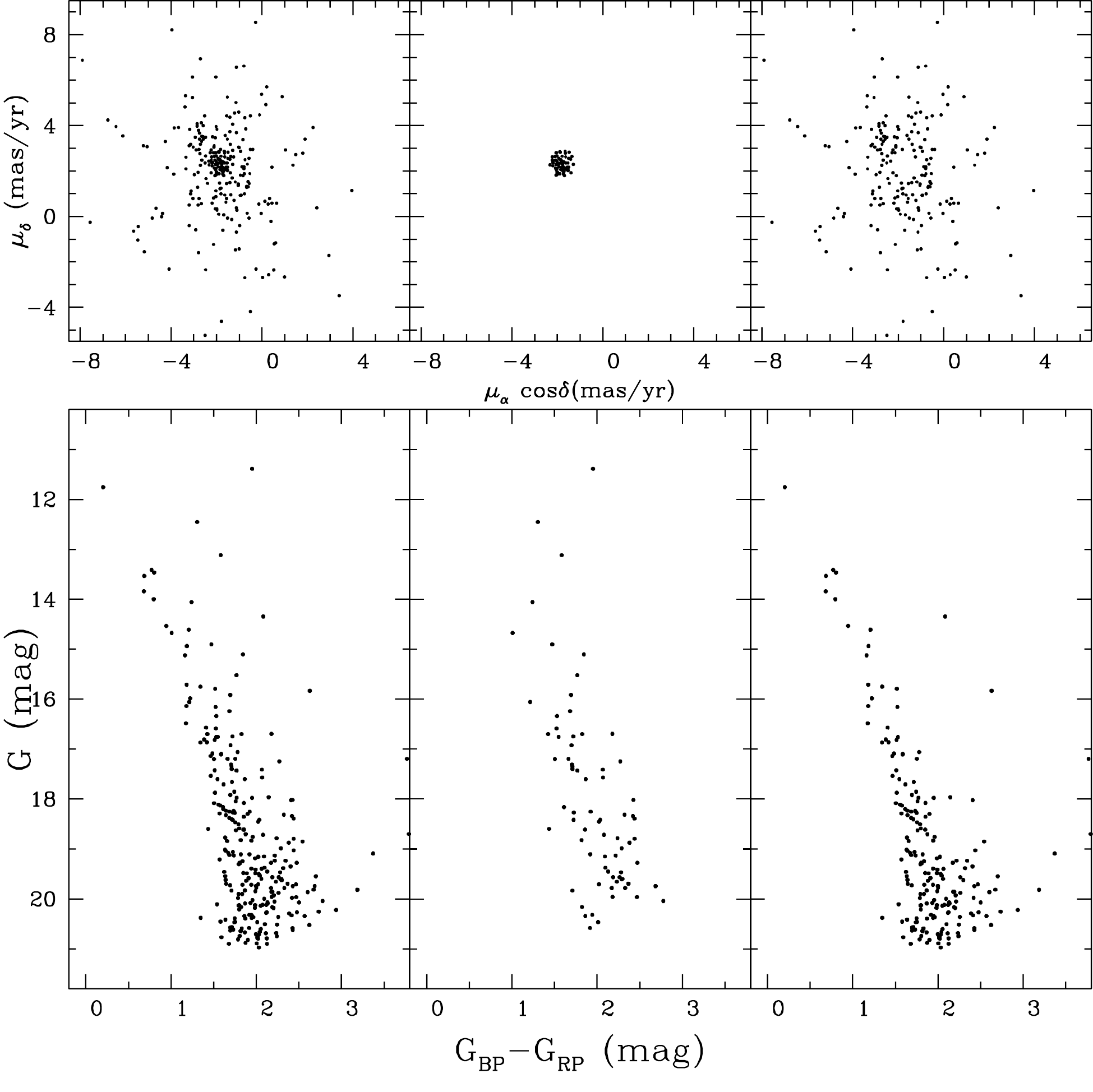}
\includegraphics[width=8cm,height=9cm]{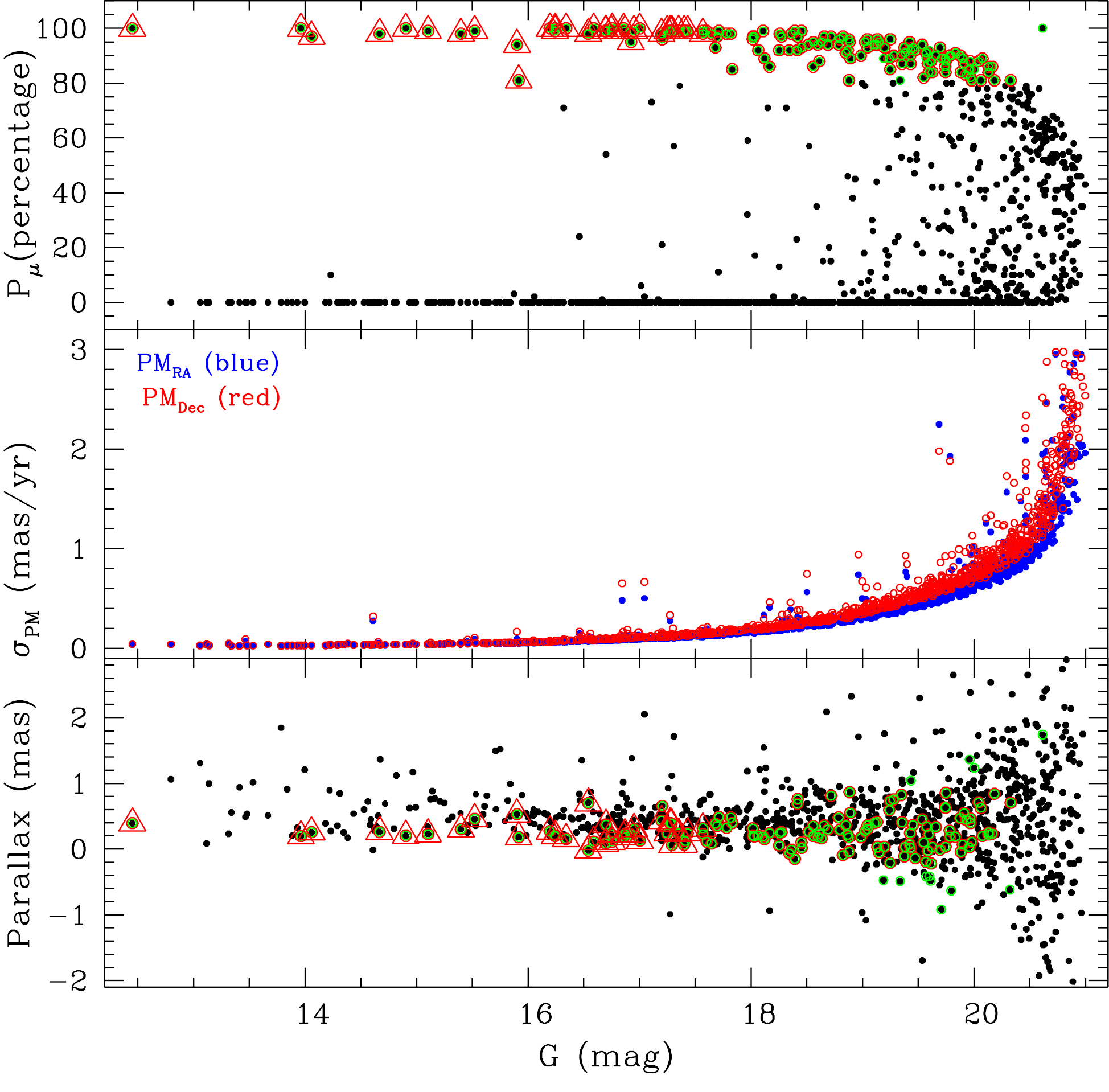}
\caption{\label{pm1} Left panel: PM vector-point diagrams (VPDs; top sub-panels) and
{\it Gaia} DR2 $G$ vs. $(G_{BP} - G_{RP})$ CMDs (bottom sub-panels) for  stars located in the Mayer 3 cluster region.
The left sub-panels show all stars, while the middle and right sub-panels show
the probable cluster members and field stars.
Right panel: Membership probability P$_\mu$, {\bf PM errors $\sigma_{PM}$} and parallax of stars as 
a function of $G$ magnitude for stars in the Mayer 3 cluster region.
The probable member stars (P$_\mu>$80 \%) are shown by green circles while 35 members of the Sh 2-305 
having parallax values with good accuracy taken from \citet{2018AJ....156...58B}, are shown by red triangles.
}
\end{figure*}

\subsection{Membership probability of stars in the Mayer~3 cluster}

Gaia DR2  has opened up the possibility of an entirely new perspective on the problem of
membership determination in cluster studies by
providing the new and precise parallax measurements upto very faint limits\footnote{https://www.cosmos.esa.int/web/gaia/dr2}.
As there is a clear clustering  of stars in the central region of Sh 2-305 (Mayer 3), 
$Gaia$  proper motion (PM) data located within this region (cf. Section 3.1, radius $<$ 2$^{\prime}$.25)
and having PM error $\sigma_{PM}<$3 mas/yr
are  used to determine membership probability of stars located in this region.
 Proper motions (PMs), $\mu_\alpha$cos($\delta$) and $\mu_\delta$, are plotted as vector-point diagrams (VPDs) in the top sub-panels of
Figure \ref{pm1} (left panel). The bottom sub-panels show the corresponding 
$G_{(330-1050 nm)}$ versus $G{_{BP (330-680 nm)}} - G_{{RP(630-1050 nm)}}$
Gaia color-magnitude diagrams (CMDs). 
The left sub-panels show all stars, while the middle and right sub-panels show
the probable cluster members and field stars. A circular area of
a radius of 0.6 mas yr$^{-1}$ around the cluster centroid in the VPD of PMs has been selected visually to define
our membership criterion. The chosen radius is a compromise between
loosing cluster members with poor PMs and including field
stars sharing mean PM.  The CMD of the most probable
cluster members is shown in the lower-middle sub-panel.
The lower-right sub-panel represents the CMD for field stars. Few cluster members are visible in
this CMD because of their poorly determined PMs.
The tight clump centering at $\mu_{xc}$ = -1.87 mas yr$^{-1}$, $\mu_{yc}$ = 2.31 mas yr$^{-1}$ and radius = 0.6 mas yr$^{-1}$
in the top-left sub-panel represents the cluster stars, and a broad distribution is seen for the probable 
field stars.
Assuming a distance of $\sim$ 3.7 kpc (cf. Section 3.3) and a
radial velocity dispersion of 1 kms$^{-1}$ for open clusters \citep{1989AJ.....98..227G}, the expected dispersion ($\sigma_c$) in PMs of the cluster would be $\sim$0.06 mas yr$^{-1}$.
For remaining stars (probable field stars), we have calculated:
 $\mu_{xf}$ = -2.02 mas yr$^{-1}$, $\mu_{yf}$ = 3.31 mas yr$^{-1}$, 
$\sigma_{xf}$ = 1.77 mas yr$^{-1}$ and $\sigma_{yf}$ = 3.57 mas yr$^{-1}$.
These values are further used to construct the frequency distributions of cluster stars ($\phi_c^{\nu}$) and field stars ($\phi_f^{\nu}$) by using the equations given in \citet{2013MNRAS.430.3350Y} and then the value of
membership probability (ratio of distribution of cluster stars with all the stars) of all the stars within Sh 2-305 (Section 3.1, radius$<$ 5$^\prime$.65), is given by using the following equation:

\begin{equation}
P_\mu(i) = {{n_c\times\phi^\nu_c(i)}\over{n_c\times\phi^\nu_c(i)+n_f\times\phi^\nu_f(i)}}
\end{equation}
where $n_c$ (=0.26) and $n_f$(=0.74) are the normalized number of stars for the cluster
and field ($n_c$+$n_f$ = 1), respectively. The membership probability estimated as above, errors in the PM, and parallax values are plotted as a function of 
$G$ magnitude in Figure \ref{pm1} (right-hand panel).
As can be seen in this plot, a high membership probability (P$_\mu >$ 80 \%)
extends down to $G\sim$20 mag. At brighter magnitudes, there is a  clear separation between cluster
members and field stars supporting the effectiveness of this technique. Errors in PM become very high at faint limits and
the maximum probability gradually decreases at those levels.
Except few outliers, most of the  stars with high membership probability (P$_\mu >$ 80 \%) are
following a tight distribution.
Finally, from the above analysis, we were able to calculate membership probability of 1000 stars in
the Sh 2-305. Out of these, 137 stars were considered as  members of the Sh 2-305 
based on their high probability P$_\mu$ ($>$80 \%)
and the parallax values (black dots with green circles around them, shown in Figure \ref{pm1} right-hand panel).
The details of these member stars are given in Table \ref{PMT}. 131 of these member stars have optical counterparts from the present photometry,
identified within a search radius of one arcsec.

\begin{figure*}
\centering
\includegraphics[width=0.45\textwidth]{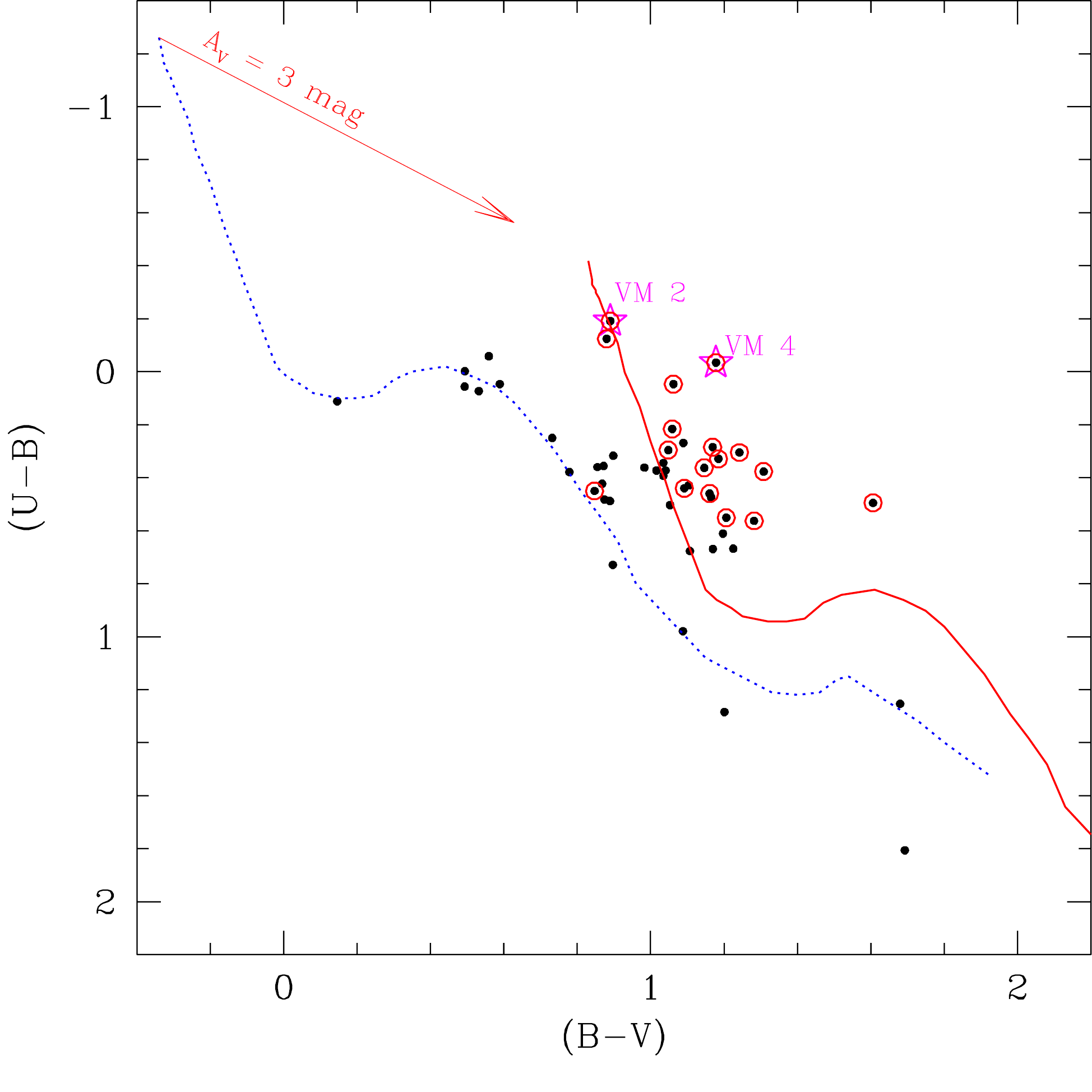}
\includegraphics[width=0.45\textwidth]{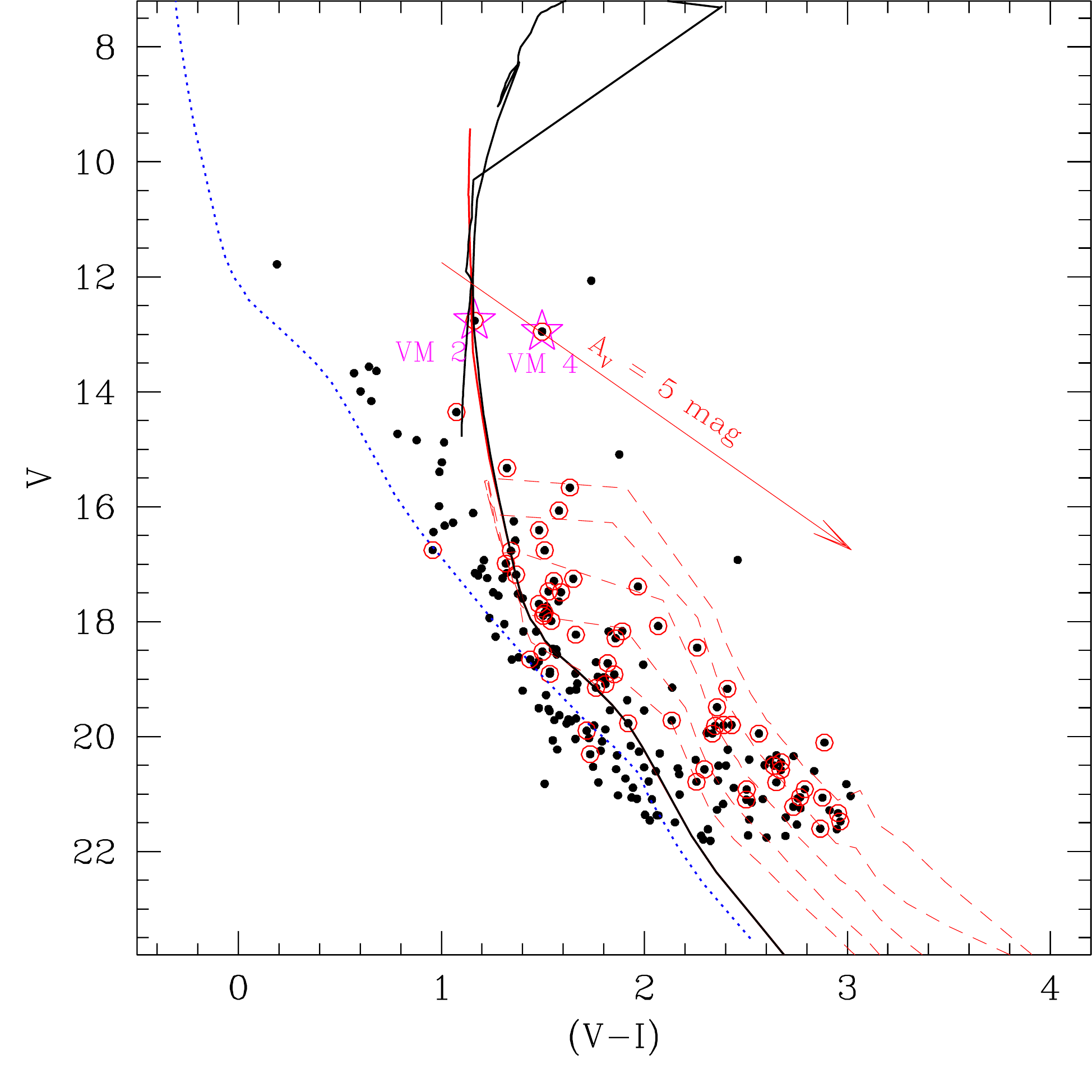}
\caption{\label{ccd} 
Left panel: $(U-B)$ vs. $(B-V)$ TCD for the sources in the Mayer 3 cluster region (radius $<$ 2$^\prime$.25, black dots).
The identified member stars using PM analysis are also plotted by red circles. 
The dotted blue curve represents the intrinsic ZAMS for $Z=0.02$ by \citet{Schmidt-Kaler1982}. The red continuous curve represents the ZAMS shifted along the reddening vector (see text for details) by $E(B-V)_{cluster}$ = 1.17 mag for the stars associated with the cluster. Right panel: $V$ vs. $(V-I)$ CMD for the same sources. The MS isochrone for 1 Myr (red curve) and 5 Myr (black curve) by \citet{2008AA...482..883M} and PMS isochrones of 0.5, 1, 2, 5, 10 Myr by \citet[][dashed red curves]{2000AA...358..593S}, corrected for the distance of 3.7 kpc and reddening $E(B-V)_{cluster}=1.17$ mag, are also shown. The blue dotted curve is the 1 Myr MS isochrone by \citet{2008AA...482..883M} corrected for the distance of 1.2 kpc and zero reddening.  We have also plotted the location of two massive stars by star symbols in the figures.
}
\end{figure*}

\subsection{Reddening, distance  and age of the Mayer 3 cluster}

The reddening and distance of a cluster can be derived quite accurately by using
the two-color diagrams (TCDs) and CMDs of their member stars \citep[cf.,][]{1994ApJS...90...31P,2006AJ....132.1669S}. 
Left-hand panel of Fig.~\ref{ccd} shows the $(U-B)$ versus $(B-V)$ TCD  with the intrinsic zero-age-main-sequence 
(ZAMS) shown in blue dotted curve, taken from  \citet{Schmidt-Kaler1982}, along with stars located within the 
boundary of the central cluster Mayer 3 (i.e., radius$<$ 2$^{\prime}$.25) denoted by black dots.
We have also over-plotted the member stars identified by using PM data as red circles.
The distribution of the stars shows a large spread along the reddening vector, indicating 
heavy differential reddening in this region.
It reveals two different populations, one (mostly black dots) distributed along the ZAMS 
and another (consisting mostly red circles) showing a large spread in their reddening value.       
The former having negligible reddening must be the foreground population and the latter could be member stars.
If we look at the MIR image of this region (Figure \ref{image1}), 
we see  several dust lanes along with enhancements
of nebular emission at several places. 
Both of them are likely responsible for the large spread in the reddening of member stars  population.
The ZAMS from \citet{Schmidt-Kaler1982} is  shifted along the reddening vector with a slope of
$E(U - B)/E(B - V)$ = 0.72 (corresponding to $R_V$ $\sim$ 3.1, cf. Appendix A) 
to match the distribution of stars showing the minimum 
reddening among the member stars population (dotted curve).  
The other member population may be embedded in the nebulosity of this H\,{\sc ii} region.
The foreground reddening value, $E(B-V)_{cluster}$, thus comes to be $\sim$ 1.17 mag 
and the  ZAMS reddened by this amount is shown by a red continuous curve. 
The approximate error in the reddening measurement `$E(B-V)$' is $\sim$ 0.1 mag, 
as has been determined by the procedure  outlined in \citet{1994ApJS...90...31P}.

In the literature, the distance estimation of Mayer 3 cluster varies from 2.5 to 5.2 kpc 
\citep[cf.][]{1975A&A....45..405V, 1984A&A...139L...5C, 1995A&AS..114..557R, 2003A&A...397..177B, 2011AJ....141..123A, 2016A&A...585A.101K}
which are derived both photometrically  and spectroscopically.
To confirm the distance to this cluster, we have used the  $V$ versus $(V - I)$ CMD,  
generated from our deep optical photometry of the stars located in the cluster region (radius $<$ 2$^{\prime}$.25, black dots), 
as shown in Figure \ref{ccd} (right-hand panel). The distribution of member stars
identified using PM data analysis (red circles) have also been shown in the CMD. Here also, the CMD reveals two different populations, one (mostly black dots) for
foreground stars having almost zero reddening value (near dotted curve) and another (mostly red circles)
for the cluster members at higher reddening value and a larger distance.  

The CMD for cluster members displays a  few main-sequence (MS) stars upto $V\sim$18 mag and
pre-main sequence (PMS) stars at fainter end. 
The dotted curve represents a ZAMS isochrone derived from \citet[][age=1 Myr]{2008AA...482..883M},
randomly corrected for a distance of 1.2 kpc, which is matching well with the foreground stars.
We can further visually fit this MS isochrone to the
distribution of member stars quite nicely, which is corrected for extinction ($E(B-V)_{cluster}$=1.17 mag)
and  distance ($\sim$ 3.7 kpc; solid red curve).
The dashed red curves are the PMS isochrones of 0.5, 1, 2, 5 and 10 Myrs by \citet{2000AA...358..593S},
corrected for the same distance (3.7 kpc) and extinction value ($E(B - V )_{cluster}$ = 1.17 mag).

An upper limit to the age of the cluster can be established 
from the most massive member star. The location of most massive star VM4 (O8.5)
in the  $V$ versus $(V - I)$ CMD is traced back along the reddening 
vector to the  turn-off point in the MS, which is equivalent to a 5 Myr old isochrone
 (cf. black curve in Figure \ref{ccd} right panel). 
Assuming a coeval star-formation event, the oldest stellar content 
in the Mayer 3 cluster must therefore be younger than 5 Myr.

To confirm further and establish the relation between the central cluster `Mayer 3' and the Sh 2-305 H\,{\sc ii} region,
we have calculated the mean of the distances of  35 members of the Sh 2-305 
having parallax values with good accuracy (i.e., error$<$ 0.1 mas) as 3.7$\pm$1.1 kpc \citep{2018AJ....156...58B}.
Clearly, this value is in agreement with distance of the central cluster Mayer 3 derived earlier,
indicating that both cluster and the H\,{\sc ii} region Sh 2-305 are associated to each other at similar distance of 3.7 kpc.

\begin{figure*}
\centering\includegraphics[width=0.3\textwidth]{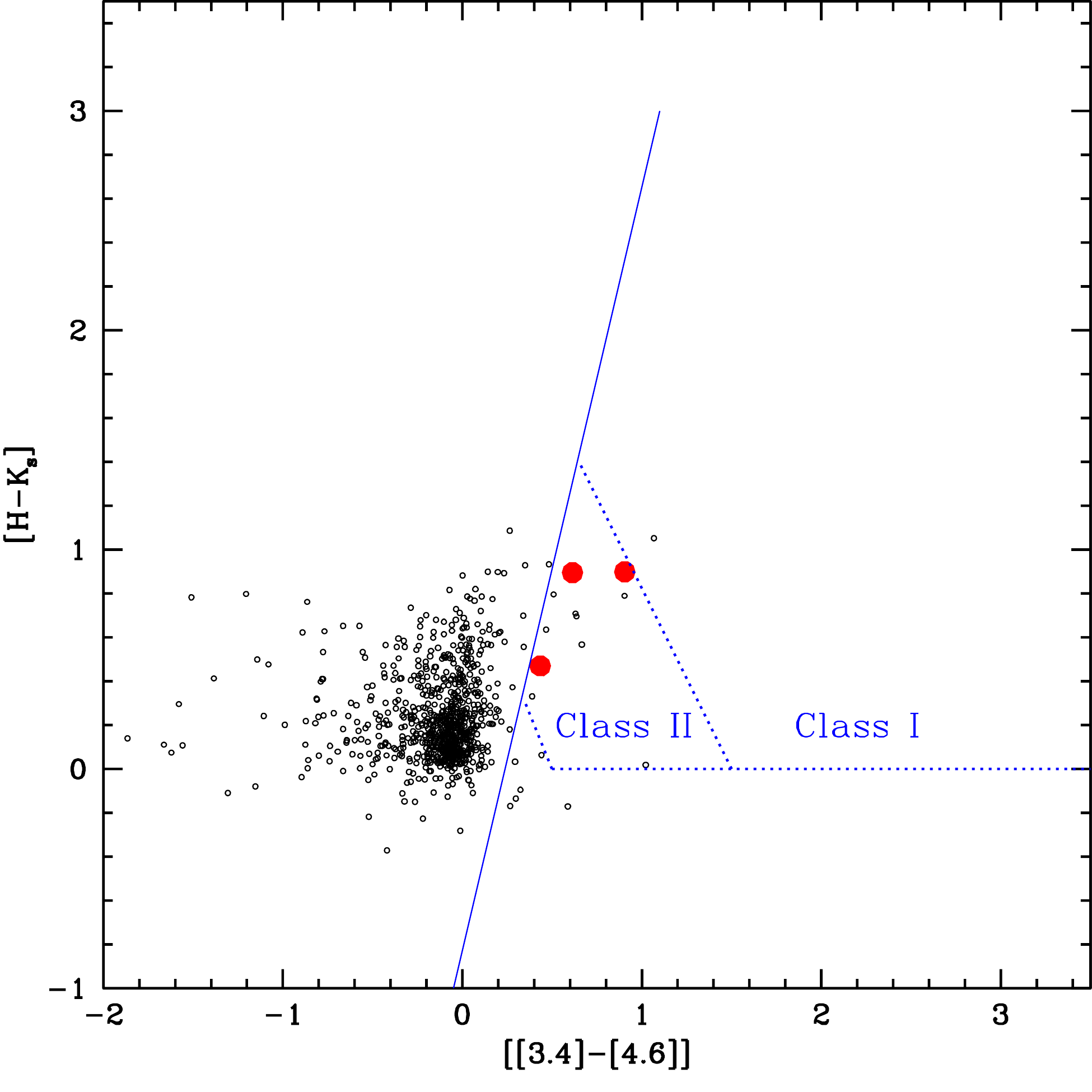}
\centering\includegraphics[width=0.3\textwidth]{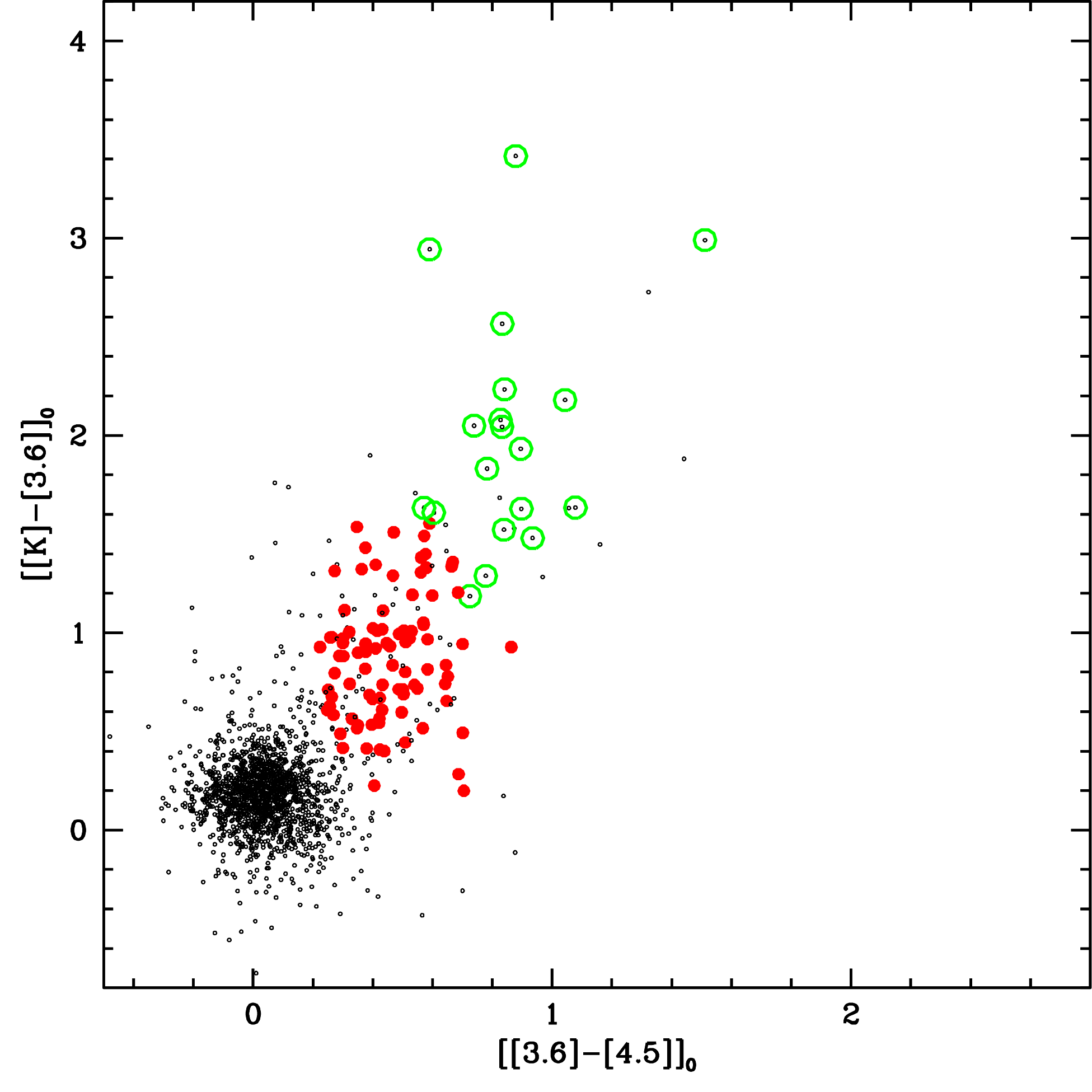}
\centering\includegraphics[width=0.3\textwidth]{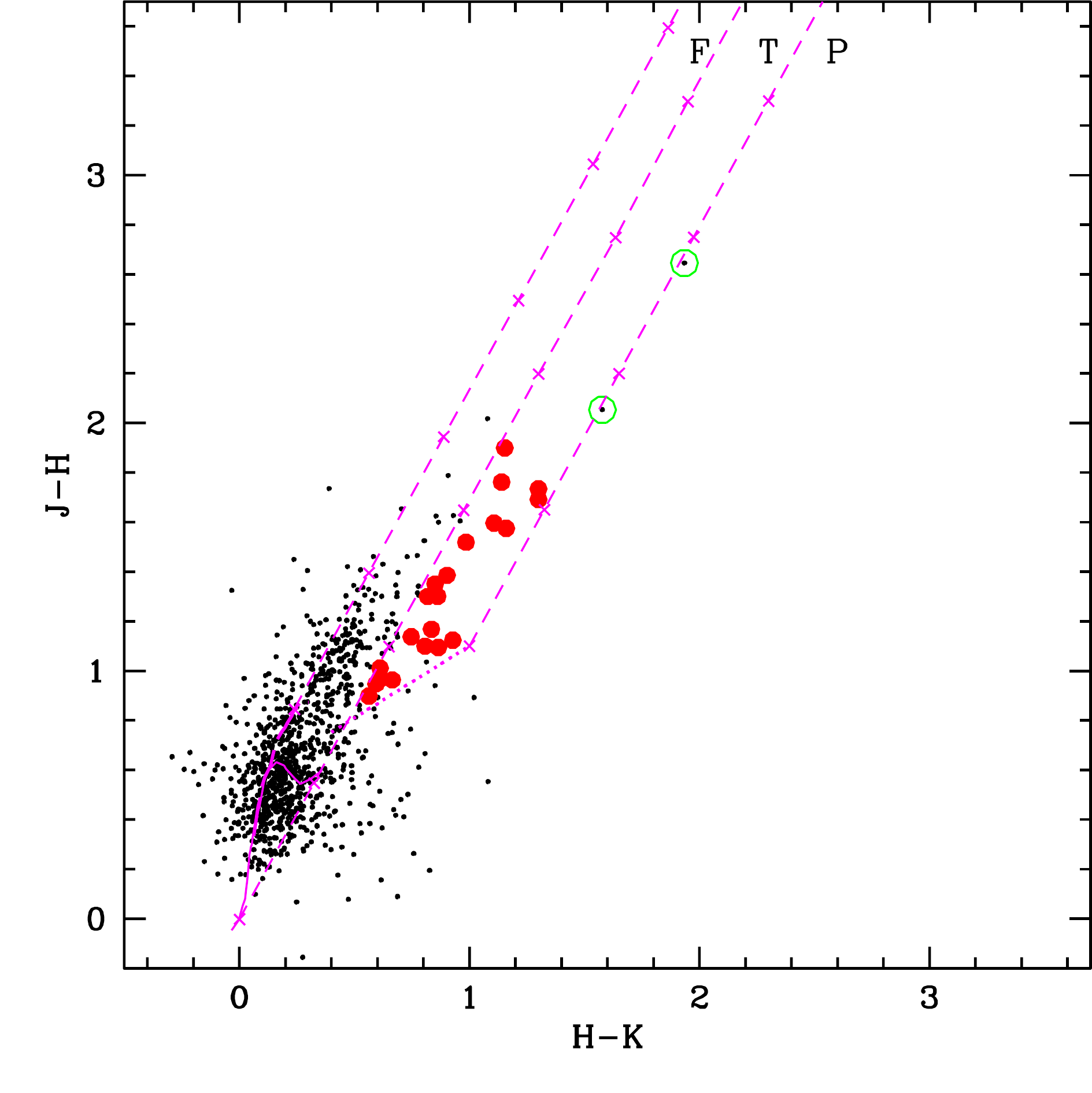}
\caption{\label{wise}  Left panel:
 ${[[3.4] - [4.5]]}$ vs. ${[H - K_s]}$ TCD of all the sources identified within the selected $\sim18^\prime.5\times18^\prime.5$ FOV
of the Sh 2-305 region. The YSOs are classified as  Class~I and Class~II based on the color criteria by \citet{2014ApJ...791..131K}.
Middle panel: ${[[K] - [3.6]]}_0$ vs. ${[[3.6] - [4.5]]}_0$ TCD for the sources in the same FOV.
The YSOs classified as Class~I and Class~II, are based on the color criteria by \citet{2009ApJS..184...18G}.
Right panel:  ${[H-K]}$ vs. ${[J - H]}$ TCD for the sources in the same FOV \citep{2004ApJ...608..797O}. 
The continuous and thick red dashed curves represent the reddened MS and giant branches \citep{1988PASP..100.1134B}, 
respectively. The dotted red line indicates the locus of dereddened CTTSs \citep{1997AJ....114..288M}. 
The parallel red dashed lines are the reddening lines drawn from the tip (spectral type M4) of the 
giant branch (left reddening line), from the base (spectral type A0) of the MS branch (middle reddening
line) and from the tip of the intrinsic CTTS line (right reddening line). 
The crosses on the reddening lines show an increment of $A_V$ = 5 mag. 
The YSOs classified as Class\,{\sc i} and Class\,{\sc ii} are shown with red dots  and  green circles, respectively.  
 }
\end{figure*}

\subsection{Identification of  YSOs \label{idf} in the Sh 2-305 region}

Identification and characterization of YSOs in 
star-forming regions (SFRs) hosting massive stars are essential stpdf to examine the
physical processes that govern the formation of the next generation stars in such regions.
In this  study, we have used NIR and MIR observations of the Sh 2-305
($\sim$ $18^\prime.5\times18^\prime.5$ FOV) to identify candidate YSOs based on their excess IR emission. The identification and classification schemes are described as below.

\begin{itemize}

\item
Using AllWISE catalog of WISE MIR data:
The location of WISE bands in the MIR matches where the excess emission from cooler 
circumstellar disk/envelope material in young stars begins to
become significant in comparison to the stellar photosphere. 
The procedures outlined in \citet{2014ApJ...791..131K} have been used to identify YSOs in this region.
We refer figure 3 of \citet{2014ApJ...791..131K} to summarize the entire scheme.
This method includes photometric quality criteria for different WISE bands 
as well as the selection of candidate extragalactic contaminants 
(AGN, AGB stars and star-forming galaxies).
Figure \ref{wise} (left panel) shows the  $[[3.4] - [4.6]]$ versus  $(H - K_s)$ TCD
for all the sources in the region, where 3 probable YSOs shown by red dots are classified as Class\,{\sc ii} source.

\item
Using $Spitzer$ MIR data:
As we have  photometric data for two IRAC bands 3.6 $\mu$m and 4.5 $\mu$m, this along with $K$ band is used
to plot the de-reddened [[3.6] - [4.5]]$_0$ versus [K - [3.6]]$_0$ TCD as shown in  Figure \ref{wise} (middle panel). 
The procedure outlined in \citet{2009ApJS..184...18G}  has
been used to de-redden and classify sources as Class\,{\sc i} (green circles, 20) and Class\,{\sc ii} (red dots, 95) YSOs.
This method also includes photometric quality criteria for different IRAC bands 
as well as the selection of candidate contaminants (PAH galaxies and AGNs).

\item
Using TIRSPEC and 2MASS NIR data:
We combined TIRSPEC NIR photometry (cf. Section 2.2) with the 2MASS catalog to 
make a final catalog covering the $18^\prime.5\times18^\prime.5$ FOV of the selected Sh 2-305 region. 
Then those stars whose corresponding
counterparts were found in $Spitzer$ or WISE catalog were removed from this  catalog
to further identify YSOs which are not detected in the  WISE or $Spitzer$ photometry, 
by using NIR TCD \citep{2004ApJ...608..797O}. 
In Figure \ref{wise} (right panel)  we have plotted NIR TCD of the above stars. The solid and thick broken curves represent the unreddened 
MS and giant branches \citep{1988PASP..100.1134B}. The dotted line indicates the locus of unreddened 
Classical T Tauri Stars (CTTS) \citep{1997AJ....114..288M}. The parallel dashed lines are the reddening vectors drawn from the tip 
of the giant branch (upper reddening line), from the base of the MS branch (middle reddening line) and from the tip of the intrinsic CTTS line (lower reddening line).
 We classified the sources according to their locations in the diagram \citep{2004ApJ...608..797O,2007MNRAS.380.1141S,2008MNRAS.383.1241P}. The sources occupying the location between the upper and middle reddening lines (`F' region) are considered to be 
either field stars or Class\,{\sc iii} sources and/or Class\,{\sc ii} sources with small NIR excess. 
The sources located between the middle and lower reddening lines (`T' region) are
 considered to be mostly CTTS (or Class\,{\sc ii} sources) with large NIR excess. 
However, we note that there may be overlap of the Herbig Ae/Be stars in the `T' region \citep{2002astro.ph.10520H}. 
Sources that are located in the region redward of the lower reddening vector (`P' region) 
most likely are Class\,{\sc i} objects. In Figure \ref{wise}, we have shown the location
of identified Class\,{\sc i} sources with green circles (2) and Class\,{\sc ii} objects with red dots (21). 
\end{itemize}

\noindent
Finally, we have merged all the YSOs identified  based on their IR excess emission by using different schemes to 
have a final catalog of 116 YSOs in the $\sim18^\prime.5\times18^\prime.5$ FOV around Sh 2-305.
The positions, magnitudes in different NIR/MIR bands of these YSOs, along with their classification are given in Table \ref{data1_yso}.
Optical counterparts for $28$ of these YSOs were also identified by using a search radius of 1$^{\prime\prime}$ and
their optical magnitudes and colors are given in Table \ref{data3_yso}.

\begin{figure*}
\centering\includegraphics[width=0.45\textwidth]{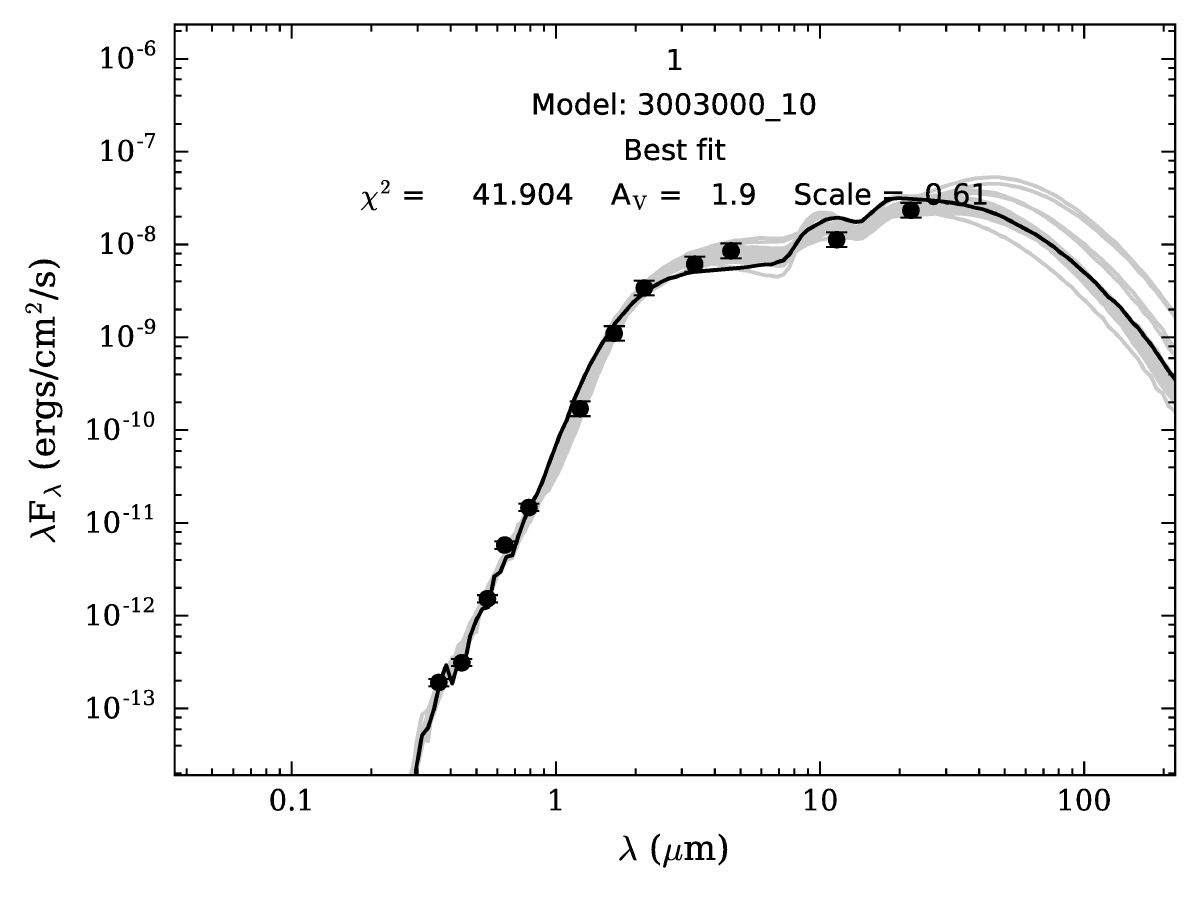}
\centering\includegraphics[width=0.45\textwidth]{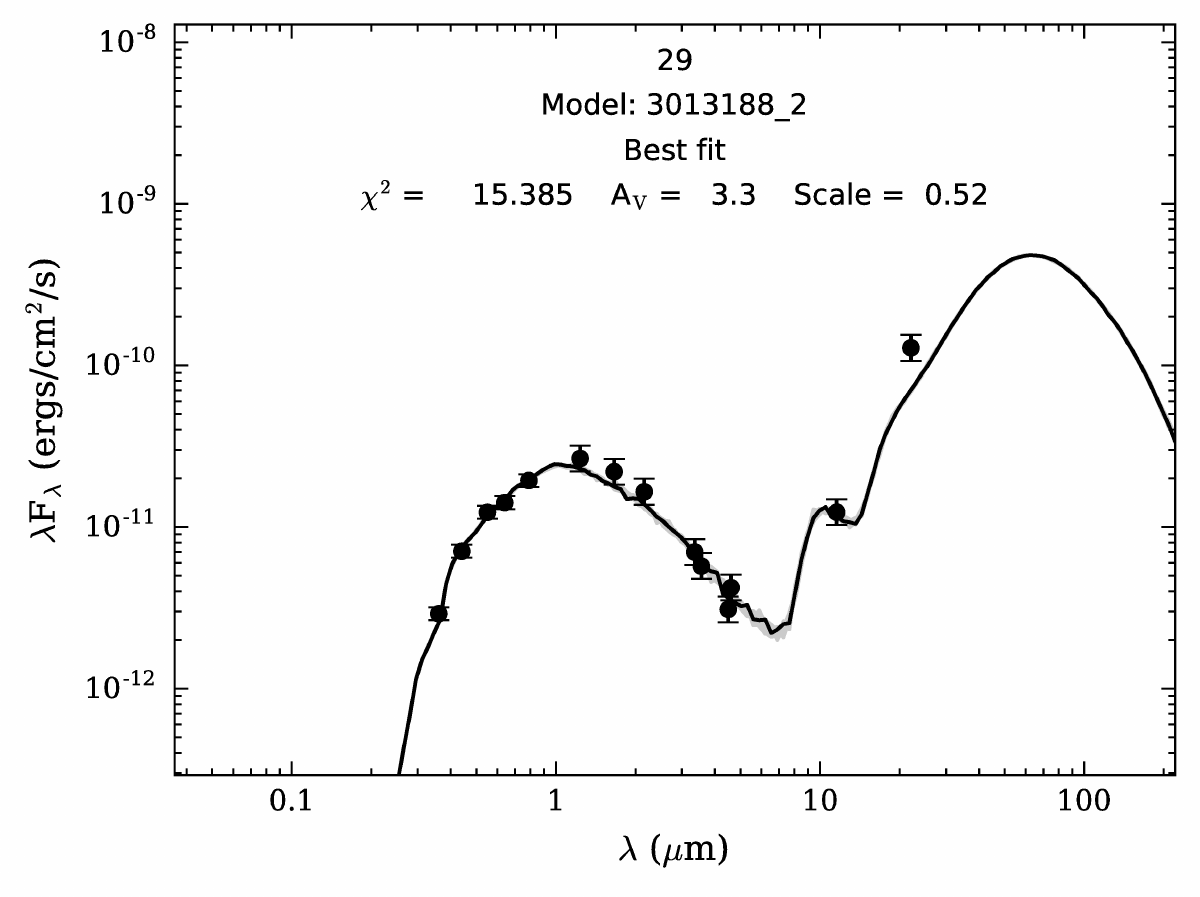}
\caption{\label{sed} Sample SEDs for Class\,{\sc i} (left-hand panel) and Class\,{\sc ii} (right-hand panel)  sources 
created by the SED fitting tools of \citet{2007ApJS..169..328R}. 
The black curve shows the best fit and the gray curves show the subsequent well fits. 
The filled circles with error bars denote the input flux values.}
\end{figure*}

\begin{figure*}
\centering\includegraphics[height=5cm,width=5.5cm,angle=0]{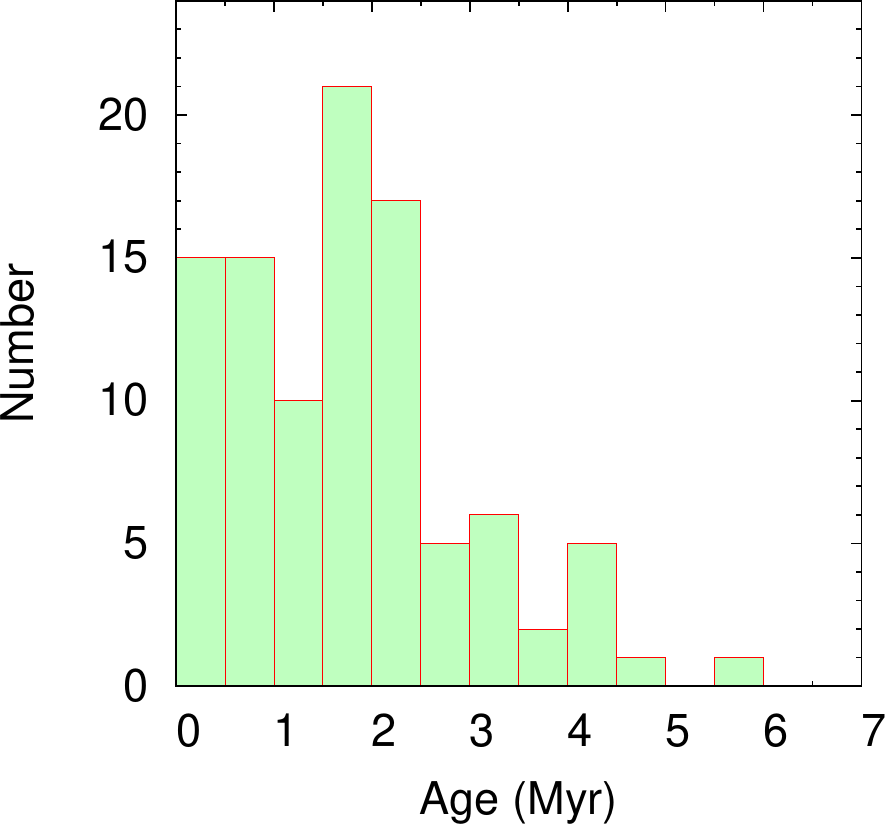}
\centering\includegraphics[height=5cm,width=5.5cm,angle=0]{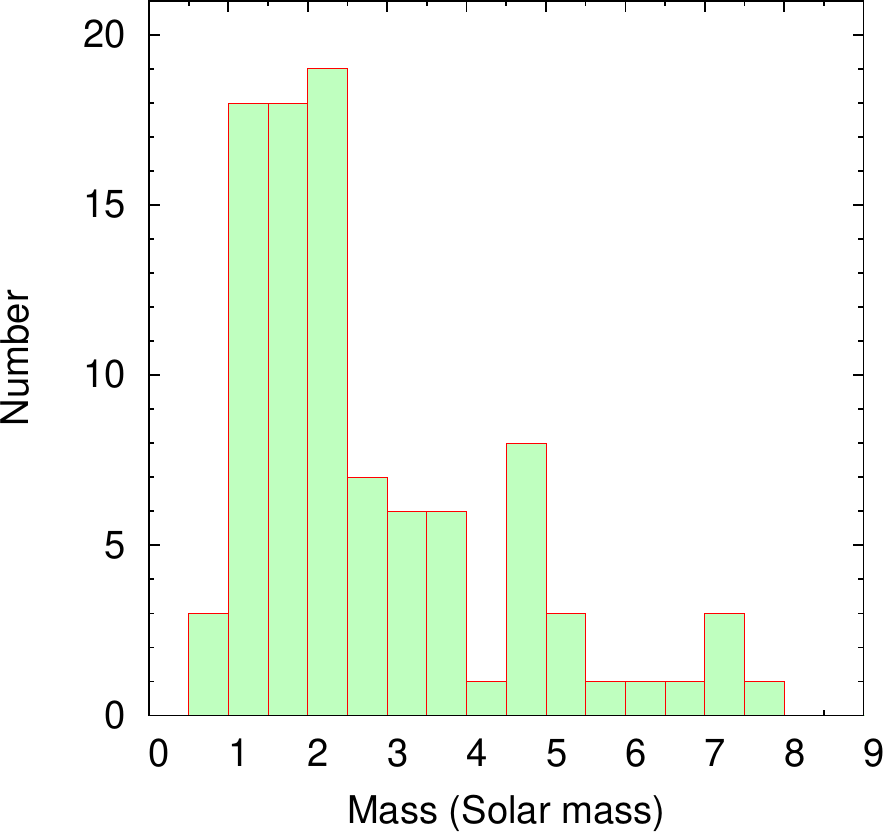}
\centering\includegraphics[height=5cm,width=5.5cm,angle=0]{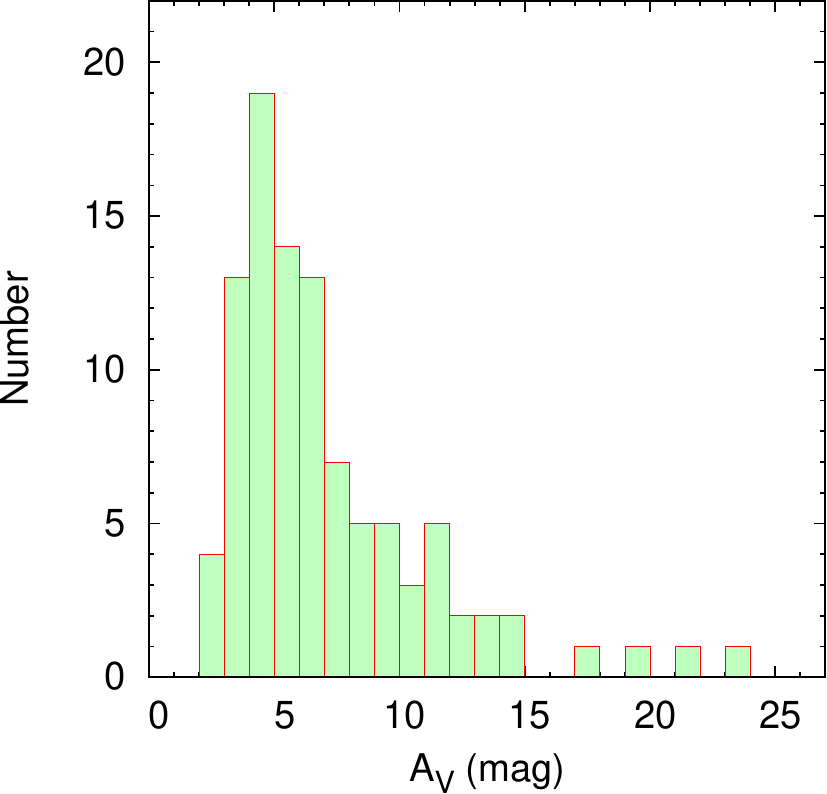}
\caption{\label{histogram} Histograms showing the distribution of the ages (left-hand panel),
masses (middle panel) and extinction values `$A_V$'  (right-hand panel) of the YSOs (98) in the Sh 2-305
as derived from the SED fitting analysis (cf. Section 3.5).
}
\end{figure*}

\begin{figure*}
\centering\includegraphics[width=0.35\textwidth]{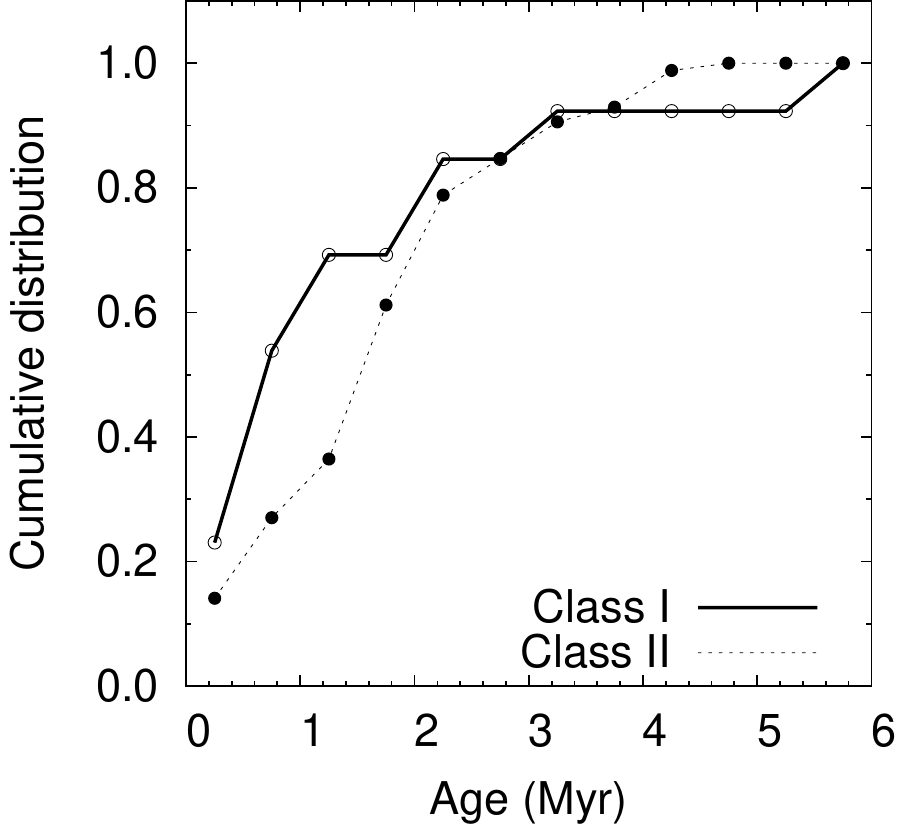}
\centering\includegraphics[width=0.35\textwidth]{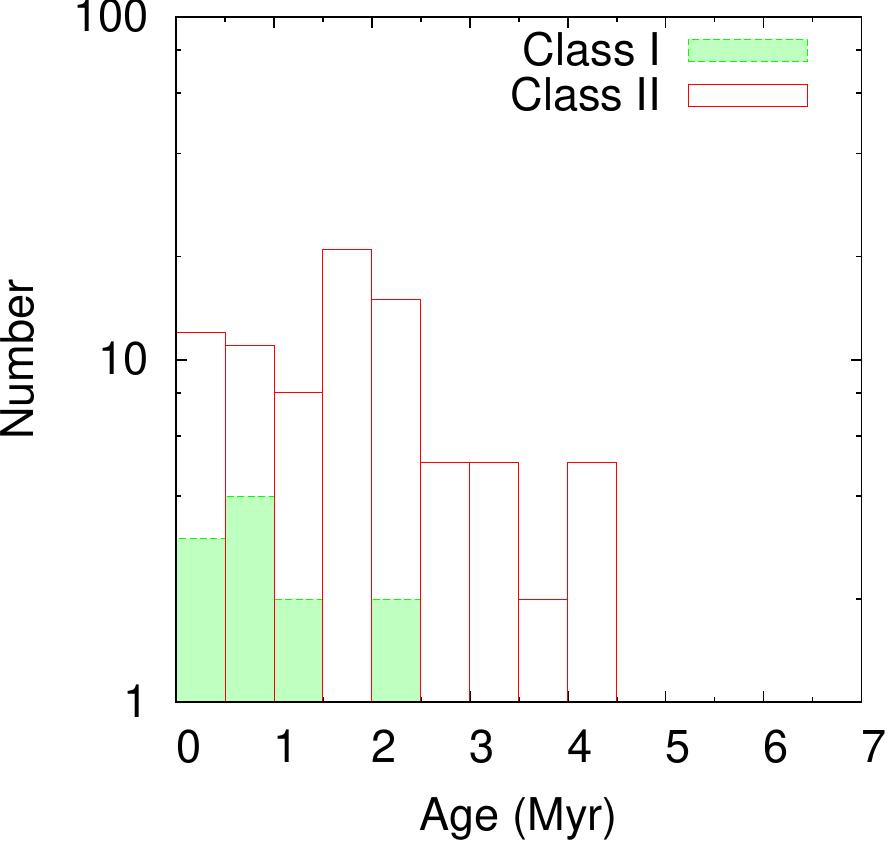}
\caption{\label{cum} 
Left panel: Cumulative distribution of Class\,{\sc i} (solid line) and Class\,{\sc ii} (dotted line) YSOs as a function of their age.
Right panel: Age distribution of Class\,{\sc i} (filled histogram) and Class\,{\sc ii} (un-filled histogram) YSOs. 
}
\end{figure*}

\subsection{Physical properties of YSOs }

Since the aim of this work is to study the star formation activities in the Sh 2-305, 
the information regarding individual properties of the YSOs is vital, 
which can be derived by using the SED fitting analysis.
We constructed SEDs of the YSOs using the grid models and fitting tools of 
\citet{2003ApJ...598.1079W,2003ApJ...591.1049W, 2004ApJ...617.1177W} and 
\citet{2006ApJS..167..256R,2007ApJS..169..328R} 
for characterizing and understanding their nature. 
This method has been extensively used in our previous studies \citep[see. e.g.,][and references therein]{2016ApJ...822...49J,2017MNRAS.467.2943S}.
We constructed the SEDs of the YSOs using the multiwavelength data 
(optical to MIR wavelengths, i.e. 0.37, 0.44, 0.55, 0.65, 0.80, 1.2, 1.6, 2.2, 3.4, 3.6, 4.5, 4.6, 12 and 22 $\mu m$) and 
with a condition that a minimum of 5 data points should be available.
Out of 116 YSOs, 98 satisfy this criterion and therefore are used in the further analysis.
The SED fitting tool fits each of the models to the data allowing the distance
and extinction as free parameters. 
The input  distance range  of the Sh 2-305 is taken as 3.3 - 4.1 kpc keeping in mind the error associated with distance, 
Since, this region is highly nebulous, we varied $A_V$ in a broader range 
i.e. from 3.6 (foreground reddening) to 30 mag
\citep[see also,][]{2012ApJ...755...20S,2013MNRAS.432.3445J,2014MNRAS.443.1614P}. 
We further set photometric uncertainties of 10\% for optical and 20\% for both NIR and MIR
data. These values are adopted instead of the formal errors
in the catalog in order to minimize any possible bias in the fitting that is caused
by underestimating the flux uncertainties. We obtained the
physical parameters of the YSOs using the relative probability
distribution for the stages of all the `well-fit' models. The
well-fit models for each source are defined by
$\chi^2 - \chi^2_{min} \leq 2 N_{data}$, where $\chi^2_{min}$ is the goodness-of-fit parameter for the
best-fit model and $N_{data}$ is the number of input data points.

In Figure \ref{sed}, we show example SEDs of Class\,{\sc i} (left panel) and Class\,{\sc ii} (right panel)
sources, where the solid black curves represent 
the best-fit and the gray curves are the subsequent well-fits. 
As can be seen, the SED of the Class~I source shows substantial MIR excess in comparison 
to the Class\,{\sc ii} source due to its optically thick disk.
From the well-fit models for each source derived
from the SED fitting tool, we calculated the $\chi^2$ weighted
model parameters such as the $A_V$, stellar mass and stellar age of each  YSO
and they are given in Table \ref {data4_yso}. 
The error in each parameter is calculated from the standard deviation of all well-fit parameters. 
Histograms of the age, mass and $A_V$ of these YSOs are shown in Figure \ref{histogram}.  
It is found that $\sim$91\% (89/98) of the sources have ages between 0.1 to 3.5 Myr.
The masses of the YSOs are between 0.8 to 16.2 M$_\odot$, a majority ($\sim$80\%) of them
being between 0.8 to 4.0 M$_\odot$. 
These age and mass ranges are comparable to the typical age and mass of TTSs.
The $A_V$ distribution shows a long tail indicating its large spread from $A_V$=2.2 - 23 mag,
which is consistent with the nebulous nature of this region.
The average age, mass and extinction ($A_V$) for this sample of YSOs are 1.8 Myr, 2.9 M$_\odot$ and 7.1 mag, respectively.

The evolutionary class of the  selected 116 YSOs, given in the Table \ref {data1_yso}, reveals that
$\sim$ 17\%  and $\sim$ 83 \%  sources are Class\,{\sc i} and Class\,{\sc ii} YSOs, respectively.
In Figure \ref{cum} (left panel), we have shown the cumulative distribution of Class\,{\sc i} and Class\,{\sc ii} YSOs as a function
of their ages, which manifests that Class\,{\sc i} sources are relatively younger than Class\,{\sc ii} sources as expected. We have performed
a  Kolmogorov-Smirnov (KS) test for this age distribution. The test indicates 
that the chance of the two populations having been drawn from the same distribution is $\sim$4\%. 
The right-hand panel of Figure \ref{cum}  plots the distribution of ages for the Class\,{\sc i} and Class\,{\sc ii} sources.
The distribution of the Class\,{\sc i} sources shows a peak at a very young age, 
i.e., $\sim$0.50-0.75 Myr, whereas that
of the Class\,{\sc ii} sources peaks at $\sim$1.50-1.75 Myr. 
Both of these figures 
show  an approximate age difference of $\sim$1 Myr between the Class\,{\sc i} and Class\,{\sc ii} sources.
Here it is worthwhile to take note that \citet{2009ApJS..181..321E} through c2d $Spitzer$ Legacy projects studied YSOs
associated with five nearby molecular clouds and concluded that the life time of the Class\,{\sc i} phase is 0.54 Myr.
The peak in the histogram of Class\,{\sc i} sources agrees well with them \citep[see also][]{2017MNRAS.467.2943S}.

\begin{figure*}
\centering
\includegraphics[height=7cm,width=10.5 cm]{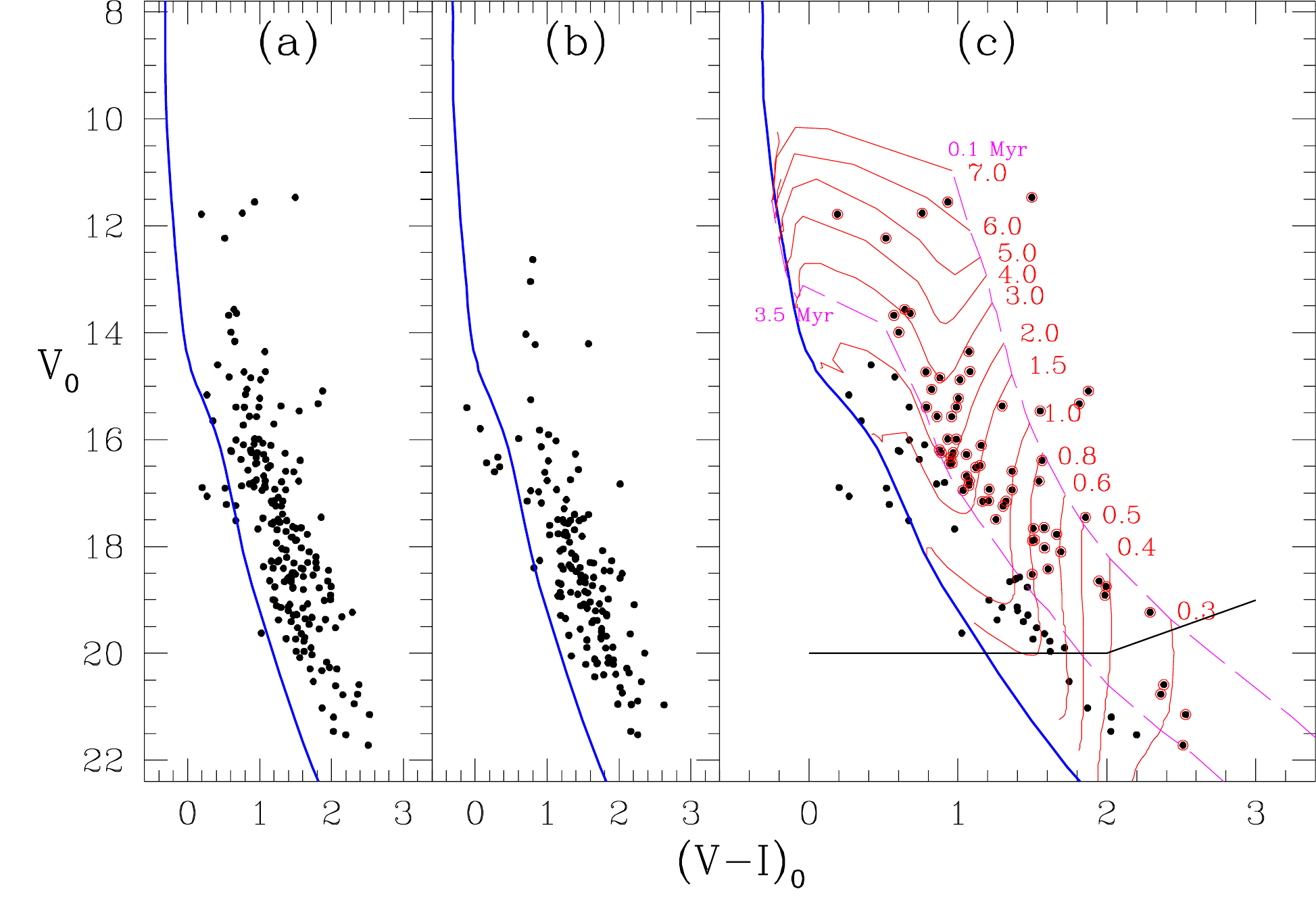}
\includegraphics[height=5.5cm,width=7cm]{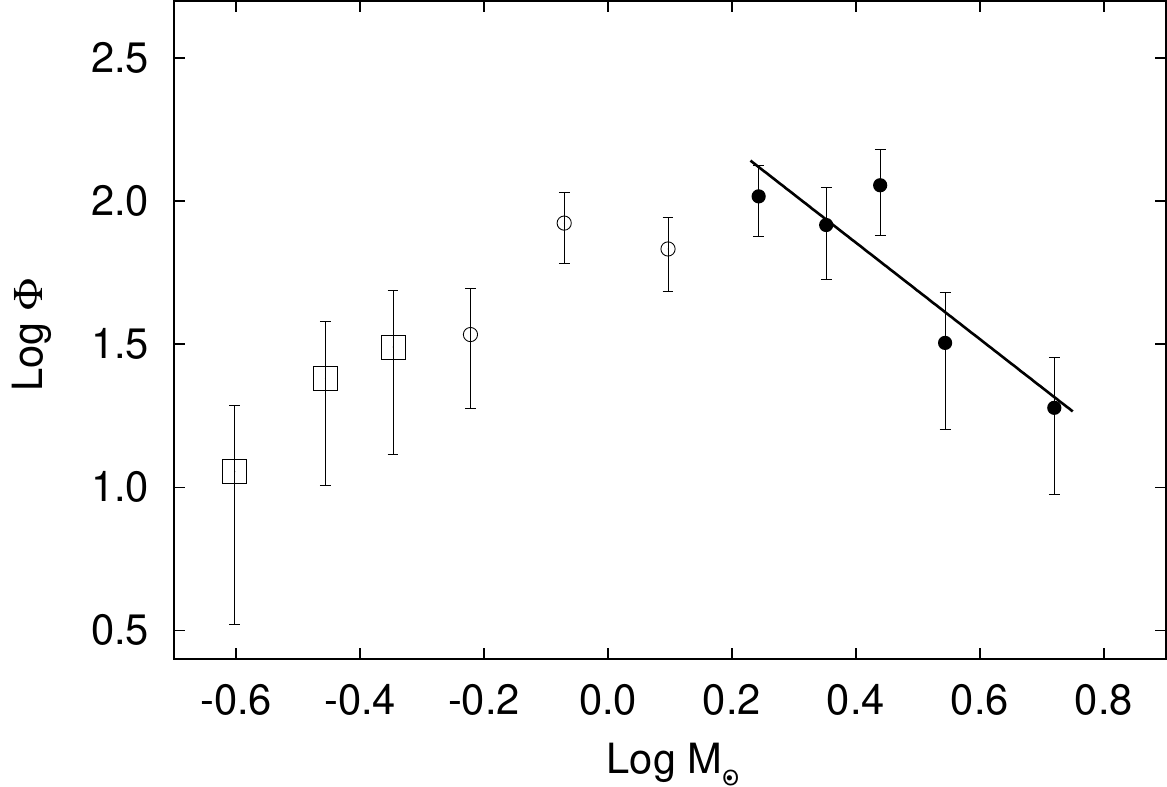}
\caption{\label{band} Left panel : $V_0$ vs. $(V-I_c)_0$ CMD for (a) stars in the Mayer 3 cluster region, and (b) stars in the reference region. 
(c) is a statistically cleaned $V_0$ vs. $(V-I_c)_0$ CMD  for stars lying in the Mayer 3 cluster region.
Red circles (ages $\leq$ 3.5 Myr) are used to estimate the MF of the region.
The isochrone of 2 Myr by \citet[blue curve,][]{2008AA...482..883M} and the PMS isochrones
of 0.1 and 3.5 Myrs along with the evolutionary tracks for different masses 
by \citet[magenta dashed and red curves,][]{2000AA...358..593S} are also shown. 
All the curves are corrected for the  distance of 3.7 kpc. The back horizontal line represents the completeness limit of the data (upto 90\%) after taking into account the average extinction of the YSOs. Right panel: A plot of the MF for the statistically cleaned CMD for the stellar sources in the Mayer 3 cluster region.
Log $\phi$ represents log($N$/dlog $m$). The error bars represent $\pm\sqrt N$ errors. The solid line shows the least
squares fit to the MF distribution (black dots).
Open squares are the data points falling below the completeness limit of 0.8 M$_\odot$. Open circles are the data points near turn-off point in the MF distribution and are not used in the fitting. 
}
\end{figure*}

\begin{figure*}
\centering
\includegraphics[width=0.49\textwidth]{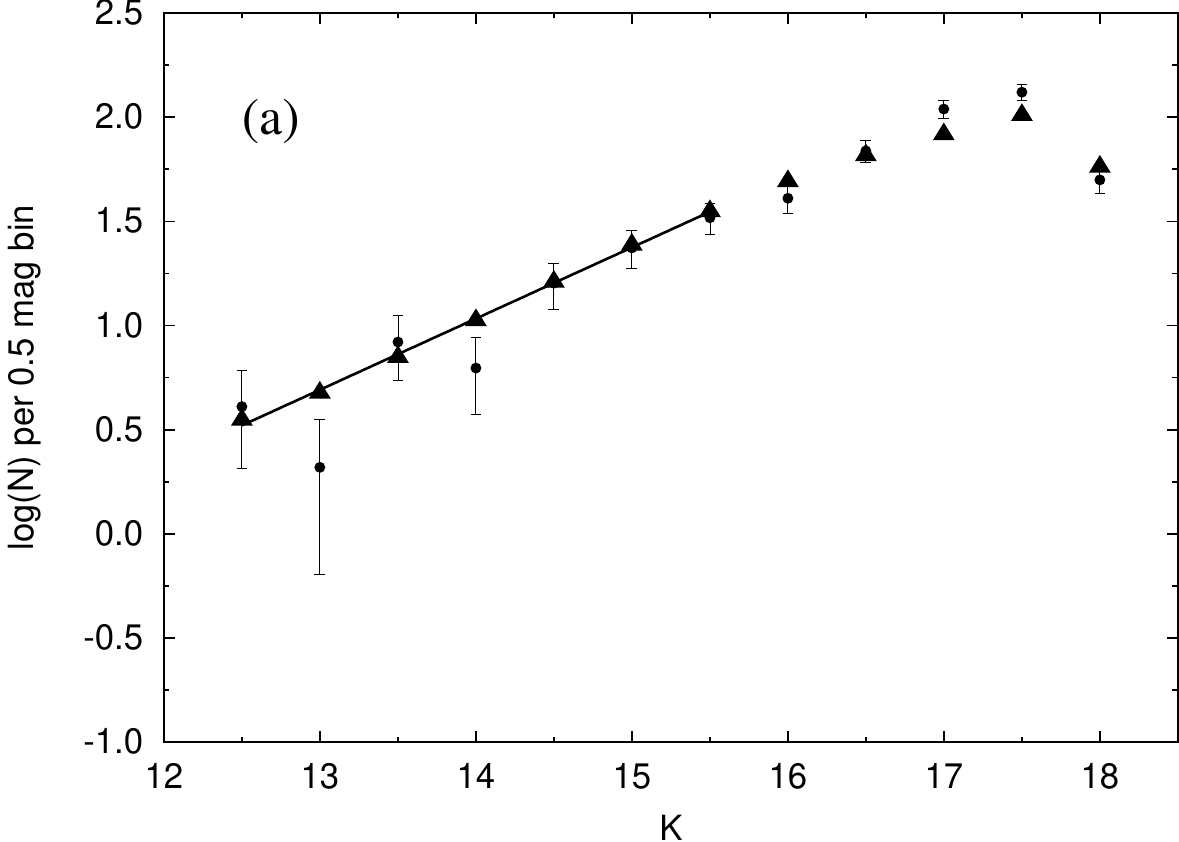}
\includegraphics[width=0.49\textwidth]{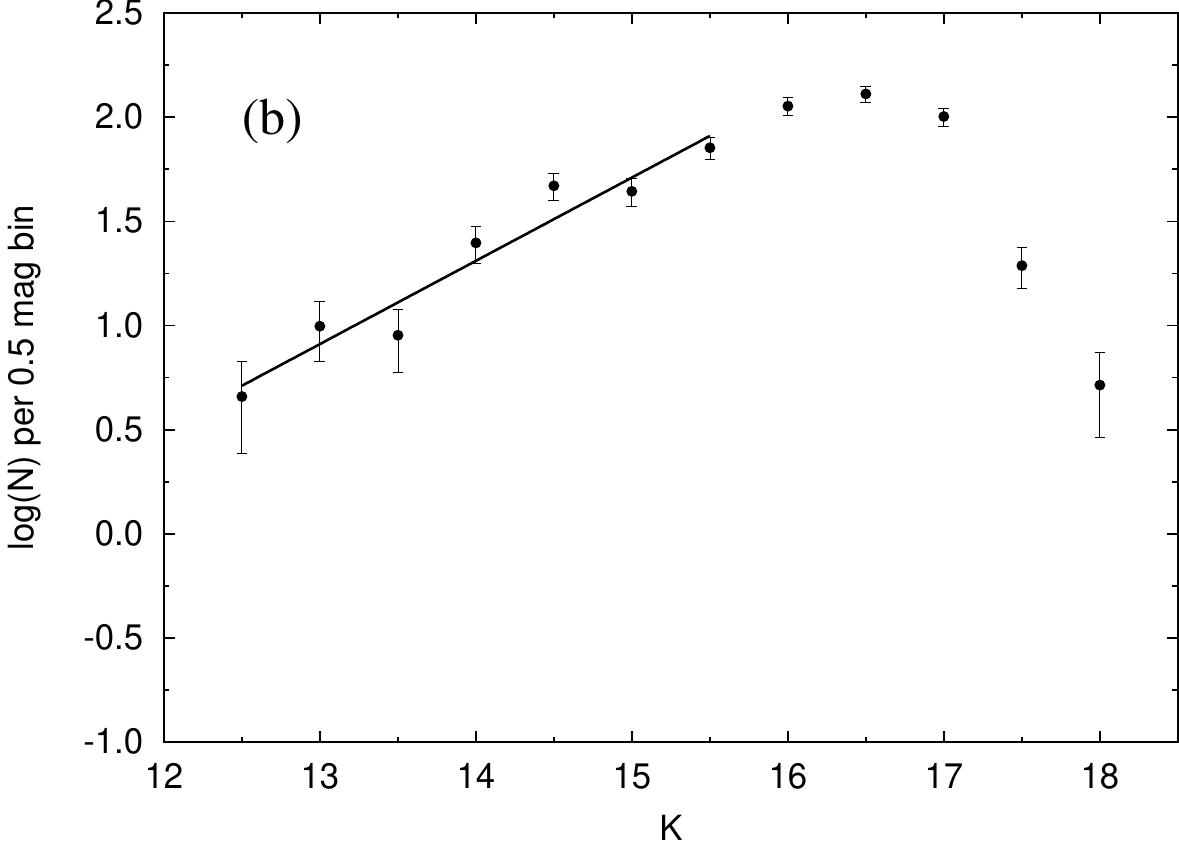}
\caption{\label{klf}(a) Comparison of the observed KLF in the reference field and 
the simulated KLF from star count modeling. The filled circles denote the
observed K-band star counts in the reference field, and triangles represent
the simulation from the Galactic model (see the text). The star counts are
number of stars per square degree and the error bars represent $\sqrt{N}$ errors. 
The KLF ($\alpha$, see Section 4.4) of the reference field (solid line) is $0.36\pm0.07$.
The simulated model is also giving the same value of slope $(0.34\pm0.01)$.
(b) The corrected KLF for the probable members in the central cluster `Mayer 3' (slope: $0.40\pm0.04$). The straight
line is the least-square fit to the data points in the range 12.5-15.5 mag.
}
\end{figure*}

\subsection{Mass Function (MF) and K-band Luminosity Function (KLF)}

The MF is an important statistical tool to understand the formation of stars
\citep[][and references therein]{2017MNRAS.467.2943S}.
The MF is often expressed by a power law,
$N (\log m) \propto m^{\Gamma}$ and  the slope of the MF is given as:
$ \Gamma = d \log N (\log m)/d \log m  $,
where $N (\log m)$ is the number of stars per unit logarithmic mass interval.
We have used our deep optical data to generate the MF of the Sh 2-305 region. 
For this, we have utilized the optical $V_0$ versus $(V-I_c)_0$ CMDs of the
sources in the target region and that of the nearby field region of equal area
and decontaminated the former sources from foreground/background
stars and corrected for data in-completeness using a statistical subtraction method
already described in detail in our previous papers
\citep[cf.][]{2007MNRAS.380.1141S,2012PASJ...64..107S,2017MNRAS.467.2943S,2008MNRAS.383.1241P,2013ApJ...764..172P,2011MNRAS.415.1202C,2013MNRAS.432.3445J}.
As an example, in  Figure \ref{band} (left panel), we have shown the $V_0$ versus $(V-I_c)_0$ CMDs for the 
stars lying within the central cluster `Mayer 3' in panel  `a' and for those in the reference field region
selected as an annular region outside the boundary of Sh 2-305 region (cf. Section 3.1) in panel `b'.
The magnitudes were corrected for the $A_V$ values derived from the reddening map (cf.  Section 3.7).
In panel `c', we have plotted the statistically cleaned $V_0$ versus $(V-I_c)_0$ 
CMD for the central cluster `Mayer 3' which is showing the presence of PMS stars in the region.
The ages and masses of the stars in this statistically cleaned CMD have been derived 
by applying the procedure described earlier in our previous papers \citep{2009MNRAS.396..964C,2017MNRAS.467.2943S}.
For reference, the post-main-sequence isochrone for 2 Myr calculated by \citet{2008AA...482..883M}
(thick blue curve) along with the PMS isochrones
of 0.1 and 3.5 Myr (purple curves) and evolutionary tracks of
different masses (red curves)  by \citet{2000AA...358..593S} are also shown in panel `c'.
These isochrones are corrected for the distance of the  Sh 2-305 (3.7 kpc, cf. Section 3.3).
The corresponding MF has subsequently been plotted in Figure \ref{band} (right panel) for the central cluster `Mayer 3'. 
For this, we have used only those sources which have ages equivalent to the average age 
of the optically identified YSOs combined with error (i.e., $\leq$3.5 Myr, cf. Table \ref{data4_yso}).
Our photometry is more than 90\% complete up to $ V=21.5$ mag,
which corresponds to the detection limit of 0.8 M$_\odot$ (cf.  Figure \ref{band} (left panel) `c') 
PMS star of $\simeq$1.8 Myr age embedded in the nebulosity of $A_V\simeq$3.0 mag 
(i.e., the average values for the optically detected YSOs, cf. Table \ref{data4_yso}).
We have applied a similar approach to derive MF of the southern clustering and the whole 
region of the Sh 2-305 (cf. Section 3.1)
and their corresponding values of MF slopes in the mass range $1.5<M_\odot<6.6$ are given in Table \ref{Tp3}.
The  MF of the northern clustering cannot be determined due to insignificant number of optically detected stars.

The KLF is also a powerful tool to investigate the IMF of young embedded clusters, 
and is related to IMF by a relation, $\alpha = \dfrac{-\Gamma}{2.5 \beta}$, where
$\alpha$, $\Gamma$ and $\beta$ are the slopes of KLF, IMF and mass-luminosity relationship, respectively  
\citep[e.g.,][]{1993ApJ...408..471L,2003ARAA..41...57L, 2004ApJ...616.1042O, 2007ApJ...667..963S, 2014MNRAS.443.3218M}. 
To take into account the foreground/background field star contamination, we used the Besan\c con Galactic model of stellar population synthesis \citep{2003AA...409..523R} and predicted the star counts in both the cluster region and in the direction of the reference field. We checked the validity of the simulated model by comparing the model KLF with that of the reference field and found that the two KLFs match rather well (Figure \ref{klf}a). An advantage of using the model is that, we can separate the foreground ($d<3.7$ kpc) and the background ($d>3.7$ kpc) field stars.
The foreground extinction towards the cluster region is found to be $A_V \sim3.63$ mag.
The model simulations with $d<3.7$ kpc and $A_V$ = 3.63 mag give the foreground contamination, 
and that with $d>3.7$ kpc and $A_V$ = 3.63 mag the background population.
We thus determined the fraction of the contaminating stars
(foreground+background) over the total model counts. This fraction was used to scale the nearby
reference region and subsequently the modified star counts of the reference region were subtracted
from the KLF of the cluster to obtain the final corrected KLF.
This KLF is expressed by the following power-law:
${{ \rm {d} N(K) } \over {\rm{d} K }} \propto 10^{\alpha K}$,
where ${ \rm {d} N(K) } \over {\rm{d} K }$ is the number of stars per 0.5 magnitude
bin and $\alpha$ is the slope of the power law. 
Figure \ref{klf}b shows the KLF for the Mayer 3 cluster region. Similarly, we have derived KLF for other clusterings  as well as the whole region (cf. Section 3.1) 
and their corresponding slope values in $K$-band completeness range  of
$12.5<K (mag)<15.5$ are given in Table \ref{Tp3}.

\subsection{Spatial distribution of  molecular gas and YSOs in the region}

\begin{figure*}
\centering
\includegraphics[width=0.95\textwidth]{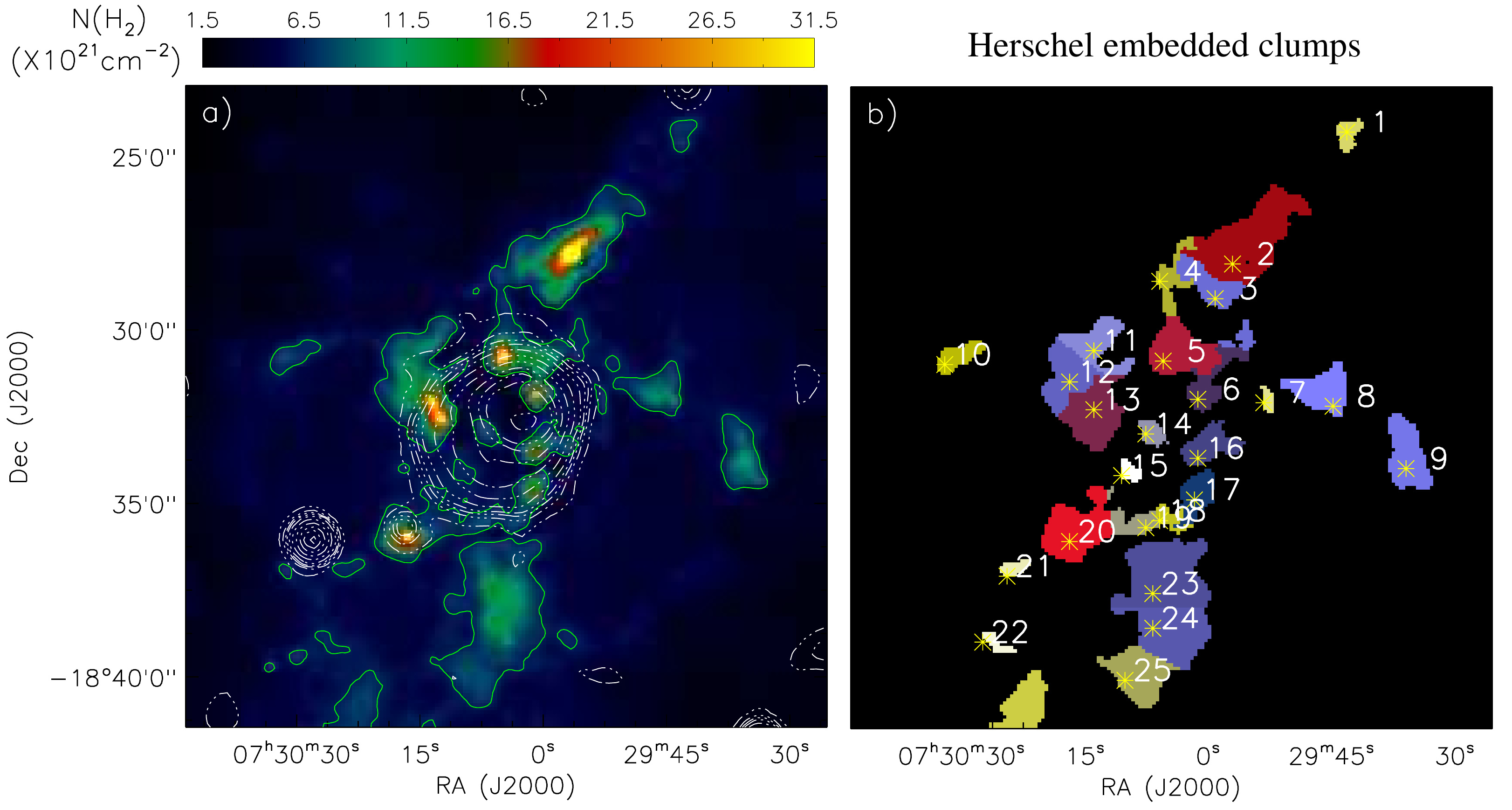}
\caption{\label{cd} Left panel: $Herschel$ column density ($N(\mathrm H_2)$) map of  $\sim18^\prime.5\times18^\prime.5$
FOV of the Sh 2-305 region (see text for details). 
The map is also superimposed with the NVSS radio continuum contours (in white) and the column 
density contour (at 6.5 $\times$ 10$^{21}$ cm$^{-2}$; in green). 
The NVSS 1.4 GHz contours are shown with levels of (0.45 mJy/beam) $\times$ (3, 10, 15, 20, 30, 50, 80, 120, 160, 200, 300).
Right panel: The boundary of each identified clump and its position are highlighted along with its corresponding clump 
ID (see Table \ref{cd1}). 
}  
\end{figure*}

In the Figure \ref{cd}a, we present {\it Herschel} 
column density ($N(\mathrm H_2)$) map\footnote{http://www.astro.cardiff.ac.uk/research/ViaLactea/}
of the Sh 2-305 region to examine the embedded structures. 
The spatial resolution of the map is $\sim$12$''$. Adopting the Bayesian {\it PPMAP} procedure operated on the {\it Herschel} data \citep{2010A&A...518L.100M} at wavelengths of 70, 160, 250, 350 and 500 $\mu$m \citep{2015MNRAS.454.4282M, 2017MNRAS.471.2730M}, the {\it Herschel} temperature and column density maps were produced for the {\it EU-funded ViaLactea project} \citep{2010PASP..122..314M}. 
The map is also overlaid with the NVSS\footnote{https://www.cv.nrao.edu/nvss/postage.shtml} 
radio continuum contours and the column density contour (at 6.5 $\times$ 10$^{21}$ cm$^{-2}$)
to enable us to study the distribution of embedded condensations/clumps against the ionized gas. In general, the $N(\mathrm H_2)$ maps shows fragmented structures with several embedded dust clumps in the target region. The peak of the NVSS radio emission appears to be surrounded by the dust clumps. 
We have used {\it clumpfind} algorithm \citep{1994ApJ...428..693W} to identify 25 clumps
in the column density map of our selected target area. The boundary and the position of each clump are also shown in Figure \ref{cd}b. 
The mass of each clump is determined and is listed in Table \ref{cd1}. 
In this connection, we employed the equation, $M_{area} = \mu_{H_2} m_H Area_{pix} \Sigma N(H_2)$, where $\mu_{H_2}$ is 
the mean molecular weight per hydrogen molecule (i.e., 2.8), $Area_{pix}$ is the area subtended by one pixel (i.e., 6$''$/pixel), and 
$\Sigma N(\mathrm H_2)$ is the total column density \citep[see also][]{2017ApJ...834...22D}. 
Table \ref{cd1} also contains an effective radius of each clump, which is provided by the {\it clumpfind} algorithm. 
The clump masses vary between 35 M$_{\odot}$ and 1565 M$_{\odot}$, and the most massive clump (ID = 2 in Table \ref{cd1}) 
lies in the northern direction of the region.

\begin{figure*}
\centering
\centering\includegraphics[height=8.0cm,width=9.0cm]{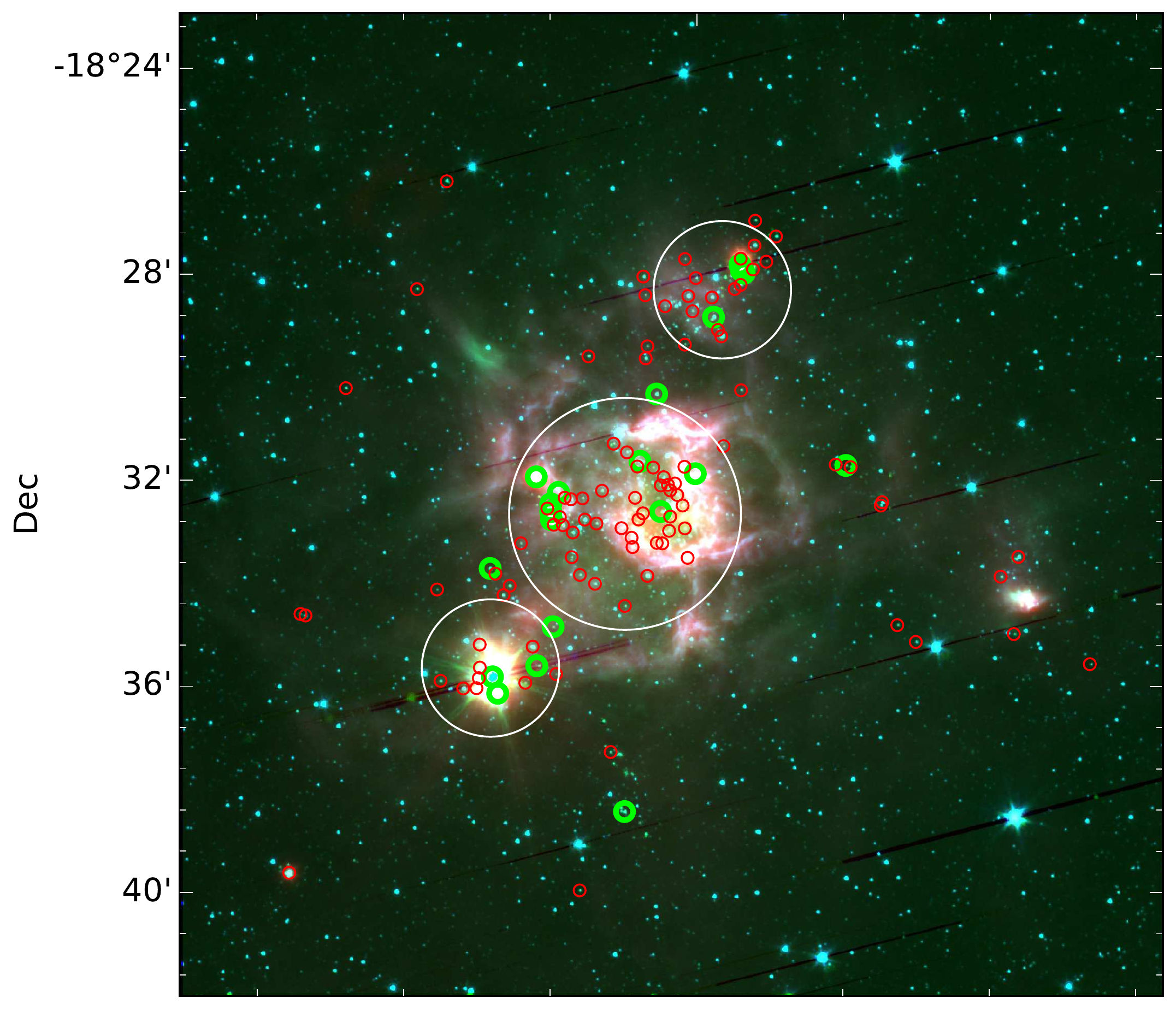}
\centering\includegraphics[height=8.0cm,width=8.0cm]{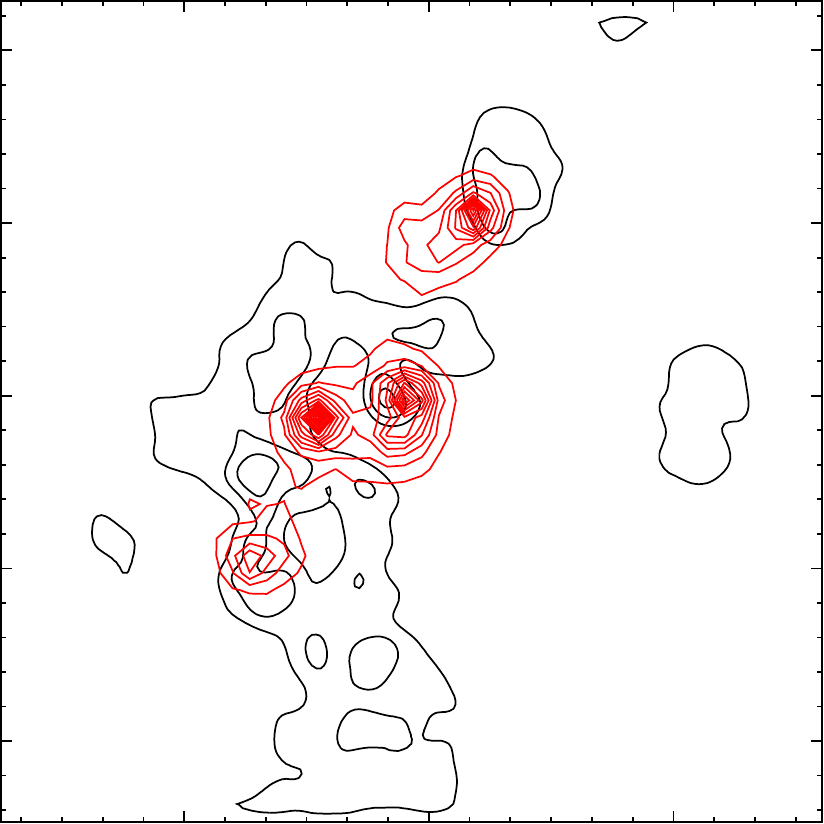}
\centering\includegraphics[height=8.5cm,width=9.0cm]{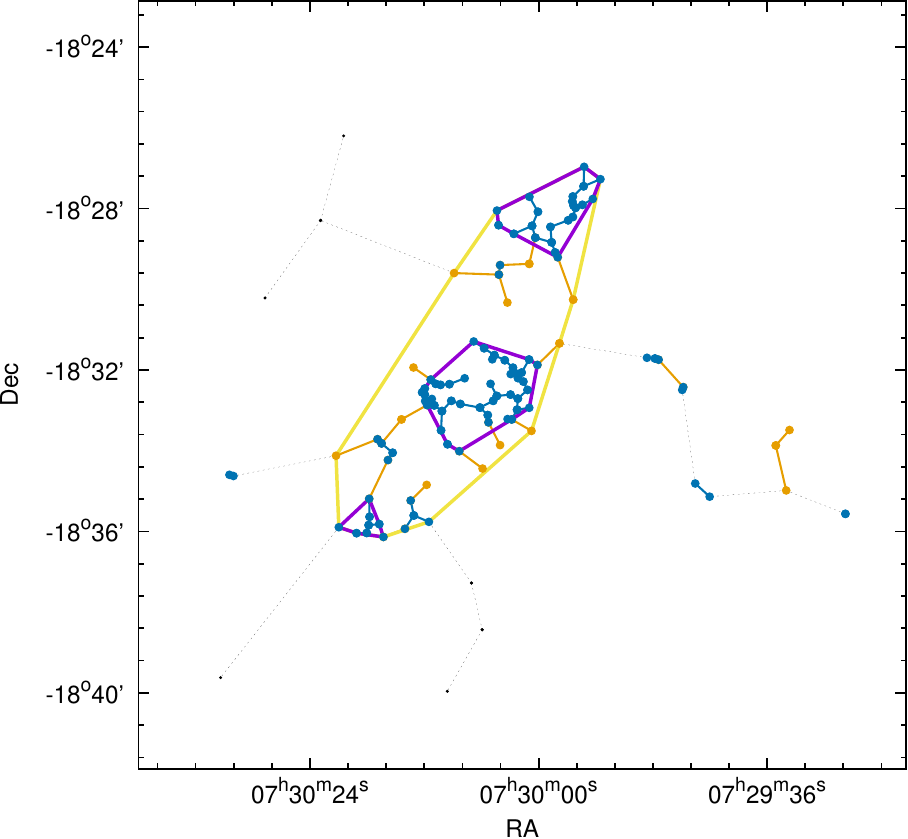}
\centering\includegraphics[height=8.5cm,width=8.0cm]{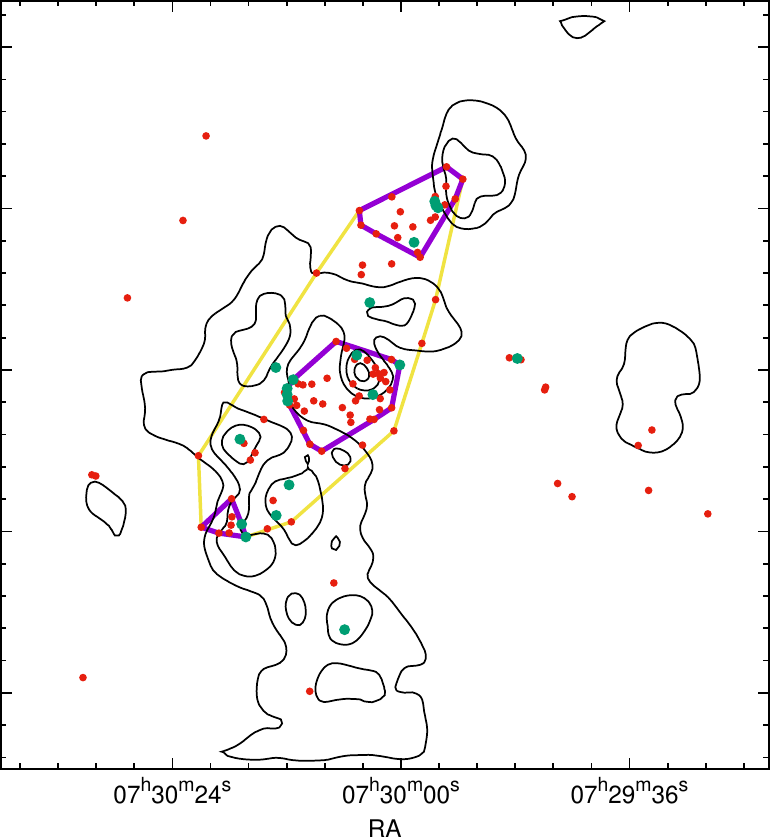}
\caption{\label{spa} Top-left panel: Spatial distribution of  YSOs
superimposed on the $\sim18^\prime.5\times18^\prime.5$ color composite (red: WISE 12 $\mu$m, green: $Spitzer$ 4.5 $\mu$m, blue:  $Spitzer$ 3.6 $\mu$m)
image of the Sh 2-305 region.
The locations of Class\,{\sc i} (green circles) and Class\,{\sc ii} (red circles) are also shown.
  The white circles represent the three clusters/clumps identified in the present analysis.
Top-right panel: Isodensity contours for the YSOs distribution (red contours) and the
reddening map (black contours) for the same region.
Bottom-left panel: Minimal spanning tree (MST) for the identified YSOs in the same
region along with the convex hull.
The blue dots connected with solid blue lines and yellow dots connected with yellow lines are the branches smaller than the critical
length for the cores and the active region, respectively.
The identified cores and the  active region are encircled with purple and yellow
solid lines (i.e. Qhull), respectively.
Bottom-right panel: Spatial correlation between the molecular material
inferred from the extinction map (black contours) and the
distribution of YSOs along with the identified cores and active region
(thick purple and yellow lines, respectively).  }
\end{figure*}

Figure \ref{spa} (top left panel) shows the  color composite image generated  using the
WISE 12 $\mu$m (red), $Spitzer$ 4.5 $\mu$m (green) and  $Spitzer$ 3.6 $\mu$m (blue) images
 of the 
Sh 2-305 region.
We have also over-plotted the identified YSOs on the figure.
The WISE 12 $\mu$m image covers the prominent PAH features
at 11.3 $\mu$m, indicative of star formation \citep[see e.g.][]{2004ApJ...613..986P}.
We can observe several filamentary structures in the 12 $\mu$m emission. 
The distribution of a majority of the YSOs belongs generally to these structures.
The distributions of the gas and dust, seen by the 
MIR emissions in the 4.5 $\mu$m and 3.6 $\mu$m images, are  well correlated with the distribution of YSOs.
Figure \ref{spa} (top-left panel) reveals that the Sh 2-305 is a site of active star formation and
there are three major groupings of YSOs, distributed from northern to southern directions in the region.

To study the density distribution of YSOs in the region, 
we have  generated the surface density map (cf. Figure \ref{spa}, red contours in the top right panel) 
using the nearest neighbor (NN) method (see Section 3.1), 
with a grid size of 6$^{\prime\prime}$ and 5$^{th}$ nearest YSOs in an area  $\sim18^\prime.5\times18^\prime.5$ of Sh 2-305.
The lowest contour is 1$\sigma$ above the mean of YSO density (i.e. 1.8 stars/pc$^2$)
and the step size is equal to the 1$\sigma$ (1.3 stars/pc$^2$).

We have also derived $A_K$ extinction map using the $(H - K)$ colors of the MS stars
to quantify the amount of extinction and to characterize the structures of the molecular 
clouds \citep[see also,][]{2009ApJS..184...18G, 2011ApJ...739...84G, 2013MNRAS.432.3445J, 2016AJ....151..126S,2019MNRAS.tmp.3221P}.
The sources showing excess emission in IR can lead to overestimation of extinction
values in the derived maps. Therefore, to improve the quality of the extinction maps,
the candidate YSOs and probable contaminating sources (cf. Section 3.4) must be excluded for the calculation of extinction.
In order to determine the mean value of A$_K$, we used the NN method as described in detail in
\citet{2005ApJ...632..397G} and \citet{2009ApJS..184...18G}.  Briefly, at each position in a uniform grid of 6$^{\prime\prime}$,
we calculated the  mean value of $(H - K)$ colors of five nearest stars.
The sources deviating above
3$\sigma$ were excluded to calculate the final mean color of each point.
To facilitate comparisons between the stellar density and the gas column density, we
adopted the grids identical to the grid size of the stellar density map for this region.
To convert $(H -K)$ color excesses into $A_K$, we used the relation  
$A_K$ = 1.82 $\times$ E$(H - K)$, where E$(H - K)$=$(H - K)$$_{obs}$ - $(H - K)_{int}$, adopting the reddening law by \citet{2007ApJ...663.1069F}.

We have assumed $(H - K)_{int}$ = 0.2 mag as an average intrinsic color for all stars in young
clusters \citep[see.][]{2008ApJ...675..491A, 2009ApJS..184...18G}. 
To eliminate the foreground contribution in the extinction measurement, we
used only those stars with $A_K >$ 0.15$\times$D, where D 
is the distance of the H\,{\sc ii} region in kpc \citep{2005ApJ...619..931I};
to generate the extinction map.  The extinction map is
 plotted in Figure \ref{spa} (top right panel) as black contours.
The lowest contour is 1$\sigma$ above the mean extinction value (i.e. $A_V$=6.8 mag)
and the step size is equal to 1$\sigma$ (1.5 mag). The extinction map displays more or less a similar morphology  
 around the region as in the $Herschel$ column density map.

Here, it is also worthwhile to note that the derived isodensity/$A_K$ values 
are the lower limits of their values as the sources with higher extinction 
may not be detected in our study.
The isodensity contours of the YSOs are clearly showing three stellar groupings 
(similar to the surface density distribution of NIR sources shown in Figure \ref{image}), 
one in the central region comprising the Mayer 3 cluster, other in the 
north direction, and another in the south direction.
The extinction map also has a north to south elongated morphology similar to the
isodensity contours, however, the peak in the stellar density is slightly off from the peak 
in extinction contours. 

\begin{figure*}
\centering\includegraphics[width=0.45\textwidth]{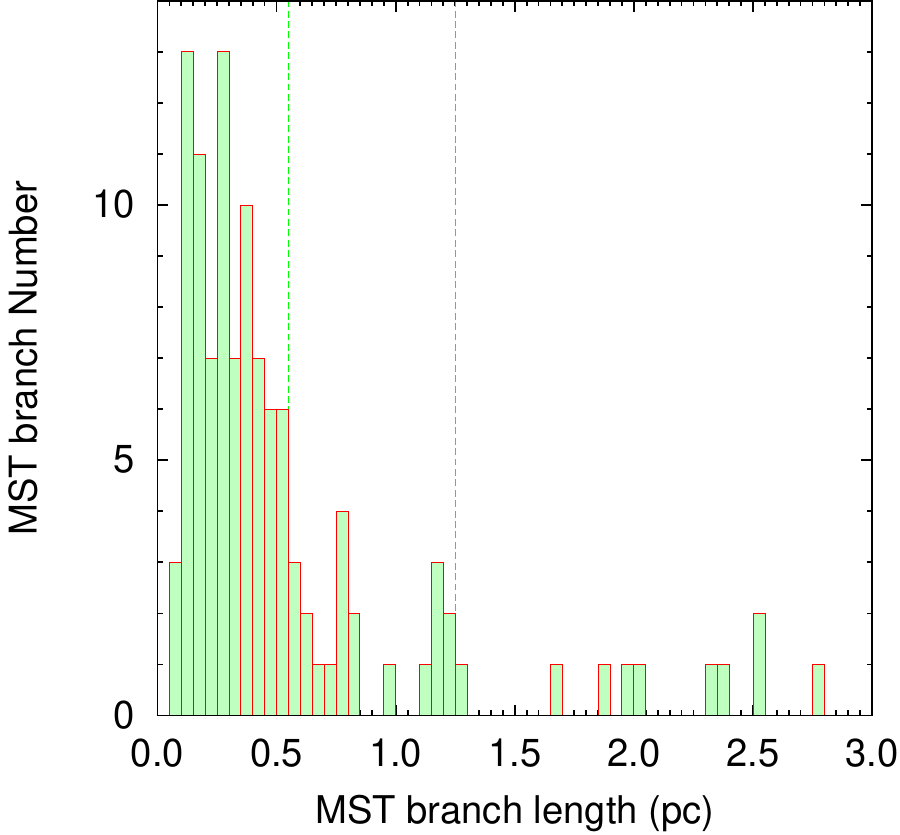}
\centering\includegraphics[width=0.45\textwidth]{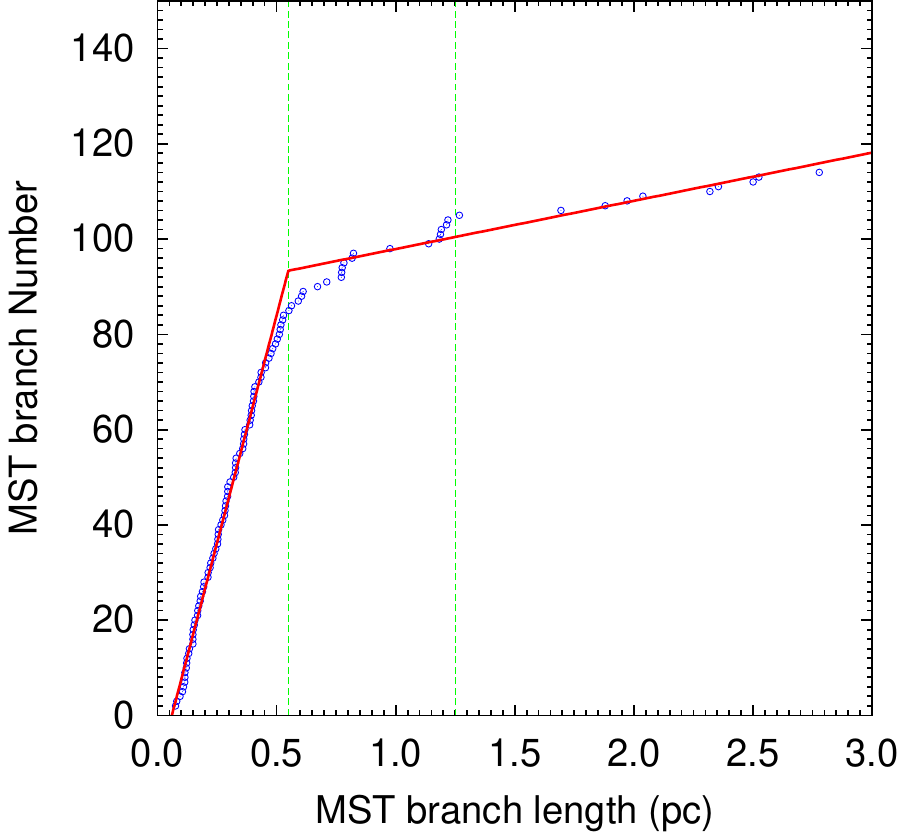}
\caption{\label{cdf}
Histogram of MST branch length (left panel) and cumulative distribution functions (CDFs, right panel), used for the critical length analysis of YSOs.
CDF plots have  sorted length values on the vertical axis and a rising integer counting index on the horizontal axis.
The red solid line is a two-line fit to the CDFs distribution.
The inner and outer vertical green lines are the critical lengths obtained
for the core and the  active region (AR), respectively.
 }
\end{figure*}

\subsection{Extraction of YSO’s cores embedded in the molecular cloud}

As discussed in previous sub-section, the Sh 2-305 contains three major sub-groups/cores of YSOs 
(cf. Figure \ref{spa}, top left panel), presumably due to fragmentation of the molecular cloud.
Physical parameters of these cores which might have formed in a single star-forming event, play a very important role in the study of star formation.
Here, we  have applied an empirical method based on the minimal sampling tree (MST)
technique to isolate groupings (cores) of the YSOs 
from their diffuse distribution in this nebulous region \citep{2009ApJS..184...18G}.
This method  effectively isolates the sub-structures without any type of smoothening
and  bias regarding  the shapes of the distribution and it preserves the underlying
geometry of the distribution 
\citep[e.g.,][]{2004MNRAS.348..589C, 2006A&A...449..151S, 2007MNRAS.379.1302B, 2009MNRAS.392..868B, 2009ApJS..184...18G,2014MNRAS.439.3719C,2016AJ....151..126S,2019MNRAS.tmp.3221P}.
In  Figure \ref{spa} (bottom left panel), we have plotted the derived MSTs for the location of YSOs.
The different color dots and lines are the positions of the
YSOs and the MST branches, respectively. A close inspection of this figure reveals that the
region exhibits different concentrations of YSOs distributed throughout the regions.
In order to isolate these sub-structures, we have adopted a surface density 
threshold expressed by a critical branch length.
In Figure \ref{cdf} (left panel), we have plotted histograms between 
MST branch lengths and MST branch numbers for the YSOs.
From this plot, it is clear that they have a peak at small spacings
and a relatively long tail towards large spacings. These peaked distance distributions
typically suggest a significant sub-region (or sub-regions) above a relatively uniform, elevated surface density.
By adopting a MST length threshold, we can isolate those sources which are closer
than this threshold, yielding populations of sources that make up local surface density
enhancements. To obtain a proper threshold distance, 
we have fitted two true lines in shallow and steep
segment of the cumulative distribution function (CDF) for the branch length of MST for YSOs  
(cf. Figure \ref{cdf}, right panel). We adopted the intersection point between these two lines as the MST critical branch length,
as shown in Figure \ref{cdf} (right panel) \citep[see also,][]{2009ApJS..184...18G, 2014MNRAS.439.3719C,2016AJ....151..126S,2019MNRAS.tmp.3221P}.
The YSO cores were then isolated from the lower density distribution by clipping
MST branches longer than the critical length found above.
Similarly, we have enclosed this star-forming region by selecting the point where the shallow-sloped segment has a gap in the length distribution of the MST branches.
This point can also be seen near a bump in the  MST branch length histogram.
We have called this region  in the Sh 2-305 as its active region (AR) where recent star formation took place or
that contains YSOs which moved out from the cores due to dynamical evolution.
The values of the critical branch lengths for the cores and the AR 
are 0.55 pc and 1.25 pc, respectively (cf. Figure \ref{cdf} right panel).
Blue dots/blue MST connections and yellow dots/yellow MST connections in Figure \ref{spa}  (bottom left panel) 
represent the locations of  YSOs in the cores and AR of the Sh 2-305 respectively, identified by using the above procedure. 
The central coordinates along with number of YSOs associated with them are given in Table \ref{Tp1}.

We have defined the enclosed area of the cores and AR by using the `convex
hull\footnote{Convex hull is a polygon enclosing all points in a grouping with internal angles between two
contiguous sides of less than 180$^\circ$.}' which is  popularly being
used in the similar studies related to embedded star clusters in  the
SFRs  \citep[][]{2006A&A...449..151S,2009ApJS..184...18G,2014MNRAS.439.3719C,2016AJ....151..126S,2019MNRAS.tmp.3221P}.
The convex hull is computed using the program Qhull\footnote{C. B. Barber, D.P. Dobkin, and H.T. Huhdanpaa, "The Quickhull Algorithm for Convex Hulls," ACM Transactions on Mathematical Software, 22(4):469-483, Dec 1996, www.qhull.org [http://portal.acm.org; http://citeseerx.ist.psu.edu].} on the positions of YSOs associated with cores and AR and their respective convex hulls are plotted in Figure  \ref{spa}
(lower panels) by solid purple and solid yellow lines, respectively.
The total number of YSOs and the number of vertices of each convex hull are given in Table \ref{Tp1}.
We call the northern YSO core as North Clump (hereafter NC), central YSO core as Central Clump (hereafter CC), 
and the southern YSO core as South Clump (hereafter SC).

\subsection{Feedback of massive stars in the Sh 2-305 region}
\label{sec:feedb}

In the literature, we find that the Sh 2-305 hosts two spectroscopically identified O-type stars (i.e., O8.5V and O9.5). 
These massive stars can interact with their surrounding environment via their different 
feedback pressure components (i.e., pressure of an H\,{\sc ii} region $(P_{HII})$, 
radiation pressure (P$_{rad}$), and stellar wind ram pressure (P$_{wind}$)) \citep[e.g.,][]{2012ApJ...758L..28B, 2017ApJ...834...22D}. 
The equations of these pressure components ($P_{HII}$, P$_{rad}$, and P$_{wind}$) are given below \citep[e.g.,][]{2012ApJ...758L..28B}:

\begin{equation}
P_{HII} = \mu m_{H} c_{s}^2\, \left(\sqrt{3N_{uv}\over 4\pi\,\alpha_{B}\, D_{s}^3}\right);   
\end{equation}

\begin{equation}
P_{rad} = L_{bol}/ 4\pi c D_{s}^2; 
\end{equation}

\begin{equation}
P_{wind} = \dot{M}_{w} V_{w} / 4 \pi D_{s}^2; 
\end{equation}

In the equations above, N$_{uv}$ is the Lyman continuum photons, c$_{s}$ is the sound speed in the 
photoionized region \citep[=11 km s$^{-1}$;][]{2009A&A...497..649B}, ``$\alpha_{B}$'' is the radiative recombination 
coefficient \citep[=  2.6 $\times$ 10$^{-13}$ $\times$ (10$^{4}$ K/T$_{e}$)$^{0.7}$ cm$^{3}$ s$^{-1}$; see][]{1997ApJ...489..284K}, $\mu$ is the mean molecular weight in 
the ionized gas \citep[= 0.678;][]{2009A&A...497..649B}, m$_{H}$ is the hydrogen atom mass, 
$\dot{M}_{w}$ is the mass-loss rate, 
V$_{w}$ is the wind velocity of the ionizing source, and L$_{bol}$ is the bolometric luminosity of 
the ionizing source. 

For O9.5V and O8.5V stars, we have considered $L_{bol}$ = 66070 L$_{\odot}$ and 93325 L$_{\odot}$ \citep[][]{1973AJ.....78..929P},
$\dot{M}_{w}$ $\approx$ 1.58 $\times$ 10$^{-9}$ M$_{\odot}$ yr$^{-1}$ and  1.6 $\times$ 10$^{-8}$ M$_{\odot}$ yr$^{-1}$ \citep[][]{2009A&A...498..837M}, 
V$_{w}$ $\approx$ 1500 km s$^{-1}$ and 3051 km s$^{-1}$ \citep[][]{2017A&A...598A..56M}, 
N$_{uv}$ = 1.2 $\times$ 10$^{48}$ and 2.8 $\times$ 10$^{48}$ s$^{-1}$  \citep[][]{1973AJ.....78..929P}, respectively.
D$_{s}$ is the projected distance from the location of the massive O-type stars. 
All of the above pressure components driven by these two massive stars are 
estimated for different D$_{s}$ values i.e., 1 pc (inner core region of Mayer 3 cluster), 
2.4 pc (extent of Mayer 3 cluster and radio emission boundary), 
5 pc (mean distance of the two newly identified clumps), 7.5 pc 
\citep[distance of a compact radio source,][]{1995A&AS..114..557R}, 
and 10 pc (outer field region) and are given  in Table~\ref{tab1}.

\section{Discussion}

\subsection{Physical properties of YSO cores and the active region}

Stellar clusterings in SFRs show a wide range of sizes, morphologies and star numbers and
it is important to quantify these numbers accurately to have clues on star formation events
\citep[cf.][]{2008ApJ...674..336G,2009ApJS..184...18G,2011ApJ...739...84G,2014ApJ...787..107K,2014MNRAS.439.3719C,2016AJ....151..126S}.
In the following  sub-sections, we will investigate the physical properties of the cores and
AR identified in the Sh 2-305.

\subsubsection{Core morphology and structural $Q$ parameter}

We estimated the area  `$A_{hull}$' of the each core and AR by using the convex hull of the
data points, normalized by an additional geometrical factor taking into account
the ratio of the number of objects inside and on the convex hull 
\citep[see,][for details]{1983ApJ...268..527H,2006A&A...449..151S,2016AJ....151..126S}.
We also define the cluster radius, $R_{hull}$, as the radius of a circle with 
the same area, $A_{hull}$, and the circular radial size, $R_{circ}$, as half
of the largest distance between any two members i.e, the radius
of the minimum area circle that encloses the entire grouping, and their derived values for 
the identified cores/AR are given in Table \ref{Tp1}.
The aspect ratio $R^2_{circ}\over{R^2_{hull}}$, which is a measurement of
the circularity of a cluster \citep{2009ApJS..184...18G}, is also given in Table \ref{Tp1}, 
for each region.
The $R_{hull}$ values of the cores range between 0.7 and 1.5 pc (cf. Table \ref{Tp1}),
which is within the range to the values reported for other embedded clusters/cores in the SFRs
\citep{2009ApJS..184...18G,2014MNRAS.439.3719C,2016AJ....151..126S}.
The NC is showing elongated morphology  with aspect ratio = 1.4, similar
to reported values for the cores in SFRs  \citep{2009ApJS..184...18G,2014MNRAS.439.3719C,2016AJ....151..126S}.
The CC and SC are showing almost circular morphology. 
The AR is showing  highly elongated morphology (aspect ratio = 2.76) with
$R_{hull}$ of 3.48 pc ($\sim3^\prime$.23). 
The total number of YSOs in the AR is 96, out of which 74 ($\sim$77\%) falls in the cores.
These numbers are similar to those given in the literature i.e., 
62\% for low mass embedded clusters \citep{2009ApJS..184...18G}, 66\% for massive embedded clusters \citep{2014MNRAS.439.3719C} and  60\% for cores in bright rimmed clouds \citep{2016AJ....151..126S}.
The YSOs in the cores have almost similar surface density of $\sim$5.15 pc$^{-2}$ (mean value), whereas
the AR is slightly less dense (2.53 pc$^{-2}$, cf., Table \ref{Tp1}). 
The peak surface densities vary between 11 - 31 pc$^{-2}$.
The CC is having higher density/ shorter NN2 MST length as compared to other cores (cf. Table \ref{Tp1}), indicating a strong clustering of YSOs in the center of this region. 

The spatial distribution of YSOs associated with a SFR can also be investigated
by their structural $Q$ parameter values.  The $Q$ parameter 
\citep{2004MNRAS.348..589C,2006A&A...449..151S} is used to measure the level of
hierarchical versus radial distributions of a set of points, and it is defined by
the ratio of the MST normalized mean branch length and the normalized mean separation between points \citep[cf.][for details]{2014MNRAS.439.3719C}.
According to \citet{2004MNRAS.348..589C}, a group of points distributed radially will
have a high $Q$ value ($Q$ $>$ 0.8), while clusters with a more fractal
distribution will have a low $Q$ value ($Q$ $<$ 0.8).
We find that for the NC, CC and AR, the $Q$ values are less than 0.8, 
whereas for the SC the $Q$ value is greater than 0.8 (cf. Table \ref{Tp1}).
The AR is showing a highly fractured distribution of stars ($Q$=0.53), 
which is obvious with the identification of three sub-clusterings in this region.
\citet{2014MNRAS.439.3719C} have found a weak trend in the distribution of $Q$ 
values per number of members, suggesting a higher occurrence of sub-clusters merging
in the most massive clusters, which decreases the value of the $Q$ parameter.
For our sample, we also found that the cores having higher number of sources have
lower $Q$ value (cf. Table \ref{Tp1}). 

\subsubsection{Associated molecular material}

In the $Herschel$ column density map (see Figure \ref{cd}a), we have observed 
 the distribution of lower column density materials in the CC region as compared to the outer regions including the SC 
(Clump ID = 20 in Table \ref{cd1}) and NC (Clump ID = 2 in Table \ref{cd1}).
We have also calculated the mean and peak $A_V$ values of the identified cores/AR (cf. Table \ref{Tp2})
using the extinction maps generated earlier in Section 3.7. 
We found that the SC and NC are the obscured clumps, 
whereas CC has a comparatively lower value of $A_V$, indicating that
the central region is devoid of molecular material may be due to the
radiative effects of massive stars in the region.
The $A_V$ value for the AR is $\sim$ 6.7 mag, 
which is lower than NC and SC but higher than CC.

We have also calculated the  molecular mass of the identified cores/AR 
using the extinction maps discussed in Section 3.7. 
First, we have converted the average $A_V$ value (corrected for the 
foreground extinction, cf. Section 3.3) in each grid of our map into $H_2$ column density 
using the relation given by  \citet{1978ApJS...37..407D} and \citet{1989ApJ...345..245C},  i.e.
$\rm N(H_2) = 1.25\times10^{21} \times A_V ~~cm^{-2}~~ mag^{-1}  $.
Then, this $H_2$ column density has been integrated over the convex hull of each region
and multiplied by the $H_2$ molecule mass to get the molecular mass of the cloud.
The extinction law, $A_K/A_V=0.090$ \citep{1981ApJ...249..481C} has been 
used to convert $A_K$ values to $A_V$.
We have also calculated dense gas mass M$_{0.8}$ in each core/AR which is the mass above a column density equivalent to A$_K$ = 0.8 mag
\citep[as explained in][]{2014MNRAS.439.3719C}.
The properties of the molecular clouds associated with the cores and ARs
are listed in Table \ref{Tp2}. In our sample of cores/AR, we can easily observe that with increase in the
molecular material, more number of YSOs are formed. 

We have also calculated the fraction of the Class\,{\sc i} objects among all 
the YSOs (cf. Table \ref{Tp2}) as an indicator of the ``star formation age'' of a region.
The SC seems to be the youngest of all selected regions, whereas the CC is having age  more than that
of the whole AR. To confirm further, we have also given the mean age and mass  of YSOs in each 
region, determined by using the SED fitting (cf. Section 3.5).
Clearly, the SC is the youngest and the most massive as compared to other regions, whereas 
the CC is the oldest (age more that the age of AR) and less massive.
Also, there is no dense gas in the CC, while the SC has most 
available dense gas in our sample of cores.
This is also in agreement with the earlier studies \citep{2009ApJS..184...18G, 2011ApJ...739...84G,2016AJ....151..126S} where 
the youngest stars are found in regions having denser molecular material as compared to more evolved  PMS stars.

\subsubsection{Jeans Length and star formation efficiency}

\citet{2009ApJS..184...18G} analyzed  the spacings of the YSOs in the 
stellar cores of 36 star-forming clusters and 
suggested that Jeans fragmentation is a starting
point for understanding the primordial structure in SFRs.
We have also calculated the minimum radius required for the gravitational collapse of a 
homogeneous isothermal sphere (Jeans length `$\lambda_J$')
in order to investigate the fragmentation scale by using the formulas given in \citet{2014MNRAS.439.3719C}.
The Jeans lengths for the cores in the Sh 2-305 are in between 0.83 pc to 1.13 pc, which is comparable
to the values given in \citet{2009ApJS..184...18G}, \citet{2014MNRAS.439.3719C}, and \citet{2016AJ....151..126S}.

We have also compared $\lambda_J$ and the mean separation `$S_{YSO}$' between cluster members
and found that the ratio $\lambda_J/S_{YSO}$ for the AR has a value of  4.7 which is also similar to the
values given in  \citet[4.9,][]{2016AJ....151..126S} and  \citet[4.3,][]{2014MNRAS.439.3719C}, respectively.
The CC has the highest value and the SC has the lowest value of this ratio (cf. Table \ref{Tp2}). 
The present results indicate a non-thermal driven fragmentation since it took place at
scales smaller than the Jeans length \citep[see also,][]{2014MNRAS.439.3719C}.
We have also found that the variations in the peak YSOs density is proportional to the
Jeans length  of the three identified stellar clusters  in the region, i.e.,
the CC has maximum YSOs density and longest Jeans length, 
whereas the SC has minimum YSOs density and the shortest Jeans length.

The wide range in observed YSO surface densities provides
an opportunity to study how this quantity is related to the
observed  star formation efficiency (SFE) and the properties
of the  associated molecular cloud \citep{2011ApJ...739...84G}. 
\citet{2009ApJS..181..321E} showed that the YSO clusterings 
tend to exhibit higher SFE (30\%) than their lower density surroundings (3\%-6\%). 
\citet{2008ApJ...688.1142K} found SFEs of $>$10\%-17\% for high surface
density clusterings and 3\% for lower density regions. 
\citet[][]{2016AJ....151..126S} found SFEs between 3 \% and 30 \% with an average of $\sim$14 \% in the
 cores associated with eight bright-rimmed clouds.	
\citet{2014MNRAS.439.3719C} have obtained the SFE of a range of 3-45 \% with an  average of $\sim$ 20 \% 
for the sample of embedded clusters.
We have also calculated the SFE, defined as the percentage of
gas  mass converted into stars by using the cloud mass derived from $A_K$  
inside the cluster convex hull area and the number of YSOs found in the same area \citep[see also][]{2008ApJ...688.1142K}, and is given in Table \ref{Tp2} for each of our selected regions.

We have assigned mass to each YSOs as determined by the SED fitting analysis and the molecular mass of the selected region as determined by the reddening map (cf. Tables \ref{Tp1} and \ref{Tp2}). We found that the SFEs vary between 15.6 \% and 36.5\% for cores and 8.5\% for the AR.
These numbers are comparable to previously determined values discussed above
and are in agreement with the efficiencies needed to go
from the core MF to the IMF
\citep[e.g., 30\% 
in the Pipe nebula and 40\% 
in Aquila, from][respectively]{2007A&A...462L..17A, 2010A&A...518L.102A}.
The SC is clearly showing the highest SFE, indicating that this is a region 
where a very active massive star formation  is going on (cf. Table \ref{Tp2}).
The SFE is not correlated with the number of the members of each region.
Similar results are also shown in \citet[][]{2016AJ....151..126S} and \citet{2014MNRAS.439.3719C}, 
indicating  that the feedback processes may impact only in the later stages of the cluster evolution.
The present SFE estimate (8.5\%) of the whole Sh 2-305 region is found to be similar to that of the Sh 2-148 region \citep[8\%,][]{1991MNRAS.249..385P} 
which also contains two ionizing massive stars (O8 V and B2V).

\begin{figure*}
\centering
\includegraphics[width=0.53\textwidth]{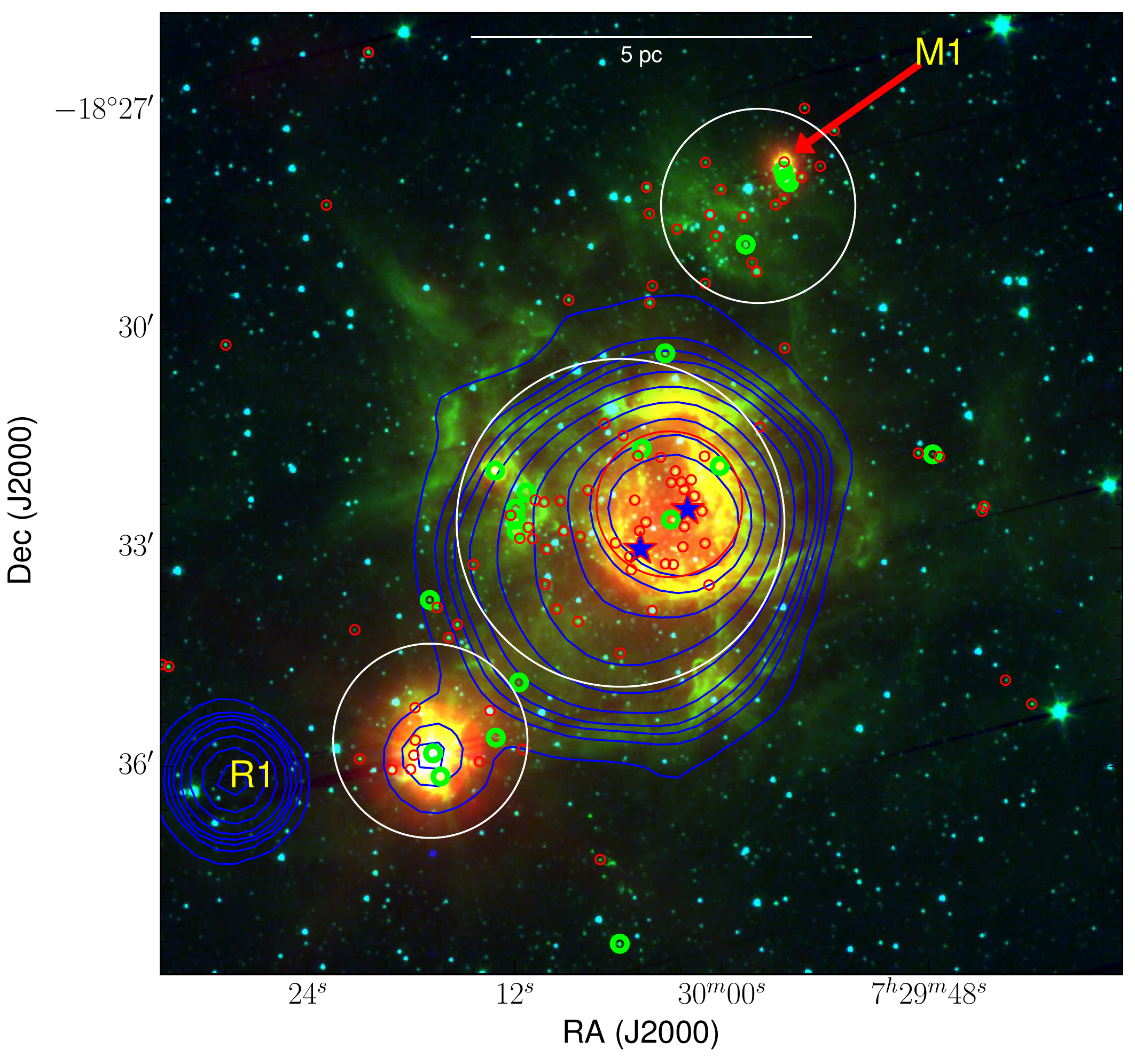}
\includegraphics[width=0.462\textwidth]{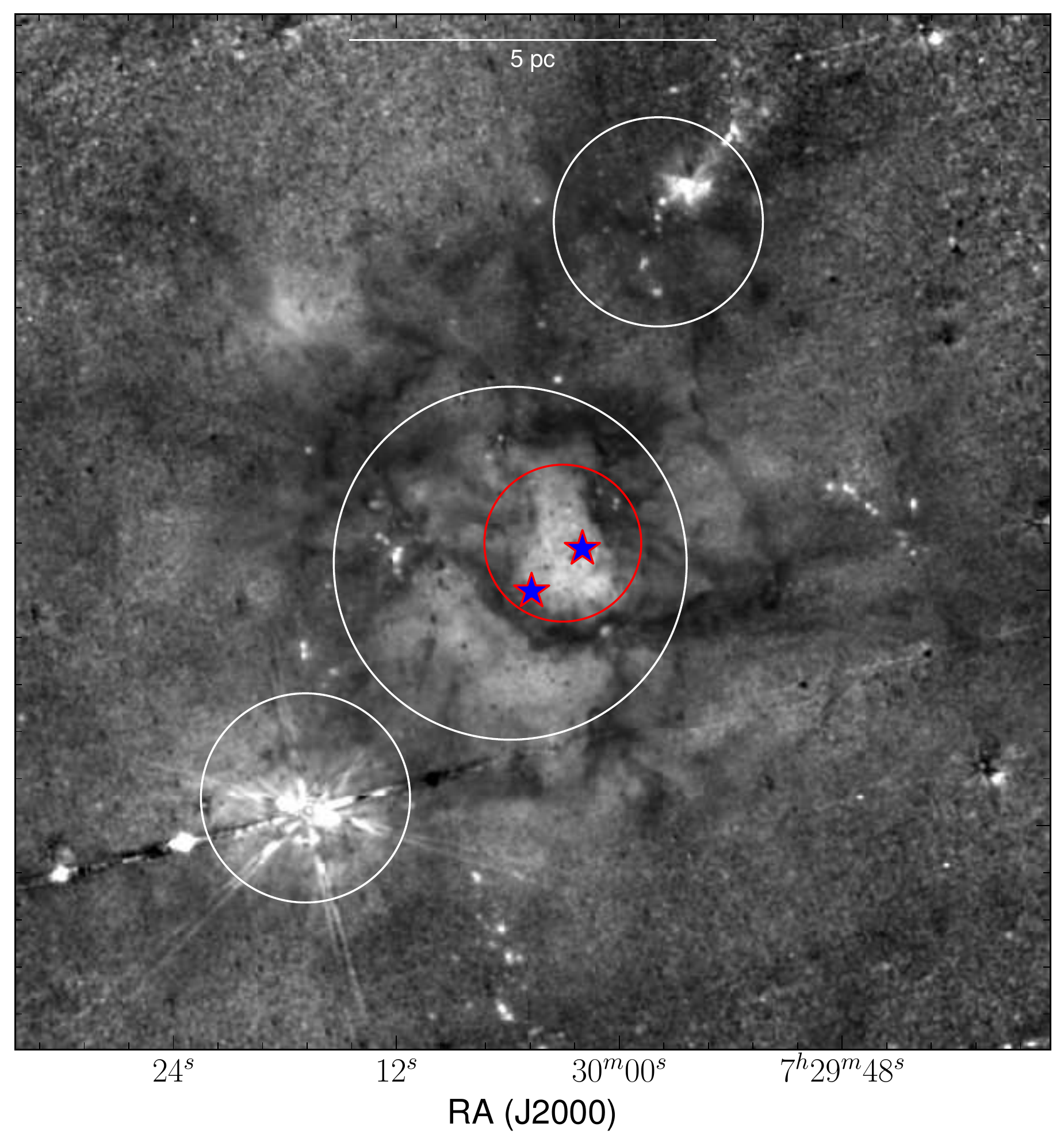}
\caption{\label{sfg} Left panel: Color composite image of Sh 2-305 obtained by using 22 $\mu$m (red), 
3.6 $\mu$m (green) $\mu$m and 2.2 $\mu$m (blue) images taken from WISE, $Spitzer$ and 2MASS, respectively.
NVSS 1.4 GHz radio continuum contours are shown by blue curves
at 10, 20, 30, 40 and 50 \% of the peak flux value (0.0145 Jy). 
Positions of the two previously identified massive stars (O8.5V:VM4 and O9.5:VM2) 
are shown by blue star symbols. The white circles represent the three sub-clusterings identified in the present analysis.
The red circle enclosing the CC is the core region of the central clustering (cf. Section 3.1).
Right panel: {\it Spitzer} ratio map of 4.5 $\mu$m/3.6 $\mu$m emission. The ratio map is smoothened 
using a Gaussian function with radius of four pixels.
}  
\end{figure*}

\subsection{MF and KLF slope}

The higher-mass stars mostly
follow the Salpeter MF \citep{1955ApJ...121..161S}. At lower
masses, the IMF is less well constrained, but appears to flatten
below 1 M$_\odot$  and exhibits fewer stars of the lowest masses \citep{2002Sci...295...82K, 2003PASP..115..763C, 2016ApJ...827...52L}.
In this study, we find a change of MF slope from the high to low mass end  with a turn-off
at around 1.5 M$_\odot$.
This truncation of MF slope at a bit higher mass bins has often been noticed in other SFRs
also under the influence of massive OB-type stars \citep{2007MNRAS.380.1141S, 2008MNRAS.383.1241P, 2008MNRAS.384.1675J,2017MNRAS.467.2943S}.
While the higher-mass domain is thought to be mostly formed through fragmentation and/or accretion onto
the protostellar core \citep[e.g.,][]{2002ApJ...576..870P, 2006MNRAS.370..488B} 
in the low-mass and substellar regime additional
physics is likely to play an important role. The density, velocity
fields, chemical composition, tidal forces in the natal molecular
clouds, and photo-erosion in the radiation field of massive stars in
the vicinity can lead to different star formation processes and
consequently some variation in the characteristic mass (turn-off point) of the IMF \citep{2002ApJ...576..870P, 2004A&A...427..299W, 2005MNRAS.356.1201B, 2009MNRAS.392.1363B}. 

We have also found that the MF slopes are  bit steeper ($\Gamma\simeq$-1.7) than the
 \citet{1955ApJ...121..161S} value i.e., $\Gamma$=-1.35 in  a mass range of $1.5<M_\odot<6.5$, 
indicating the abundance of low-mass stars,  probably 
formed due to the positive feedback of the massive stars  in this region.
Most of the sensitive studies of massive SFRs
\citep[see, e.g.][]{2009MNRAS.396.1665L, 2009A&A...501..563E,2011AA...530A..34P} 
found large numbers of low-mass stars in agreement with the expectation from
the ``normal" field star MF and supports the notion that
OB associations and  massive star clusters are the
dominant supply sources for the Galactic field star population, as
already suggested by \citet{1978PASP...90..506M}.

The KLF slope value  ($\alpha = 0.40\pm0.04$) for the Mayer 3 cluster is  similar to  the average
slopes ($\alpha \sim 0.4$) for embedded clusters \citep{1991ASPC...13....3L,2003ARAA..41...57L}. 
For north and south sub-clusterings, the KLF slope is lower  ($\alpha \simeq 0.2$). Low KLF values (0.27 - 0.31) has been  found for some of the very young star clusters 
(Be 59: \citet{2008MNRAS.383.1241P}; Stock 8: \citet{2008MNRAS.384.1675J}; W3 Main: \citet{2004ApJ...608..797O}).
This indicate that the north and south sub-clusterings are bit younger as compared to the central cluster  Mayer 3.

\subsection{Star formation in the Sh 2-305}

\citet{1995A&AS..114..557R} reported the velocity of H$\alpha$ emission to be 38 km s$^{-1}$ in Sh 2-305, which was noticeable different as compared to the radial velocities from CO emission (i.e., $V_{LSR}\sim$43 km s$^{-1}$).
They suggested that  the champagne flow mechanism \citep{1979A&A....71...59T}
might be responsible for this discrepancy between the ionized and molecular material.
In this mechanism, if the H\,{\sc ii} region is on the near side of the molecular cloud,
the ionized gas flows away from the molecular material, more or less in the direction of an observer.
Massive stars can influence their surrounding and trigger the formation of new generation of stars,
either by sweeping the neighboring molecular gas into a
dense shell which subsequently fragments into pre-stellar cores
\citep[e.g.,][]{1977ApJ...214..725E,1994MNRAS.268..291W,1998ASPC..148..150E}
or by compressing pre-existing dense clumps
\citep[e.g.,][]{1982ApJ...260..183S,1989ApJ...346..735B,1994A&A...289..559L}.
The former process is called `collect and collapse' and the later
`radiation driven implosion'.

There are many examples in literature in which the massive stars have triggered
the formation of YSOs in the GMCs \citep[][and references therein]{2016MNRAS.461.2502Y,2017MNRAS.467.2943S,2017MNRAS.472.4750D}.
YSOs are mostly found embedded in the GMCs with large variation in their numbers 
imprinting the fractal structure of the GMCs. Thus, the density variations of the YSOs (i.e., embedded clusters)
provide a direct observational signature of the star formation processes.
In this study also, we have identified a population of YSOs distributed in the 
Sh 2-305 in three major clusterings, i.e.,
the central cluster of YSOs (CC) lies within the boundary of cluster Mayer 3 and other two (NC \& SC)
in north and south directions at a distance of $\sim$5 pc from the CC. 
The CC lies in a region with lowest column density, which 
can be attributed to the dispersion of the gas by two massive stars 
located in this region (VM2 and VM4, cf. Figure \ref{image1}). 
The SC is found to be highly obscured and have the highest value of SFE and the fraction of Class\,{\sc i} sources
(cf. Table \ref{Tp2}), suggesting an active star formation in this core.

To further investigate the influence of massive stars, 
in the left panel of Figure \ref{sfg} we show the color composite image of the
Sh 2-305 obtained by using 22 $\mu$m (red),
3.6 $\mu$m (green) and 2.2 $\mu$m (blue) images.
Spatial distribution of the radio continuum emission (NVSS 1.4 GHz), massive stars, and the
YSOs are also shown.
This region also has a compact radio source (R1 in the left panel of Figure \ref{sfg}), possibly a young ultra-compact 
H\,{\sc ii} region lying at about $\sim7^\prime$ from the Sh 2-305 center \citep{1993ApJS...86..475F,1995A&AS..114..557R}
and an H$_2$O maser coinciding with the infrared source IRAS~07277-1821 (M1 in the left panel Figure \ref{sfg}) \citep{1988A&A...191..323W}. 
SC and NC appears to host radio continuum emission and H2O maser, respectively, 
which indicates that high (or massive) stars are forming inside these structures.
Though, in general radio counterpart traces more evolved state than H2O maser, 
but in the present study, it is difficult to differentiate the 
evolution status of NC and SC considering the errors on the derived parameters. The ionized region (radio continuum emission) and the heated dust grains (22 $\micron$ emission) 
are distributed towards the CC in the Sh 2-305.
The CC is also surrounded by 3.6 $\mu$m emission which covers the prominent 
PAH features at 3.3 $\mu$m, indicative of photon dominant region (PDR)
under the influence of feedback from massive stars \citep[see e.g.][]{2004ApJ...613..986P}.
The massive stars, VM2 and VM4, are located near the
center of CC and their high energy feedback might be responsible for these emissions.

The right panel of Figure \ref{sfg} shows the {\it Spitzer} ratio map of 4.5 $\mu$m/3.6 $\mu$m emission, revealing the existence of
dark and bright regions. The {\it Spitzer} band at 4.5 $\mu$m contains a prominent molecular
hydrogen line emission ($\nu$ = 0--0 $S$(9); 4.693 $\mu$m) and the Br-$\alpha$ emission (at 4.05 $\mu$m).
The {\it Spitzer} band at 3.6 $\mu$m includes the PAH emission at 3.3 $\mu$m.  
Note that both these {\it Spitzer} images have the same PSF, allowing to remove point-like sources
as well as continuum emission \citep[see][for more details]{2017ApJ...834...22D}.
In the ratio map, the bright regions indicate the excess of 4.5 $\mu$m emission, while the
black or dark gray regions show the domination of 3.6 $\mu$m emission.
The regions with the excess of 4.5 $\mu$m emission (i.e., bright regions) are seen in the direction of the NVSS radio continuum emission, suggesting the
presence of the Br-$\alpha$ emission. Furthermore, the bright regions ($\sim$ $\alpha$$_{2000}$ =07$^{h}$29$^{m}$55.7$^{s}$, $\delta$$_{2000}$ = -18$\degr$28$\arcmin$03$\arcsec$) away from the NVSS radio continuum emission appear to trace the outflow activities in the Sh 2-305, where YSOs are also investigated.
Considering the PAH feature at 3.3 $\mu$m in the 3.6 $\mu$m band, several dark regions trace 
the presence of PDRs in the  Sh 2-305, indicating the impact of massive O-type stars (i.e., VM2 and VM4).
Overall, the {\it Spitzer} map displays the signatures of outflow activities and the impact of massive stars in the Sh 2-305.
Hence, star formation activities in Sh 2-305 seem to be influenced by  massive O-type stars.

To quantify the impact of massive O-type stars (i.e., VM2 and VM4) on their surroundings, 
we have calculated the total pressure exerted
by these two  sources (see Section 3.9 and Table \ref{tab1}).
The total pressure (P$_{total}$) at D$_{s}$ = 1 pc (Mayer 3 cluster core), 2.4 pc (Mayer 3 cluster extent), 5 pc (location of two YSOs clumps i.e. NC and SC),
and 7.5 pc (location of the radio source R1) is 6.4$\times$10$^{-10}$,  1.5$\times$10$^{-10}$,  4.8$\times$10$^{-11}$, and 2.6$\times$10$^{-11}$ (dynes\, cm$^{-2}$), respectively.
We have also calculated the   P$_{total}$  at far distance of D$_{s}$ = 10 pc as 1.6$\times$10$^{-11}$ (dynes\, cm$^{-2}$), which
is comparable to the pressure of a typical cool molecular cloud 
($P_{MC}$$\sim$10$^{-11}$--10$^{-12}$ dynes cm$^{-2}$ for a temperature $\sim$20 K 
and particle density $\sim$10$^{3}$--10$^{4}$ cm$^{-3}$) \citep[see Table 7.3 of][]{1980pim..book.....D}.

From these calculations, we can easily see that the pressure near massive stars (a core region having CC) 
is significantly higher as compared to that of a molecular
cloud and then it starts to decrease with the distance.
At the location of NC and SC, the pressure exerted
by ionizing sources is still significant.
Also, the YSOs in the NC and SC are found to be younger than that of the CC (cf. Table \ref{Tp2}), 
implying that the star formation there might have started after CC.
This is in line to the feedback effects from central massive stars.

Therefore, based on the  distribution of warm dust/ionized region near the CC which is surrounded by the PDRs and YSOs,
pressure calculations, age gradient etc., we conclude that massive O-type stars associated with the CC 
might have triggered the formation of younger stars located in the region including  NC and SC.
This argument is also supported with the slopes of MF/KLF, SFEs, mean age of YSOs, fraction of Class\,{\sc i} sources, and dense gas mass,
found in these regions.

Regarding the formation of the central massive stars itself along with the origin of 
radio peak powered by the ultra-compact H\,{\sc ii} region, it requires further detailed analysis.
It has been observationally reported that the merging/collisions of
filamentary structures can form the dense massive star-forming clumps, where the most massive stars
form \citep[e.g.,][]{2012A&A...540L..11S,2014ApJ...791L..23N}. 
Identification of embedded filaments and
characterizing their physical properties (e.g. temperature and column density, velocity profile, etc)
by using high-resolution multiwavelength (from radio to mm, i.e., ionized region, molecular distribution, 
distribution of cold and warm dust) investigation can help us to explore this scenario as well.

\section{Conclusion}

In the present work we studied a Galactic H\,{\sc ii} region Sh 2-305 to understand the 
star formation using deep optical and NIR photometry (V$\sim$22 mag, K$\sim$18.1 mag),
along with multiwavelength archival data.  Following  are the conclusions made from the above study.

\begin{itemize}

\item
Stellar density distribution generated by using NIR data has been used to study the structure of the molecular cloud and clusterings in the Sh 2-305. We have found three stellar sub-clusterings in this region, i.e, one in the central region (Mayer 3) and one each towards north and south directions of the Sh 2-305. One hundred thirty seven cluster members on the basis of Gaia DR2 proper motion data are found to be associated with Mayer 3. These member stars indicate a normal reddening law towards this region. The foreground reddening and distance to the cluster come out to be $E(B-V)$ = 1.17 mag and  3.7 kpc, respectively.

\item 
We identified $\sim$116 YSOs in the $\sim18^\prime.5\times18^\prime.5$ FOV of Sh 2-305 on the basis of excess IR emission. Out of
them 87\% (96) are  Class\,{\sc ii} and 13\% (20) are Class\,{\sc i} sources. Age and mass of 98 YSOs has been estimated  using the SED fitting analysis. The masses of the YSOs range between 0.8 to 16.2 
M$_\odot$, however, a majority ($\sim$80\%) of them ranges between 0.8 to 4.0 M$_\odot$ . It is found that $\sim$91\% (89/98) of the sources have ages between 0.1 to 3.5 Myr. The region indicates a differential reddening which vary between $A_V$=2.2 to 23 mag, indicating a clumpy nature of gas and dust in this region. The average age, mass and $A_V$ for this sample of YSOs are 1.8 Myr, 2.9 M$_\odot$ and 7.1 mag, respectively.

\item
MST analysis of YSO's location yields three cores viz. North Clump (NC), Central Clump (CC) and South Clump (SC) which match well with the  stellar density distribution. The active region contains 96 YSOs, out of which 74 (77\%) belong to these cores. The average MST branch length in these cores is found to be $\sim$0.3 pc.

\item
Basic structural parameters  of these cores have been estimated. The core size and aspect ratio vary between 1.4 - 3 pc and 0.73 - 1.40, respectively. The CC is showing higher YSOs density as compared to other cores. Mean extinction value for the NC, CC and SC is 7.4 mag, 5.9 mag, 10.1 mag, respectively. The molecular mass associated with NC, CC , SC and AR is 379.8 M$_\odot$, 453.8 M$_\odot$, 
96.1 M$_\odot$  and 3199.7 M$_\odot$,  respectively. We have found that the SC has the highest value of dense gas mass (55.8 M$_\odot$), whereas the CC has no dense gas mass associated with it. The Jeans length `$\lambda_J$' is calculated as 1.13, 1.35 and  0.83 pc for NC, CC and SC, respectively. SFE has the highest value (36.5 \%) for SC, 21.4 \% for CC and the lowest value (15.6 \%) for NC. 

\item
The slope of the MF, $\Gamma$, in the  mass range $1.5<M_\odot<6.5$ is found to be $\sim$ -1.7, which is steeper than the \citet{1955ApJ...121..161S} value ($\Gamma$=-1.35) and suggests an abundance of low-mass stars, probably formed due to the positive feedback of the massive stars in this region. The KLF slope values for Mayer 3 cluster ($\alpha = 0.40\pm0.04$) and the north and south sub-clusterings ($\alpha \simeq 0.2$) indicate that the north and south sub-clusterings are bit younger as compared to the central cluster Mayer 3.

\item
It is found that the two massive O-type stars (VM2 and VM4) located in the center of Sh 2-305 might have triggered the formation of younger stars. This argument is also supported with the  distribution of warm dust/ionized region surrounded by the PDRs and YSOs, pressure calculations, age gradient, slopes of MF/KLF, SFEs, mean age of YSOs, fraction of Class\,{\sc i} sources, and dense gas mass, for the sub-clusterings found in the region.

\end{itemize}

\section*{Acknowledgments}

We thank the anonymous referee for the helpful comments. The observations reported in this paper were obtained by using the 1.3m telescope, Nainital, India and  the 2 m HCT at IAO, Hanle, the High Altitude Station of Indian Institute of Astrophysics, Bangalore, India. We also acknowledge TIFR Near Infrared Spectrometer and Imager mounted on 2 m HCT  using which we have made NIR observation. This publication makes use of data from the Two Micron All Sky Survey, which is a joint project of the University of Massachusetts and the Infrared Processing and Analysis Center/California Institute of Technology, funded by the National Aeronautics and Space Administration and the National Science Foundation. This work is based on observations made with the $Spitzer$ Space Telescope, which is operated by the Jet Propulsion Laboratory, California Institute of Technology under a contract with National Aeronautics and Space Administration. This publication makes use of data products from the Wide-field Infrared Survey Explorer, which is a joint project of the University of California, Los Angeles, and the Jet Propulsion Laboratory/California Institute of Technology, funded by the National Aeronautics and Space Administration. We acknowledge support of the Department of Atomic Energy, Government of India, under project no. 12-R\&D-TFR-5.02-0200.

\bibliography{ms-S305-R1}{}
\bibliographystyle{aasjournal}


\begin{table}
\centering
\caption{\label{log}  Log of  observations.}
\begin{tabular}{@{}rrr@{}}
\hline
Telescope/Instrument & Comments& Exp. (sec)$\times$ No. of frames\\
$[$Date of observations$]$& $[$Filter$]$   &\\
\hline
 1.3 m DFOT/2K CCD& Optical imaging of Sh 2-305&\\
$[$ 2014 Jan 29$]$&$U$   &  $300\times4$\\
$"$&$B$   &  $300\times3,30\times3$\\
$"$&$V$   &  $300\times3,60\times3,10\times3$\\
$"$&$R_c$ &  $300\times3,60\times3,10\times3$\\
$"$&$I_c$ &  $300\times3,60\times3,10\times3$ \\
$[$2017 Mar 19 $]$&$I_c$    & $1800\times4$\\
$[$2017 Mar 23$]$&$V$ &    $1800\times2$\\
& Optical imaging of Standard field (SA98)&\\
$[$2014 Jan 29$]$&$U$   &  $300\times5,120\times1$\\
$"$&$B$   &  $120\times2,40\times3,60\times1$\\
$"$&$V$   &  $30\times3,60\times3$\\
$"$&$R_c$ &  $30\times3,20\times3$\\
$"$&$I_c$ &  $30\times3,20\times3$\\
2m HCT/TIRSPEC&NIR imaging of Sh 2-305 (five pointings)&\\
$[$2016 Feb 1  \& 2017 Oct 9$]$&$J$   &  $20\times35$\\
$"$&$H$   &  $20\times35$\\
$"$&$K$   &  $20\times35$\\
\hline

\end{tabular}
\end{table}

\begin{table}
\centering
\caption{\label{cftt} Completeness of the photometric data}
\begin{tabular}{@{}rcccc@{}}
\hline
Band& Number of & Detection & Completeness &  Completeness Limit (up to 80 \%)  \\
    &   sources & Limit (mag)& Limit (up   80 \%) (mag)&  Mass ($M_\odot$$^c$)\\   
\hline
$U$    &712                 &20.0&$-$&$-$\\
$B$    &1249                &20.9&$-$&$-$\\
$V$    &2646                &21.9&20.5&0.7\\
$R_c$  &2467                &20.9&$-$&$-$\\
$I_c$  &2641                &20.0&22.0&0.1\\
$J$    &1062$^a$+ 1104$^b$  &19.2&17.0&0.7\\
$H$    &1796$^a$+ 1134$^b$  &18.7&16.0&0.8\\
$K$    &1796$^a$+ 1134$^b$  &18.1&15.5&1.1\\
$[3.6]$&2246                &16.3&14.2&1.4\\
$[4.5]$&2246                &15.7&13.5&1.5\\
\hline
\end{tabular}

$^a$: data from $TIRSPEC$;
$^b$: data from  2MASS;
$^c$: for distance = 3.7 kpc and $E(B-V)$ = 1.17 mag.
\end{table}

\begin{table*}
\centering
\caption{\label{PMT} Sample of 137 stars identified as a member of the Sh 2-305. 
The complete table is available in the electronic form only.}
\begin{tabular}{ccrcccccc}
\hline
ID& $\alpha_{(2000)}$&$\delta_{(2000)}$& Parallax$\pm\sigma~~~~~$&$\mu_\alpha\pm\sigma$&$\mu_\delta\pm\sigma$ & $G$ & $G_{BP}-G_{RP}$ & Probability\\
& {\rm $(degrees)$} & {\rm $(degrees) $} & (mas)&  (mas/yr)& (mas/yr) & (mag) & (mag) & (Percentage)\\
\hline
1 & 112.506538&-18.626102&$  0.108\pm 0.500$&$-1.759\pm 0.591$&$1.351\pm 0.805$&$19.979$&  2.700  &  81  \\
2 & 112.543953&-18.622734&$  0.082\pm 0.113$&$-1.769\pm 0.128$&$2.049\pm 0.166$&$17.645$&  1.569  &  98  \\
3 & 112.491783&-18.620470&$  0.866\pm 0.247$&$-1.729\pm 0.294$&$1.907\pm 0.352$&$18.883$&  1.977  &  95  \\
4 & 112.505150&-18.618370&$  0.198\pm 0.545$&$-1.536\pm 0.618$&$2.235\pm 0.741$&$20.075$&  1.849  &  89  \\
\hline
\end{tabular}
\end{table*}

\begin{table*}
\tiny
\centering
\caption{\label{data1_yso}  A sample table containing information for 116 YSOs   
identified in the Sh 2-305. 
The complete table is available in an electronic form only. }
\begin{tabular}{@{}r@{ }c@{ }r@{ }l@{ }l@{ }l@{ }l@{ }l@{ }l@{ }l@{ }l@{ }l@{ }l@{ }c@{ }}
\hline
ID& $\alpha_{(2000)}$ & $\delta_{(2000)}$ & $J\pm\sigma$ & $H\pm\sigma$ &   $K\pm\sigma$ &  $[3.6]\pm\sigma$ & $[4.5]\pm\sigma$ & $[3.4]\pm\sigma$& $[4.6]\pm\sigma$& $[12]\pm\sigma$& $[22]\pm\sigma$& Class\\
      &$(degrees)$    & $(degrees)$       & (mag)  &(mag)&  (mag)       &  (mag)         & (mag)             &    (mag)         & (mag)           &(mag)            &(mag)           &                        \\
\hline
1& 112.539932 &  -18.666079 &$-$               &$15.787  \pm0.171 $  &$14.997\pm0.150$ &$13.843\pm0.082$ &$13.189\pm0.051$ &$13.837\pm0.026$ &$12.933\pm0.029$ &$-$               &$- $              & II \\
2& 112.639084 &  -18.660336 &$12.038 \pm0.027$ &$11.469  \pm0.034 $  &$10.863\pm0.024$ &$9.555\pm0.109$  &$9.033\pm0.079$  &$9.623\pm0.022$  &$ 8.912\pm0.018$ &$5.412\pm 0.012$  &$3.68  \pm0.024$  & II \\
3& 112.524623 &  -18.640574 &$-$               &$15.332  \pm0.130 $  &$14.344\pm0.092$ &$12.871\pm0.059$ &$12.016\pm0.048$ &$12.309\pm0.022$ &$11.024\pm0.021$ &$8.505\pm0.081 $  &$5.862 \pm0.047$  & I \\   
4& 112.529347 &  -18.621298 &$-$               &$15.837  \pm0.176 $  &$14.785\pm0.112$ &$14.013\pm0.089$ &$13.362\pm0.044$ &$13.699\pm0.054$ &$12.632\pm0.050$ &$-$               &$5.883  \pm0.170$ & II \\

\hline
\end{tabular}
\end{table*}

\begin{table*}
\centering
\caption{\label{data3_yso}  A sample table containing information for 28 optically identified YSOs. IDs are the same as in Table \ref{data1_yso}. 
The complete table is available in an electronic form only.
 }
\begin{tabular}{@{}rcccccc@{}}
\hline
ID &  $V\pm \sigma$  & $(U-B) \pm \sigma$ &  $(B-V) \pm \sigma$ &  $(V-R_c)\pm \sigma$& $(V-I_c)\pm \sigma$  \\
  &  (mag)& (mag) &  (mag)& (mag)&  (mag) \\
\hline
 $  2$ &$14.534\pm0.003$&   $0.382\pm0.013$   &$ 1.047\pm0.009 $&  $0.541\pm0.004$&     $1.280\pm0.005 $ \\
 $  7$ &$17.699\pm0.007$&   $-$               &$ 1.737\pm0.019 $&  $0.976\pm0.011$&     $2.142\pm0.011$ \\
 $ 11$ &$17.821\pm0.047$&   $-0.222\pm0.170$  &$ 2.089\pm0.056 $&  $1.769\pm0.063$&     $3.237\pm0.076$\\
 $ 16$ &$15.555\pm0.004$&   $0.233\pm0.018$   &$ 0.963\pm0.009 $&  $0.475\pm0.017$&     $1.272\pm0.012$\\
\hline
\end{tabular}
\end{table*}

\begin{table*}
\centering
\caption{\label{data4_yso}  A sample table containing stellar parameters of selected 98 YSOs derived using the SED fitting analysis. IDs are the same as in Table \ref{data1_yso}. The complete table is available in an electronic form only.}
\begin{tabular}{@{}c@{ }c@{ }c@{ }c@{ }c@{ }c@{}}
\hline
 ID &   $N_{data}$ & $\chi^2_{min}$  &  $A_V$$\pm \sigma$     &  Age$\pm \sigma$ &  Mass$\pm \sigma$ \\
       &    &            &   (mag)       & (Myr)      &($M_{\odot}$) \\
\hline
 1 &  9 & 2.2 &$6.2\pm3.8$&$0.4\pm1.2$&$0.8\pm0.8$\\
 2 & 14 & 19.1 &$2.7\pm0.3$&$0.6\pm0.1$&$4.9\pm0.3$\\
 4 &  9 & 11.0 &$5.0\pm3.6$&$0.6\pm1.6$&$1.6\pm1.4$\\
 6 &  5 &  0.4 &$9.6\pm3.1$&$1.7\pm2.5$&$3.0\pm1.2$\\
\hline
\end{tabular}
\end{table*}

\begin{table*}
\centering
\caption{\label{Tp3} The values of mass function (MF)  and $K$-band luminosity function (KLF) slopes
for the stellar sub-clusterings identified in the Sh 2-305 (cf. Section 3.1).
}
\begin{tabular}{@{}lccc@{}}
\hline
Region &MF Slope &  KLF Slope\\
    & ($\Gamma$) & $(\alpha)$ \\    
\hline
North sub-clustering  & $-$           & $0.24\pm0.14$ \\
Mayer 3           & $-1.69\pm0.49$& $0.40\pm0.04$ \\
South sub-clustering  & $-1.78\pm0.48$& $0.21\pm0.04$ \\
&&&\\                         
Whole region      &$-1.76\pm0.30$& $0.33\pm0.04$\\
\hline
\end{tabular}
\end{table*}

\begin{table*}
\centering
\caption{\label{cd1} Properties of the  clumps identified using the {\it Herschel} column density map.
Center coordinates of the identified clumps along with the radius (column 4) and mass (column 5) 
of the clumps are given in the table.}
\begin{tabular}{@{}lcccrc@{}}
\hline
ID & $\alpha_{(2000)}$&$\delta_{(2000)}$&Radius& Mass \\
 & {\rm $(^h:^m:^s)$} & {\rm $(^o:^\prime:^{\prime\prime)} $} &  (pc) & (M$_\odot$) \\
\hline
    1 &   07:29:43.1 & -18:24:15 & 0.42   &   85  \\
    2 &   07:29:57.0 & -18:28:03 & 1.32   & 1565  \\
    3 &   07:29:59.1 & -18:29:03 & 0.81   &  410  \\
    4 &   07:30:05.9 & -18:28:33 & 0.60   &  175  \\
    5 &   07:30:05.5 & -18:30:51 & 0.91   &  700  \\
    6 &   07:30:01.2 & -18:31:57 & 0.66   &  320  \\
    7 &   07:29:53.2 & -18:32:03 & 0.28   &   35  \\
    8 &   07:29:44.8 & -18:32:09 & 0.79   &  355  \\
    9 &   07:29:35.9 & -18:33:57 & 0.85   &  415  \\
   10 &   07:30:32.0 & -18:30:57 & 0.53   &  140  \\
   11 &   07:30:13.9 & -18:30:33 & 0.77   &  335  \\
   12 &   07:30:16.8 & -18:31:27 & 0.84   &  465  \\
   13 &   07:30:13.9 & -18:32:15 & 0.96   &  845  \\
   14 &   07:30:07.6 & -18:32:57 & 0.40   &   85  \\
   15 &   07:30:10.5 & -18:34:09 & 0.33   &   50  \\
   16 &   07:30:01.2 & -18:33:39 & 0.69   &  305  \\
   17 &   07:30:01.7 & -18:34:51 & 0.64   &  270  \\
   18 &   07:30:05.9 & -18:35:27 & 0.40   &   80  \\
   19 &   07:30:07.6 & -18:35:39 & 0.55   &  145  \\
   20 &   07:30:16.8 & -18:36:03 & 1.00   &  715  \\
   21 &   07:30:24.4 & -18:37:03 & 0.33   &   50  \\
   22 &   07:30:27.4 & -18:38:57 & 0.32   &   45  \\
   23 &   07:30:06.7 & -18:37:33 & 1.25   &  930  \\
   24 &   07:30:06.7 & -18:38:33 & 1.05   &  670  \\
   25 &   07:30:10.1 & -18:40:03 & 0.90   &  415  \\
\hline
\end{tabular}
\end{table*}

\begin{table*}
\centering
\caption{\label{Tp1} Properties of the identified cores and ARs.
Center coordinates of the identified cores and ARs
along with the total number of the YSOs are given in columns 2, 3 and 4, respectively.
The  vertices of the convex hull, hull radius and circle radius along
with the aspect ratio are given in columns 5, 6, 7 and 8, respectively.
Columns 9 and 10 represent the mean and peak stellar density obtained using the isodensity contours.
Columns 11, 12 and 13 are the mean MST branch length, NN distances and $Q$ parameters, respectively.}
\begin{tabular}{@{}lccrrccccccccc@{}}
\hline
Region& $\alpha_{(2000)}$&$\delta_{(2000)}$&N$^a$&V$^b$& $R_{\rm hull}$& $R_{\rm cir}$& Aspect & $\sigma_{\rm mean}$&$\sigma_{\rm peak}$ & MST$^c$ & NN2$^c$ & $Q$\\
 & {\rm $(^h:^m:^s)$} & {\rm $(^o:^\prime:^{\prime\prime)} $} &  &&  (pc)& (pc)& Ratio & (pc$^{-2}$)& (pc$^{-2}$)  & (pc) &  (pc) &  & \\
\hline
NC &  07:29:58.14 &  -18:28:07.9& 22& 7 &    1.26 & 1.48 & 1.40 & 4.44 & 21.9&0.39 & 0.32 & 0.72 \\
CC &  07:30:06.22 &  -18:32:32.5& 44&11 &    1.54 & 1.59 & 1.07 & 5.94 & 31.3&0.26 & 0.18 & 0.66 \\
SC  &  07:30:18.06 &  -18:35:49.6&  8& 5 &    0.71 & 0.61 & 0.73 & 5.08 & 11.1&0.28 & 0.27 & 0.84 \\
AR  &  07:30:06.05 &  -18:31:53.5& 96&12 &    3.48 & 5.77 & 2.76 & 2.53 & 31.3&0.33 & 0.25 & 0.56 \\
\hline
\end{tabular}

a: Number of YSOs enclosed in the group;
b: Vertices of the convex hull;
c: median branch length;
\end{table*}

\begin{table*}
\centering
\caption{Combined pressure components driven by two massive stars (i.e., O8.5V and O9.5V) at different D$_{s}$ (cf. Section 3.9) .}
\label{tab1}
\begin{tabular}{@{}lccccccr@{}}
\hline
   Pressure    &   D$_{s}$ = 1 pc  &     D$_{s}$ = 2.4 pc&         D$_{s}$ = 5 pc  & D$_{s}$ = 7.5 pc &  D$_{s}$ = 10 pc  \\
(dynes\, cm$^{-2}$)&               &                     &                         &                 &  \\
\hline
 P$_{HII}$     & 4.6$\times$10$^{-10}$  & 1.2$\times$10$^{-10}$ & 4.1$\times$10$^{-11}$& 2.3$\times$10$^{-11}$ & 1.4$\times$10$^{-11}$ \\
 P$_{rad}$     & 1.7$\times$10$^{-10}$  & 3.0$\times$10$^{-11}$ & 6.8$\times$10$^{-12}$& 3.0$\times$10$^{-12}$ & 1.7$\times$10$^{-12}$ \\
 P$_{wind}$    & 2.6$\times$10$^{-12}$  & 4.7$\times$10$^{-13}$ & 1.1$\times$10$^{-13}$& 4.8$\times$10$^{-14}$ & 2.7$\times$10$^{-14}$ \\
&&&&\\                                                                                                        
 P$_{total}$   & 6.4$\times$10$^{-10}$  & 1.5$\times$10$^{-10}$ & 4.8$\times$10$^{-11}$& 2.6$\times$10$^{-11}$ & 1.6$\times$10$^{-11}$ \\
\hline
\end{tabular}
\end{table*}

\begin{table*}
\centering
\scriptsize
\caption{\label{Tp2} Properties of the identified cores and ARs.
Columns  2, 3, 4 represent the Jeans length,  ratio of Jeans
length with mean separation of YSOs, and star formation efficiency (SFE), respectively.
The mean and peak extinction values are given in columns 5 and 6, respectively.
Column 7 represents the cloud mass in the convex hull derived using the extinction maps.
Column 8 represents the mass of the dense gas having $A_K$ greater than 0.8 mag.
Columns 9, 10 and 11 represent the number of Class\,{\sc i}, Class\,{\sc ii} YSOs along with the fraction of Class\,{\sc i} 
YSOs over total YSOs. Columns 12, 13 and 14 represent the mean age and mean mass, derived from the SED analysis along
with the number of YSOs used in that analysis.}
\begin{tabular}{@{}lccccccccccccc@{}}
\hline
Region& $\lambda_J$ & $\lambda_J$/$S_{YSO}$& SFE &  $A_{V_{mean}}^d$ & $A_{V_{peak}}$ &  Mass       & Mass$_{0.8}$ & Class & Class & Frac$^e$ &  Age  & Mass        & N \\
     &(pc)&& (\%) &     (mag)          & (mag)          & (M$_\odot$) & (M$_\odot$)  & (I)   & (II)  & $(\%)$   & (Myr) & (M$_\odot$) & (SED)\\
\hline
NC &  $1.13\pm0.22  $  &   3.4 &$15.6\pm4$&$ 7.4\pm1.5$ &  12.0  & $ 379.8\pm81$ & $  17.8\pm2 $& 4& 18& 18  & $1.7\pm1.2$& $3.2\pm1.6$& 19   \\
CC &  $1.35\pm0.27  $  &   5.2 &$21.4\pm5$&$ 5.9\pm1.4$ &  10.5  & $ 453.8\pm97$ & $  $-$    $& 7& 37& 16  & $1.8\pm1.2$& $2.8\pm1.6$& 39   \\
SC &  $0.83\pm0.16  $  &   2.5 &$36.5\pm6$&$10.1\pm1.4$ &  14.0  & $  96.1\pm17 $ & $  55.8\pm8 $& 2&  6& 25  & $1.2\pm0.9$& $6.9\pm4.7$&  5   \\
&&&&&&&&&&\\ 
AR &  $1.86\pm0.34$  &   4.7 &$ 8.5\pm5$&$ 6.7\pm1.6$ &  17.9  & $3199.7\pm800$ & $ 219.3\pm30$&18& 78& 19  & $1.7\pm1.2$& $3.1\pm2.2$& 78   \\
\hline
\end{tabular}

d = from NIR+2MASS $(H-K)$ map;
e: Class\,{\sc i}/(Class\,{\sc i} + Class\,{\sc ii})
\end{table*}

\clearpage
\newpage
\appendix

\section{Reddening Law}

\begin{figure*}[h]
\centering
\includegraphics[width=0.45\textwidth]{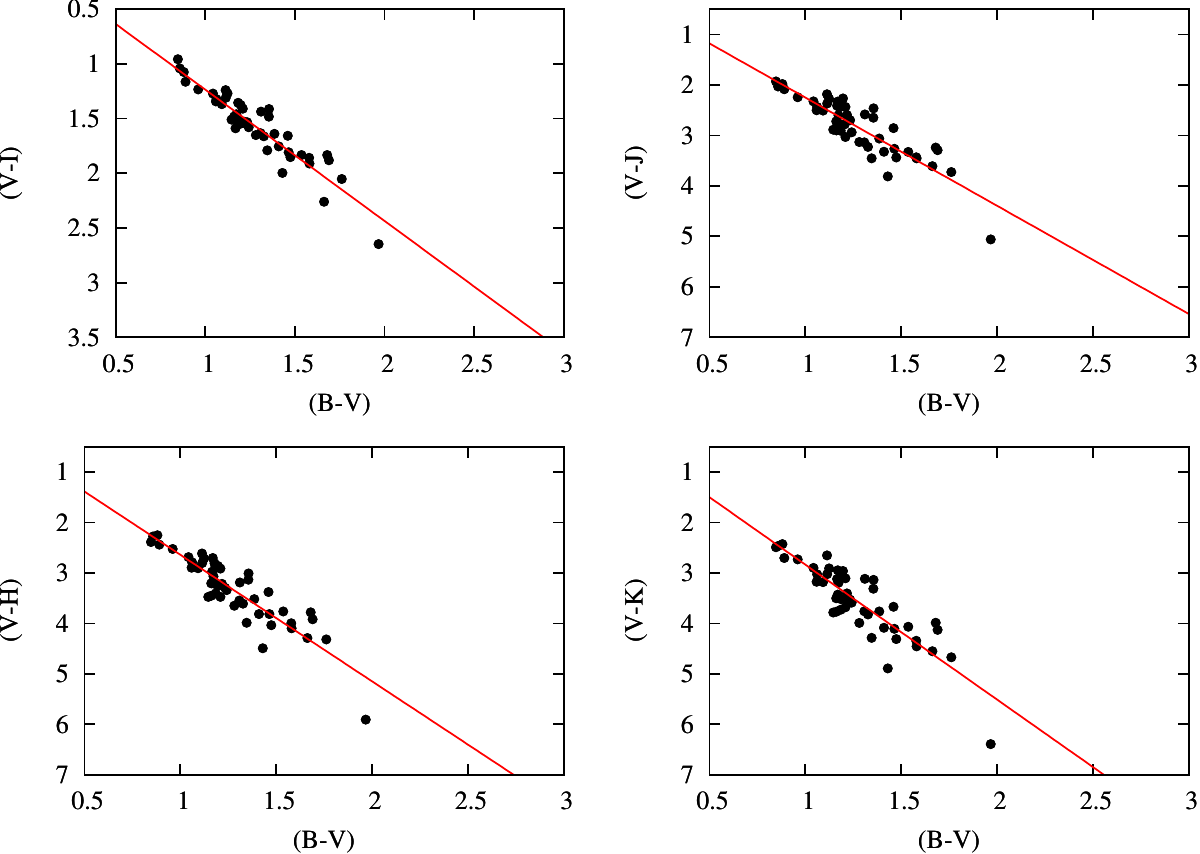}
\caption{\label{2color} $(V-I_c),(V - J), (V -H), (V - K)$ vs. $(B - V)$ TCDs
for the stars associated with the Sh 2-305 region. 
Straight lines show the least-square fit to the distribution of  stars.}
\end{figure*}

We have used the technique as described by \citet{2003AA...397..191P} to study the nature of the
diffuse interstellar medium (ISM) associated with the Sh 2-305 region.
This can be represented by the ratio of total-to-selective extinction $R_V$ = $A_V$/$E(B-V)$.
Though in the solar neighborhood, the normal reddening law gives the 
value $R_V$ = 3.1$\pm$ 0.2 \citep{2003dge..conf.....W, 1989AJ.....98..611G, 2011JKAS...44...39L},
in the case of several SFRs, it is found to be
anomalously high \citep[see e.g.,][]{2000PASJ...52..847P,2008MNRAS.383.1241P,2012AJ....143...41H,2013ApJ...764..172P,2014A&A...567A.109K}. 
The two-color diagrams (TCDs) of the form of $(V - \lambda)$ versus $(B - V)$,
where $\lambda$ indicates one of the wavelengths of the broad-band filters ($R, I, J, H, K, L$), 
provide an effective method for separating the influence of the normal extinction produced by
the general ISM from that of the abnormal extinction arising within regions
having a peculiar distribution of dust sizes \citep[cf.][]{1990A&A...227L...5C, 2000PASJ...52..847P}.
We have selected all member stars of Sh 2-305 
having optical and NIR detections and plotted
their  $(V - \lambda)$ versus $(B - V)$ TCDs in Figure \ref{2color}.
Since YSOs show excess infra-red (IR) emission, their positions can deviate from those of the MS stars in the TCDs, therefore
they  have not been used in the calculation of the reddening law.
The slopes of the least squares fit to the distribution of the MS member 
stars (black dots) in the $(V-I_c),(V-J),(V-H)$ and $(V-K)$ versus $(B-V)$ TCDs are found to be
$1.20\pm0.07, 2.14\pm0.16, 2.51\pm0.18$ and $2.67\pm0.21$, respectively, which are within error similar
to those found for the general ISM \citep[1.10, 1.96, 2.42 and 2.60; cf.,][]{2003AA...397..191P},
indicating normal reddening law `$R_V$ = 3.1' in this region.

\end{document}